\documentclass[numberedappendix,12pt,preprint,appendixfloats]{emulateapj}
\usepackage{longtable}
\usepackage{array}

\shorttitle{MACS J0416.1-2403 from HFF and GLASS}
\shortauthors{Hoag et al. (2016)}

\usepackage{color}			            
\definecolor{midgray}{gray}{0.4}		
\definecolor{orange}{rgb}{1,0.5,0}      

\usepackage[colorlinks=true,citecolor=midgray,linkcolor=midgray]{hyperref}        

\newcommand{\simgt}{\,\rlap{\lower 3.5 pt \hbox{$\mathchar \sim$}} \raise 1pt \hbox {$>$}\,}
\newcommand{\simlt}{\,\rlap{\lower 3.5 pt \hbox{$\mathchar \sim$}} \raise 1pt \hbox {$<$}\,}

\newcommand{\fstar}{$f_{\star}$}
\newcommand{\OIII}{[O\,{\sc iii}]$\lambda\lambda$4959,5007}
\newcommand{\OII}{[O\,{\sc ii}]$\lambda\lambda$3727,3729}

\newcommand{\NimgTOT}{272}   
\newcommand{\NsysTOT}{92}   
\newcommand{\NimgBronze}{58} 
\newcommand{\NimgUSE}{72}   
\newcommand{\NsysUSE}{26}   
\newcommand{\Nsysconfirm}{16}   
\newcommand{\Nimgconfirm}{31}  
\newcommand{\Nsysnew}{6}   
\newcommand{\Nimgnew}{13}   
\newcommand{\NELQthree}{4}     
\newcommand{\NELQfour}{166}      
\newcommand{\HST}{\emph{HST}}
\newcommand{\Spitzer}{\emph{Spitzer}}

\newcommand{\MJ}{MACSJ0416}

\begin{document}


\title{The Grism Lens-Amplified Survey from Space (GLASS). VI. Comparing the Mass and Light in MACSJ0416.1-2403 using Frontier 
Field imaging and GLASS spectroscopy}


\author{
A.~Hoag$^1$, K.~Huang$^1$, T.~Treu$^{2}$,  M.~Brada\v{c}$^1$, K.~B.~Schmidt$^3$, X.~Wang$^{2,4}$, G.~B.~Brammer$^5$, A.Broussard$^6$, R.~Amorin$^7$, M.~Castellano$^7$, A.~Fontana$^7$, E.~Merlin$^7$, T.~Schrabback$^8$, M.~Trenti$^9$, B.~Vulcani$^{9,10}$}
\affil{$^1$ Department of Physics, University of California, Davis, CA, 95616, USA}
\affil{$^2$ Department of Physics and Astronomy, UCLA, Los Angeles, CA, 90095-1547, USA }
\affil{$^3$ Leibniz-Institut fŸr Astrophysik Potsdam (AIP), An der Sternwarte 16, 14482 Potsdam, Germany}
\affil{$^4$ Department of Physics, University of California, Santa Barbara, CA, 93106-9530, USA}
\affil{$^5$ Space Telescope Science Institute, 3700 San Martin Drive, Baltimore, MD, 21218, USA}
\affil{$^6$ Department of Physics and Astronomy, Texas A\&M, College Station, TX 77843, USA}
\affil{$^7$ INAF - Osservatorio Astronomico di Roma Via Frascati 33 - 00040 Monte Porzio Catone, 00040 Rome, Italy}
\affil{$^8$ Argelander-Institut f\"ur Astronomie, Auf dem H\"ugel 71, D-53121 Bonn, Germany}
\affil{$^9$ School of Physics, University of Melbourne, Parkville, Victoria, Australia}
\affil{$^{10}$ Kavli Institute for the Physics and Mathematics of the Universe (WPI), Todai Institutes for Advanced Study, the University of Tokyo, Kashiwa, 277-8582, Japan}

\begin{abstract}
We present a strong and weak gravitational lens model of the galaxy cluster MACSJ0416.1-2403, constrained using spectroscopy from the \emph{Grism Lens-Amplified Survey from Space} (GLASS) and \emph{Hubble Frontier Fields} (HFF) imaging data. We search for emission lines in known multiply imaged sources in the GLASS spectra, obtaining secure spectroscopic redshifts of \Nimgconfirm{} multiple images belonging to \Nsysconfirm{} distinct source galaxies. The GLASS spectra provide the first spectroscopic measurements for \Nsysnew{} of the source galaxies. The weak lensing signal is acquired from 884 galaxies in the F606W HFF image. By combining the weak lensing constraints with 15 multiple image systems with spectroscopic redshifts and 9 multiple image systems with photometric redshifts, we reconstruct the gravitational potential of the cluster on an adaptive grid.
The resulting total mass density map is compared with a stellar mass density map obtained from the deep \Spitzer\ Frontier Fields imaging data to study the relative distribution of stellar and total mass in the cluster. We find that the projected stellar mass to total mass ratio, \fstar{}, varies considerably with the stellar surface mass density. The mean projected stellar mass to total mass ratio is $\langle f_{\star} \rangle= 0.009 \pm 0.003 $ (stat.), but with a systematic error as large as $0.004-0.005$, dominated by the choice of the IMF. We find agreement with several recent measurements of \fstar{} in massive cluster environments. The lensing maps of convergence, shear, and magnification are made available to the broader community in the standard HFF format. 

\end{abstract}



\section{Introduction}
\label{sec:intro}

Gravitational lensing by clusters of galaxies is now a commonplace
tool in astrophysics and cosmology (see \citealp{Treu+Ellis14} for a recent review). The magnification of background objects produced by cluster lenses has been used to find extremely distant and faint galaxies \cite[e.g.][]{Zheng:2012p33879,Coe:2013p26313,Zit++14}. As a result, clusters of galaxies are becoming increasingly popular tools for studying the epoch of reionization. Cluster-scale lensing has also been used to determine the spatial distribution of the total cluster mass, revealing insights into physics of dark matter and structure formation 
\citep[e.g.][]{clowe06,Bra++06,San++08,New++13,Sha++14,2015ApJ...806....4M}. 

The Hubble Frontier Fields (HFF) program (PI Lotz; Lotz et al. in prep.) is imaging
six clusters of galaxies and six parallel fields to
extreme depths in seven optical and near-infrared (NIR) bands, using
the Advanced Camera for Surveys
(ACS) and the Wide Field Camera 3 (WFC3). A principal objective of the HFF initiative is to search for
magnified objects behind the six clusters. Lens models providing accurate magnification maps
are needed to determine the unlensed (intrinsic) properties of the background objects.
Lens models are primarily constrained by multiply imaged galaxies and weakly sheared sources. The added depth 
of the HFF images allows one to identify more multiply imaged galaxies, thus 
increasing the number of constraints and therefore the quality of the lens models. The CLASH\footnote{The Cluster Lensing And Supernova Survey with Hubble (CLASH); \url{ http://www.stsci.edu/~postman/CLASH/Home.html}} images 
\citep[limiting magnitude $\sim$ 27 AB mag for a 5$\sigma$ point
source][]{Postman:2012p27556} revealed $\sim10$ candidate multiple image
systems per cluster \cite[e.g.][]{Zitrin+15}. The release of the HFF images (limiting magnitude $\sim$ 29 AB mag for a 5$\sigma$ point source) has approximately tripled 
the number of known multiply imaged galaxies for the four clusters analyzed so far 
\citep[e.g.][]{Jauzac+15b,Jauzac+14,Diego+15b,Wang+15,Treu+16}.

While the added depth of the HFF imaging has enabled the photometric
identification of a greater number of multiply imaged systems,
the redshifts of the new systems remain uncertain without
spectroscopic follow up. The redshifts must be well-constrained in order to be useful 
for the lens models. It has recently been shown that the number of \textit{spectroscopic} systems has a strong influence on the quality of the lens model \cite[]{Rod++15,Rodney+15b}. Photometric redshifts are useful when spectroscopy 
is lacking, but they can be prone to catastrophic errors, 
especially for sources at or near the limiting magnitude of the image. 
An alternative approach to photometric redshifts is to estimate 
the redshift of new multiply imaged systems using an existing model \cite[e.g.][]{Jauzac+14}. 
This method can potentially introduce confirmation bias in the
modeling process. That is, unless a correct lens model has already been
obtained, the predicted redshifts of the multiply imaged galaxies may be incorrect, 
and the uncertainties may be underestimated. 
Unless decided upon in advance, different approaches to determining redshifts in the absence of spectroscopic
data can lead to different constraints among teams modeling the same cluster.

It is therefore paramount to obtain spectroscopic redshifts for the multiply imaged
systems in the HFF. The Grism Lens-Amplified Survey from Space (GLASS) data for MACSJ0416.1-2403 (\MJ{} hereafter) is one such effort. GLASS\footnote{\url{http://glass.astro.ucla.edu}}\textsuperscript{,}\footnote{\url{https://archive.stsci.edu/prepds/glass/}}
is a large \HST{} program that has obtained deep NIR grism spectroscopy in the fields of ten clusters,
including all six HFF clusters. For details on the observation strategy and data products of GLASS, see \citet{Sch++14} and \citet{Treu+15a}. 

In this paper, we present new spectroscopic redshifts from GLASS and combine them with redshifts from the literature to model the HFF images of \MJ{}. When spectroscopy is not available, we use photometric redshifts tested against GLASS spectra of singly-imaged objects. The paper is organized as follows. In Section~\ref{sec:data} we give
an overview of the optical and NIR HFF and mid-IR \Spitzer\ Frontier Fields imaging data, as well as the GLASS NIR spectroscopic data. In Section~\ref{sec:glass} we
briefly cover the reduction and analysis of the GLASS data. In
Section~\ref{sec:mul} we describe the process of selecting the set of multiple images used to constrain the lens model and 
discuss the GLASS spectroscopic measurements. In Section~\ref{sec:mass} we present our lens
model and compare it to other lens models of the cluster using similar constraints.
In Section~\ref{sec:smass}, we study the relative distribution of stellar and total mass. 
Finally, we summarize our results in Section~\ref{sec:conc}. 
We adopt a standard concordance cosmology with $\Omega_m=0.3$,
$\Omega_{\Lambda}=0.7$ and $h=0.7$. All magnitudes are given in the AB
system \citep{Oke74}.

\section{Data}
\label{sec:data}

Discovered by the MACS survey \citep{Ebeling+01} as a result of
its large X-ray luminosity ($\sim10^{46}$ erg $s^{-1}$;
\citealp{Mann+2012}), \MJ{} was found to likely be a binary head-on
merging cluster system. The first optical and NIR \HST\ imaging data of \MJ{}
were obtained by the CLASH survey. The ESO VIMOS large program CLASH-VLT (186.A-0798; PI: P. Rosati), a spectroscopic campaign designed to obtain thousands of
optical spectra in the CLASH cluster fields with VIMOS on the
Very Large Telescope (VLT), recently found further evidence for the
merging state of the cluster \citep{Balestra+15}. 

Here, we present lens modeling
and spectroscopy results using the deepest optical and NIR imaging and spectroscopy
data newly acquired with \emph{Hubble}, as part of the HFF initiative (\ref{sec:HFF}) and the GLASS program (\ref{sec:glass}), following our study of Abell 2744 in \citet{Wang+15}.
We also use mid-IR imaging data acquired with the \Spitzer\ Space Telescope \citep[\ref{sec:spitz};][]{Werner+04} obtained by the DDT program \#90258 (Capak et al. in prep).

\subsection{Hubble Frontier Fields Imaging}
\label{sec:HFF}

Ultra deep \HST\ images of six clusters (Abell 370, Abell 2744, MACSJ2129, \MJ{}, MACSJ0717, and MACSJ1149) are being obtained as
part of the HFF 
\footnote{\url{http://www.stsci.edu/hst/campaigns/frontier-fields}}. The 5$\sigma$ point source limiting magnitudes are
roughly 29 ABmag in both the ACS/optical (F435W, F606W, F814W) and
WFC3/IR filters (F105W, F125W, F140W, F160W). Observations of \MJ{} were completed in September, 2014.

\subsection{\Spitzer\ Frontier Fields}
\label{sec:spitz}

As a part of the HFF campaign, deep \emph{Spitzer}/IRAC images
in channels 1 and 2 (at $3.6\ \mu$m and $4.5\ \mu$m, respectively) were taken 
through the {\it Spitzer} Frontier Fields
program\footnote{\url{http://ssc.spitzer.caltech.edu/warmmission/scheduling/approvedprograms/ddt/frontier/}}.
In this work, we use the full-depth \emph{Spitzer}/IRAC mosaics for MACS0416 released by Capak et al. (in prep). The IRAC mosaics reach $\sim\!50$~hr
depth per channel in the HFF {\it primary} field and the {\it parallel}
field ($\sim\!6'$ to the west of the primary field). Due to \emph{Spitzer}
roll angle constraints and low-background requirements, six additional
{\it flanking} fields exist around the HFF primary and parallel fields
with uneven coverage in channels 1 and 2. The exposures were drizzled onto a $0\farcs6$~pix$^{-1}$ grid, and within the \emph{HST} primary field footprint there are $\sim\!1800$~frames (with \texttt{FRAMETIME} $\sim\!100$~s) per output pixel. In the primary and parallel fields, the nominal 5$\sigma$ depth of a point source reaches 26.6 mag at 3.6 $\mu$m and 26.0 mag at 4.5 $\mu$m. However, this sensitivity might not be reached near the cluster center due to blending with cluster members and the diffuse intra-cluster light (ICL).

\section{GLASS Observation and Data Reduction}
\label{sec:glass}
GLASS \citep[GO-13459; PI:
Treu;][]{Sch++14,Treu+15a}
observed 10 massive galaxy clusters with the \HST\ WFC3-IR 
G102 and G141 grisms between December 2013 and January 2015. Each of the
clusters targeted by GLASS has deep multi-band \HST\ imaging from the
HFF~(\ref{sec:HFF}) and/or from CLASH. Each cluster is observed at two
position angles (P.A.s) approximately 90 degrees apart to facilitate
deblending and extraction of the spectra. Short exposures are taken
through filters F105W or F140W during every visit to help calibrate the
spectra, model the background, model the contaminating objects, and identify
supernovae by difference imaging. The total exposure time per cluster
is 14 orbits, distributed as to provide approximately uniform
sensitivity across the entire wavelength coverage 0.8$\mu$m to
1.7$\mu$m. Parallel observations are taken with the ACS F814W direct image and
G800L grism to map the cluster infall regions. Here we focus on the NIR data on the \MJ\ cluster core.

The two P.A.s of GLASS data analyzed here were 
taken on November 23 and 30, 2014 (P.A. = 164$^{\circ}$)   
and January 13 and 18, 2015 (P.A. = 247$^{\circ}$), respectively. 
The resulting total exposure times for the individual grism observations are 
shown in Table~\ref{tab:glassobs}. Prior to reducing the complete GLASS data, He Earth-glow is
removed from individual exposures \citep{Brammer:2014p34990}. 
\begin{deluxetable}{lccc}
\tablecolumns{3}
\tablewidth{0pt}
\tablecaption{GLASS Grism and Imaging Exposure times for \MJ{} \label{tab:glassobs}}
\tablehead{
  \colhead{Filter} &
  \colhead{P.A./(deg.)} &
  \colhead{$\mathrm{t_{exp.}}$/(s)}
} 
\startdata
G102 & 164 & 10929 \\ 
G102 & 247 & 10929 \\ 
G141 & 164 & 4312 \\ 
G141 & 247 & 4312 \\ 
F105W & 164 & 1068 \\ 
F105W & 247 & 1068 \\
F140W & 164 & 1573 \\ 
F140W & 247 & 1423 \\ 
\enddata
\tablecomments{Exposure times are for the cluster core of \MJ{} only. }
\end{deluxetable}

\begin{figure*}
\centering
\includegraphics[width=\textwidth]{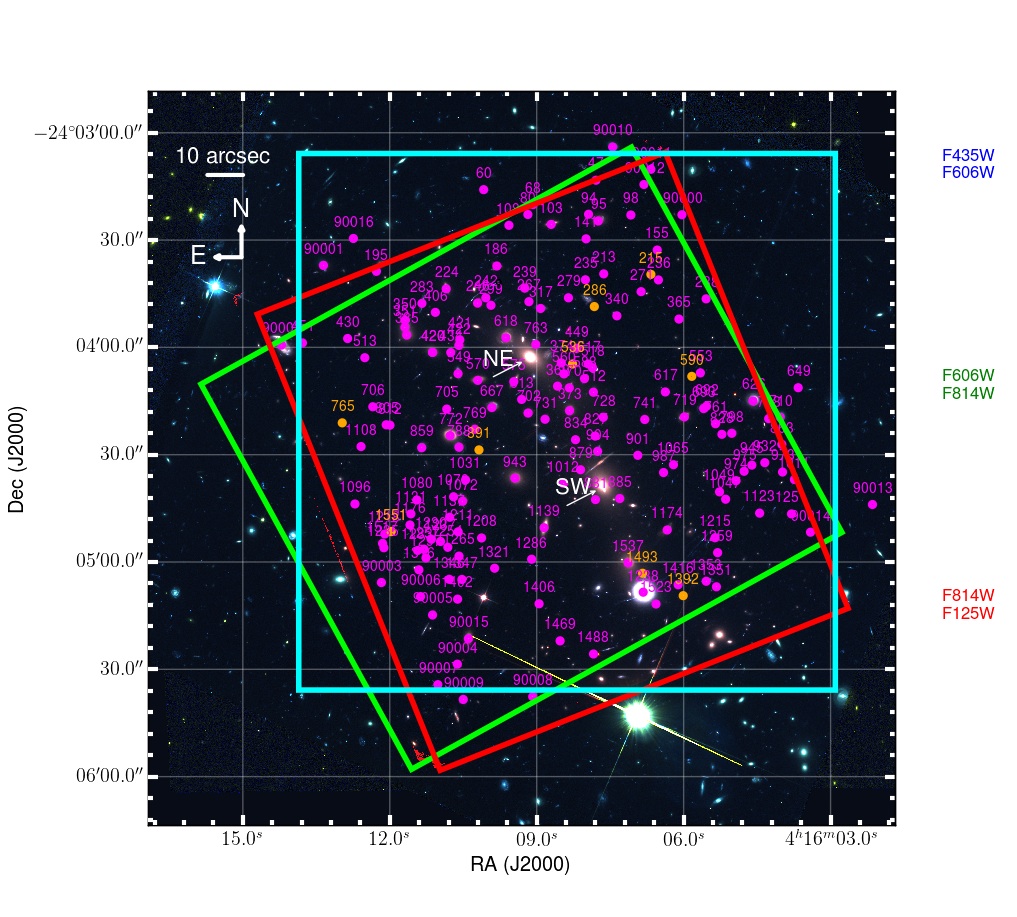}
\caption{Color composite image of \MJ{} based on HFF (Lotz et al. in prep.), CLASH (\citealp{Postman:2012p27556}) and GLASS \citep[]{Sch++14,Treu+15a} imaging. The blue, green, and red channels are composed by the labeled filters on the right. The white arrows point to the two BCGs (NE and SW) which are separated by $\Delta V_{\mathrm{r.f.}} \sim 900\,\mathrm{km\,s^{-1}}$ \citep{Balestra+15}, and are likely in a merging state \citep{Mann+2012,Jauzac+15,Ogrean+15}. The two distinct P.A.s of the spectroscopic GLASS pointings are shown by the green (P.A.=164$^\circ$) and red (P.A.=247$^\circ$) squares. The cyan square outlines the area show in Figure~\ref{fig:arcs_image}. The locations 
of the emission line objects from Table~\ref{tab:ELtot} are marked by circles, with color coding reflecting the GLASS 
spectroscopic redshift quality (cf. ``Quality'' in Table~\ref{tab:ELtot}; 
3=orange, 4=magenta) and labels given by GLASS IDs. There are a few objects that fall outside of the direct image FOV. Their spectra were dispersed onto the chip, so they were still able to be extracted and analyzed (see Figure~\ref{fig:grisms}). }
\label{fig:image}
\end{figure*}

\begin{figure*}
\includegraphics[width=0.49\textwidth]{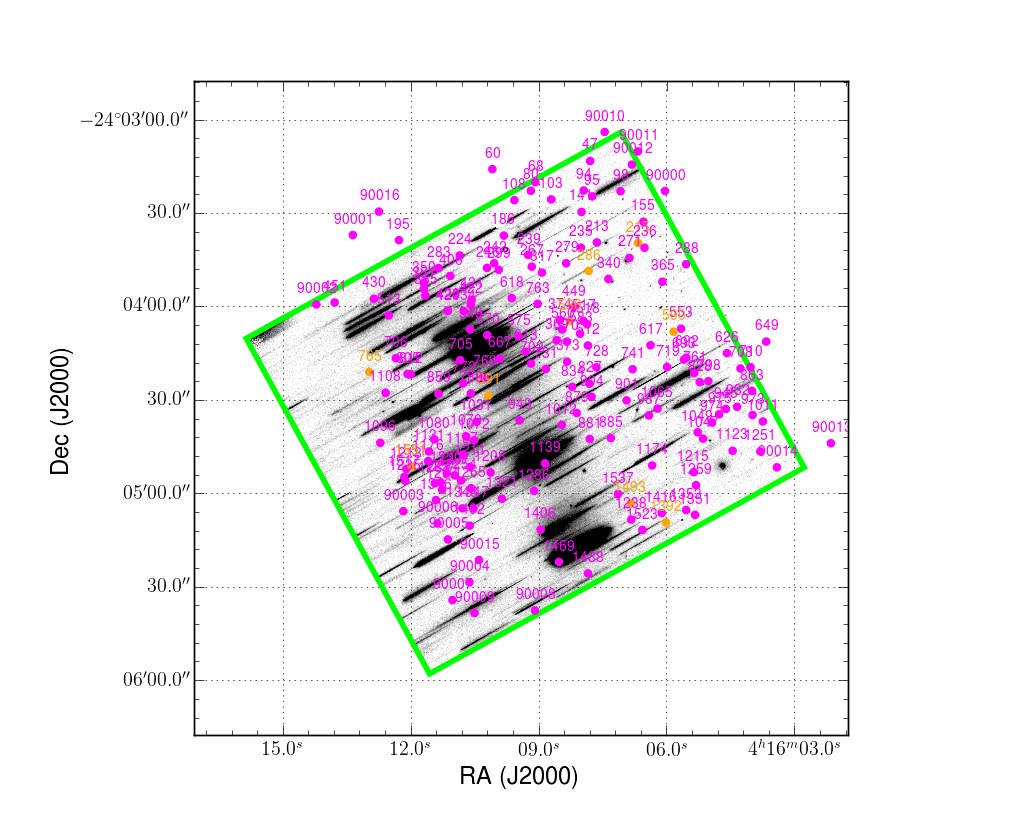}
\includegraphics[width=0.49\textwidth]{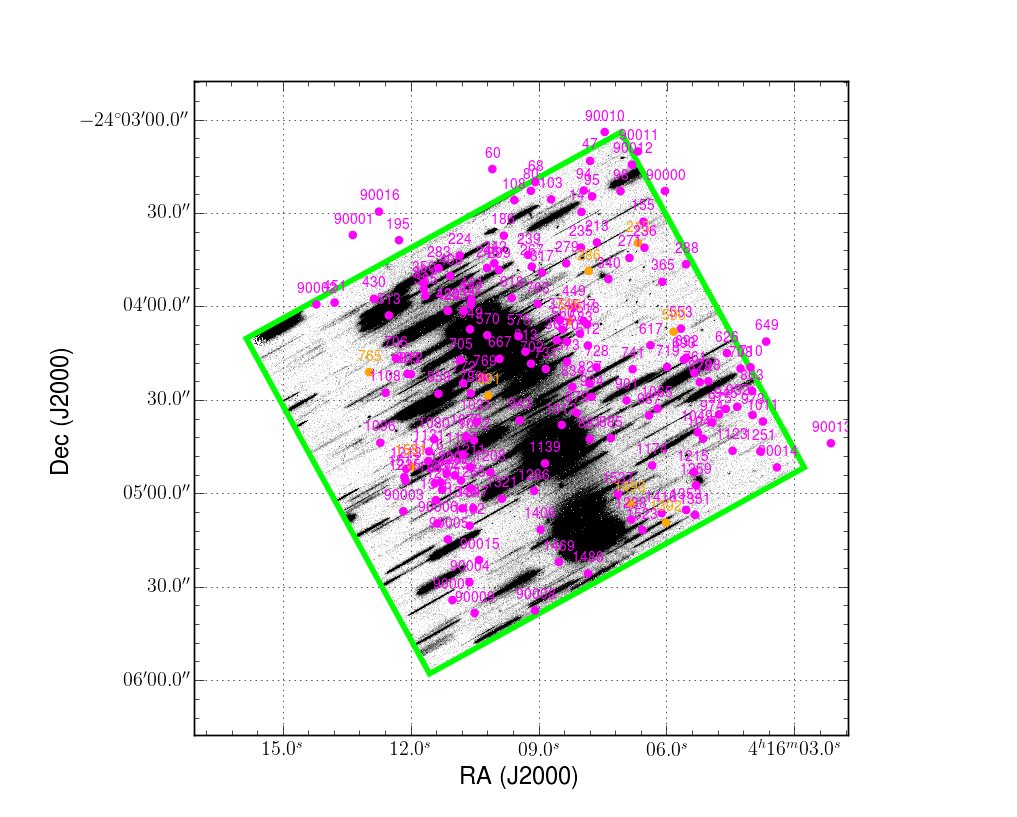}\\
\includegraphics[width=0.49\textwidth]{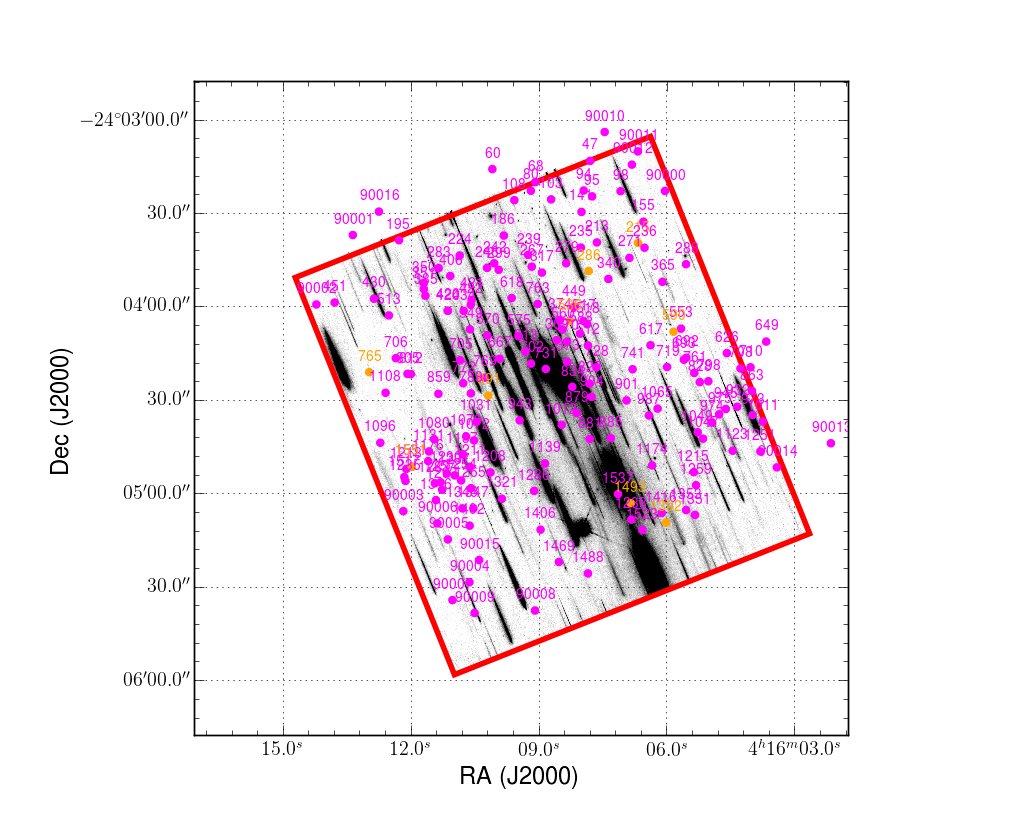}
\includegraphics[width=0.49\textwidth]{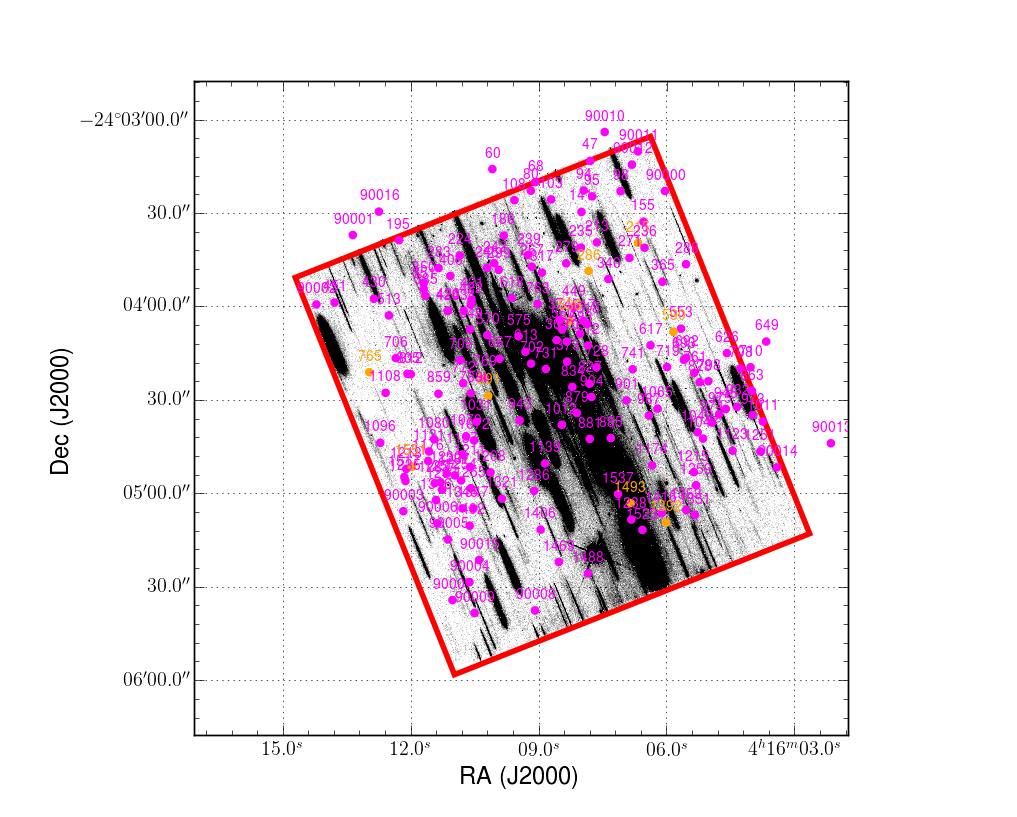}\\
\caption{The GLASS G102 (left) and G141 (right) grism pointings of \MJ{} at two distinct P.A.s, with field-of-view shown by the 
green (P.A.=164$^\circ$) and red (P.A.=247$^\circ$) squares. The circles in all panels denote the positions of the emission line objects identified in this work (Table~\ref{tab:ELtot}), with color coding and labels following the conventions adopted in Figure~\ref{fig:image}. The circles that fall outside of the grism pointings are identified in HST images with a larger FOV than the individual grism FOVs. These objects can still be observed in the grism data because their first order spectra are dispersed onto the chip. }
\label{fig:grisms}
\end{figure*}

The GLASS observations are designed to follow the 3D-HST observing strategy \citep{Brammer:2012p12977} 
and were processed with an updated version of the 3D-HST reduction pipeline\footnote{\url{http://code.google.com/p/threedhst/}} (\citealp{Momcheva+15}).
Below we summarize the main steps in the reduction process of the GLASS data but refer to 
\cite{Brammer:2012p12977} and \cite{Momcheva+15} and the GLASS survey paper \citep{Treu+15a} for further details. The GLASS data were taken in a 4-point dither pattern identical to the one
shown in Figure~3 of \cite{Brammer:2012p12977} to reduce susceptibility to bad pixels and cosmic rays
and to improve sub-pixel sampling of the WFC3 point spread function.
At each dither position, a direct and a grism exposure were taken.
The direct images are commonly taken in the filter with passband overlapping in wavelength with the grism, i.e., F105W for G102 and F140W for G141.  However, to accommodate searches for supernovae and the characterization of their light curves in GLASS clusters, each individual visit is designed 
to have imaging in both filters. Hence several pairs of F140W+G102 observations exist in the GLASS data. 
This does not affect the reduction or the extraction of the individual GLASS spectra.

The individual exposures were combined into mosaics using AstroDrizzle from the DrizzlePac software package \citep{Gonzaga:2014p26307}. All  
direct image exposures were aligned using \verb+tweakreg+, with background subtracted from the exposures by fitting a second  
order polynomial to each of the source-subtracted exposures. We subtracted the background of the grism exposures using the master sky images and algorithm presented by \cite{Brammer:2015}. The individual sky-subtracted exposures were combined using a pixel scale of $0\farcs06$ per  
pixel ($\sim$half a native WFC3 pixel). In Figure~\ref{fig:image} we show a color composite image of \MJ{},
using the optical and NIR coadded imaging from HFF and CLASH combined with the NIR imaging from GLASS.
The green (red) square shows the GLASS footprint for P.A. = 164$^{\circ}$ (P.A. = 247$^{\circ}$). 
The fiducial cluster redshift used by the HFF lens modeling team and adopted throughout the paper is $z_{\mathrm{cluster}}=0.396$. The two brightest cluster galaxies (BCGs) (NE and SW) are labeled in the figure. The NE BCG and the SW BCG are at redshifts $z=0.395$ and $z=0.400$, respectively (\citealp{Balestra+15}). Figure~\ref{fig:grisms} shows the full field-of-view 
mosaics of the two NIR grisms (G102 on the left and G141 on the right) at the two GLASS position angles for \MJ{}.  The contamination model was computed and one- and two-dimensional spectra were extracted from 2$\times$2 ``interlaced'' versions of the grism mosaics following the procedure outlined in detail by \cite{Momcheva+15}, where the source positions and extent were determined with the \verb+SExtractor+ \citep{Bertin:1996p12964} software run on the corresponding direct image mosaics.

\section{Identification of multiple images}
\label{sec:mul}

In this section we describe how we identify and vet multiple image
candidates using the HFF imaging (\ref{subsec:photometry}) and GLASS spectroscopic
(\ref{subsec:targeted} and \ref{subsec:visual}) data.

\subsection{Imaging data: identification and photometric redshifts}
\label{subsec:photometry}

\begin{figure*}
\begin{center}
\includegraphics[width=0.75\textwidth]{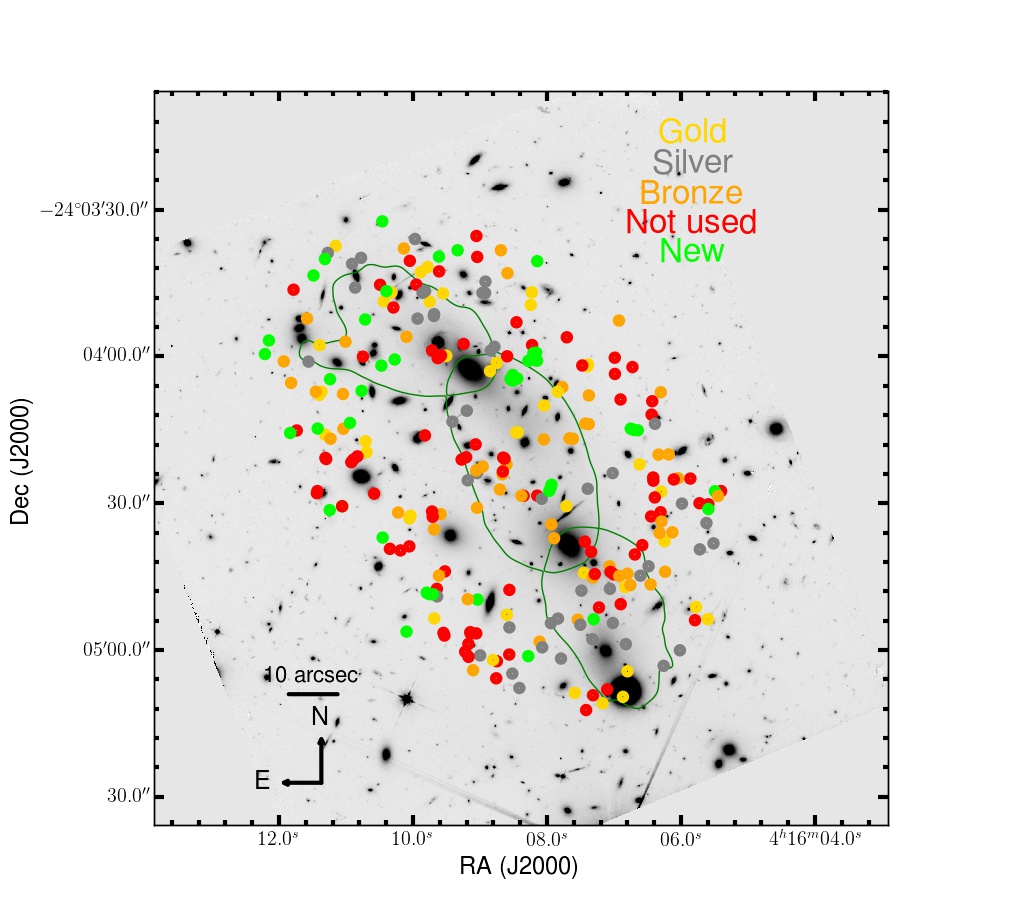}\\
\caption{All multiple images discovered to date in \MJ{}. As indicated in the key: gold, gray and orange colored circles correspond to the multiple images in the Gold, Silver and Bronze samples, respectively (Section~\ref{sec:mul}). Red circles correspond to multiple images which the HFF modeling teams deemed were less secure than those in the bronze sample so were not used in the lens models. Green circles represent new multiple image candidates discovered after the grading effort took place. The dark green line is the critical curve from our best-fit lens model at $z=2.36$, the mean source redshift of the multiple images that were used in our lens model. Shown is the co-added CLASH+HFF+GLASS F105W image.}
\label{fig:arcs_image}
\end{center}
\end{figure*}

\citet{Zitrin+13} published the first detailed strong lensing analysis of \MJ{}. They identified over 70 multiple 
images (23 source galaxies) in the CLASH imaging data, making \MJ\ the most prolific CLASH cluster in terms of multiply imaged galaxies (\citealp{Zitrin+15}). With the addition of the much deeper HFF data, 
{\NimgTOT} multiple images ({\NsysTOT} source galaxies) have been identified in \MJ\ \citep{Jauzac+14,Diego+15,Kawamata+15}. We list all of the multiple images in Table~\ref{tab:mult_images} and show their positions behind the cluster in Figure~\ref{fig:arcs_image}.

Several efforts have been made to spectroscopically confirm the redshifts of the strongly-lensed galaxy candidates in \MJ{} \citep[Rodney et al. in prep.]{Zitrin+13,Jauzac+14,Richard+14,Johnson+14,Grillo+15,Balestra+15}. Images belonging to systems 1, 2, 3, 4, 5, 7, 10, 13, 14, 16, 17, 23, 26, and 28 have been targeted by multiple authors, leading to agreement in the spectroscopic redshift within the uncertainties, with the exception of system 14. This system will be discussed in more detail in Section~\ref{subsec:targeted}. When spectroscopy is lacking, confirming that images belong to the same source is more difficult. Inspired by a previous collaborative effort to model the HFF clusters as well as the recent rigorous vetting procedure developed by \citet{Wang+15}, a new collaborative effort was undertaken by seven teams simultaneously modeling the HFF galaxy clusters to assign quality grades to the multiple image candidates. The grading was done independently by each of the seven teams and focused on ensuring consistency of the morphologies and colors of multiple images of the same source galaxy. Each team assigned each multiple image candidate a grade on a scale from 1-4, where 1 meant secure and 4 meant untrustworthy. The results from the grading process were divided into the categories: ``Gold,'' ``Silver,'' and ``Bronze.'' The Gold category was reserved for spectroscopically confirmed multiple images that also received an average grade of $< 1.5$. The multiple images confirmed using GLASS spectroscopy that are presented in this work were included in the Gold sample. The Silver category corresponded to images lacking spectra but receiving a unanimous vote of 1. Bronze was assigned to images lacking spectra and received an average grade of $> 1$ but $ < 1.5$. Images lacking spectra and receiving an average grade of $>=1.5$ were not used by the modeling teams. 
 
For our lens model, we decided to only consider multiple images in the Gold and Silver categories, which included 80/31 total multiple images/systems. In order to use a Silver multiple image in the lens model, it was necessary to estimate its redshift. Because the grading was done via visual inspection, it was not guaranteed that every multiple image would be detected in our photometric catalog. Of the 31 systems comprising Gold and/or Silver images, only \NsysUSE\ were included in our lens model. Five systems (systems 8, 33, 40, 41, and 51, all are graded Silver) are not included in our model due to problems with photometry and redshift. Neither of the two Silver images in system 8 was detected in the photometric catalog, most likely due to their faintness and proximity to bright cluster members. Systems 33, 40, and 51 had poorly constrained photometric redshift PDFs. Only a single image in system 41 was detected in the photometric catalog, and its redshift was too poorly constrained to use in the lens model. The redshift PDFs for the multiple images in systems 33, 40, 41, and 51 are shown in Appendix~\ref{sec:apA}. For the remaining Silver images, it was necessary to compute photometric redshifts to include them in the lens model. For these systems we used photometric redshifts obtained by the ASTRODEEP team (Castellano et al., 2015, submitted; Merlin et al. 2015, submitted). The ASTRODEEP photometric redshifts were obtained through $\chi^2$ minimization over the observed \HST{}+IR HFF bands using PEGASE 2.0 \citep{Fioc+97}. For more details see Castellano et al., (2015, submitted) and Merlin et al. (2015, submitted). The ASTRODEEP catalogs were built after subtracting ICL emission and the brightest foreground galaxies from the images in order to maximize the efficiency of high-redshift source detection and to obtain unbiased photometry.  The catalogs employed the seven HFF filters: F435W, F606W, F814W, F105W, F125W, F140W, and F160W, a HAWK-I $K$-band image (G. Brammer, in prep), and \Spitzer{} IRAC [3.6] and [4.5] channels. The ASTRODEEP catalogs used two different detection images to create separate photometric catalogs. One catalog used F160W as the detection image, whereas the other used a stacked NIR image. To maximize the number of multiply imaged galaxies that we could detect, we merged the two catalogs into a single photometric catalog for this work. 

Combining the photometric redshift information for multiple images of the same source provides a tighter constraint
than a single measurement. We used a hierarchical Bayesian method similar to that used in \citet{Wang+15} (see also
\citealt{1997upa..conf...49P} and \citealt{Dahlen:2013p33380}) to combine the multiple photometric redshift probability
density functions for each image of a source $P_i(z)$ into one $P(z)$ for each source. We used the peak of the combined $P(z)$ (hereafter referred 
to as $z_{\textrm{Bayes}}$) as input to the lens model. In summary, the hierarchical Bayesian method considers the concept 
of the probability that each input $P_i(z)$ is unreliable ($p_{\mathrm{bad}}$). It uses an input $P_i(z)$ in the calculation
of the combined $P(z)$ for the system if it is reliable, otherwise it uses a flat (noninformative) $P_i(z)$.
The method then marginalizes over all values of $p_{\mathrm{bad}}$ using an assumed prior on $p_{\mathrm{bad}}$
to calculate the posterior $P(z)$ for the entire system. We assume a flat prior in $p_{\mathrm{bad}}$ for
$p_{\mathrm{bad}} \leq 0.5$, i.e., that each $P(z)$ has at least 50\% chance of being informative. This has the effect that
for some systems with two images, the posterior $P(z)$ of the system has a small but non-zero floor due to the contribution
from a noninformative $P_i(z)$. Because the floor inflates the photometric redshift uncertainty, we subtract the floor from
all posterior $P(z)$'s before calculating confidence intervals. Subtracting the floor does not change the peak redshift of the posterior
$P(z)$.

\subsection{GLASS spectroscopy}
\label{subsec:targeted}

The GLASS spectroscopic data were carefully examined for a total of \NimgTOT{}
multiple image candidates in the attempt to measure spectroscopic redshifts. Each spectrum was visually inspected by multiple investigators
(A.H., T.T., and A.B.) using the GLASS Graphical User Interfaces (GUIs) dubbed the GLASS
Inspection GUI (GiG) and GLASS Inspection GUI for redshifts (GiGz\footnote{Available at \url{https://github.com/kasperschmidt/GLASSinspectionGUIs}}; \citealp{Treu+15a}). Both P.A.s were inspected individually and then again once stacked together. The
results were then combined to form a list of multiple images with
identified emission lines.  Following GLASS procedure \citep{Sch++15,Treu+15a}, a quality flag
was given to the redshift measurement: Q=4 is secure; Q=3 is probable;
Q=2 is possible; Q=1 is likely an artifact. As described in \citet{Treu+15a},
these quality criteria take into account the signal to noise ratio of
the detection, the probability that the line is a contaminant, and the
identification of the feature with a specific emission line. For
example, Q=4 is given for spectra where multiple emission lines are
robustly detected; Q=3 is given for spectra where either a single
strong emission line is robustly detected and the redshift
identification is supported by the photometric redshift, or when more
than one feature is marginally detected; Q=2 is given for a single
line detection of marginal quality. As shown in Table~\ref{tab:ELtot},
new spectroscopic redshifts were obtained for Quality 4 and 3 measurements only, consisting of
\Nimgconfirm{} images in total, corresponding to \Nsysconfirm{} systems. 
The spectra of these objects are shown in Figs.~\ref{fig:ELarc2.1}--\ref{fig:ELarc27.2}. 

The uncertainty in our spectroscopic redshift measurements is limited by the grism 
wavelength resolution of approximately 50\AA\ and by uncertainties in the zero point of the
wavelength calibration. By comparing multiple observations of the same
object we estimate the uncertainty of our measurements to be in the order of
$\Delta z\lesssim 0.01$, similar to what was reported by \citet{Wang+15}. 

Some multiple images that we confirm with GLASS spectroscopy were previously spectroscopically confirmed by other authors. We compare the redshifts obtained in this work to those previously obtained, seeing agreement within the uncertainties for all multiple images except system 14. Images 14.1, 14.2, and 14.3 were originally reported to be spectroscopically confirmed at $z=2.0531$ by \citet{Richard+14}, using incomplete CLASH-VLT data. \citet{Grillo+15}, using the complete CLASH-VLT data set for \MJ{}, recently reported redshifts of $z=1.6370$ for images 14.1 and 14.2. We confirm the updated redshift measurements of 14.1 and 14.2 by \citet{Grillo+15} by identifying strong \OII{} and \OIII{}+H$\beta$ emission at $z_{\mathrm{grism}}=1.63 \pm 0.01$ in the GLASS spectra of both images. In addition, we confirm 14.3 at $z_{\mathrm{grism}}=1.63 \pm 0.01$ by identifying the same lines observed in the grism spectra of 14.1 and 14.2 (see Figures~\ref{fig:ELarc14.1}--\ref{fig:ELarc14.3}). We note that all v1 lens models of the cluster that used system 14 used the incorrect spectroscopic redshift of $z=2.0531$.

Using the GLASS spectroscopy, we confirm the redshifts of \Nimgnew{} multiple images for the first time. These are multiple images 5.1, 5.2, 5.3, 5.4, 12.3, 15.1, 23.1, 23.3, 26.1, 27.2, 28.1, 28.2, and 29.3 in Table~\ref{tab:mult_images}. We show the GLASS spectra for these objects in Appendix~\ref{sec:apB}, with the exception of 12.3, 28.1 and 28.2 which are available in an online grism catalog\footnote{Available at \url{http://www.stsci.edu/~brammer/GLASS_zcat/}}. All spectra are also available in an online catalog\footnotemark[8]. Before the GLASS data were analyzed, none of the images in system 5 had published spectroscopic redshifts. We confirm all four images in the system as belonging to the same source galaxy to $z=2.09\pm0.01$ based on the detection of \OIII{} in both P.A.s of the G141 spectra of all four images. H$\beta$ and \OII{} were also detected in 5.1, 5.2, and 5.3, consistent with $z=2.09\pm0.01$. The CLASH-VLT team published a consistent redshift of $z=2.092$ (\citealp{Balestra+15}) for image 5.2 after the HFF modeling team determined its samples using the GLASS spectra. 12.3 was the only multiple image candidate inspected with significant continuum emission in the GLASS spectra, but no emission lines were apparent. It was confirmed to $z=1.96 \pm 0.02$ by fitting its bright continuum emission in the GLASS spectra to template SEDs using the method described by G. Brammer, (in prep.). 15.1 was confirmed to $z=2.34$ by the detection of \OIII{} in both P.A.s of the G141 spectra. 23.1 and 23.3 were confirmed to $z=2.09\pm0.01$ by the detection of strong \OIII{} in both P.A.s of the G141 spectra. We confirmed 26.1 to $z=2.18\pm0.01$ by the detection of \OII{} and \OIII{} in both P.A.s of the G141 spectra. We detected \OIII{} at $z=2.11\pm0.01$ in both P.A.s of the G141 spectra of image 27.2. 28.1 and 28.2 were blended in the segmentation map used to extract grism objects. Like multiple image 12.3, 28.1 and 28.2 were confirmed by fitting the bright continuum emission in the GLASS spectra to template SEDs, finding a best-fit redshift of $0.938_{-0.002}^{+0.001}$. 29.3 was confirmed by detecting \OIII{} at $z=2.28\pm0.01$ in both G141 P.A.s. The catalog of multiply imaged objects with GLASS redshifts is publicly available\footnotemark[3]\textsuperscript{,}\footnotemark[8]
\subsection{Visual search in GLASS data}
\label{subsec:visual}

We also conducted a search for line emitters among known sources within
the entire grism field-of-view. Two co-authors (A.B. and T.T.) visually inspected all of the 2D grism
spectroscopic data products, the contamination models, and residuals after
contamination subtraction. The grism spectra extraction is based on a public
HFF photometric catalog available on the STScI MAST archive\footnote{\url{https://archive.stsci.edu/prepds/frontier/}},
supplemented by our own photometric catalogs based on CLASH images and GLASS direct images. 
We attempted to identify new multiple systems among the galaxies with 
the same grism spectroscopic redshifts but did not find any. Some of them are ruled out
because of the relative positions of the multiple images in the cluster, while others are ruled out
because their distinct colors and morphologies are inconsistent with being the same source.

\begin{figure}[ht]
\includegraphics[width=\columnwidth]{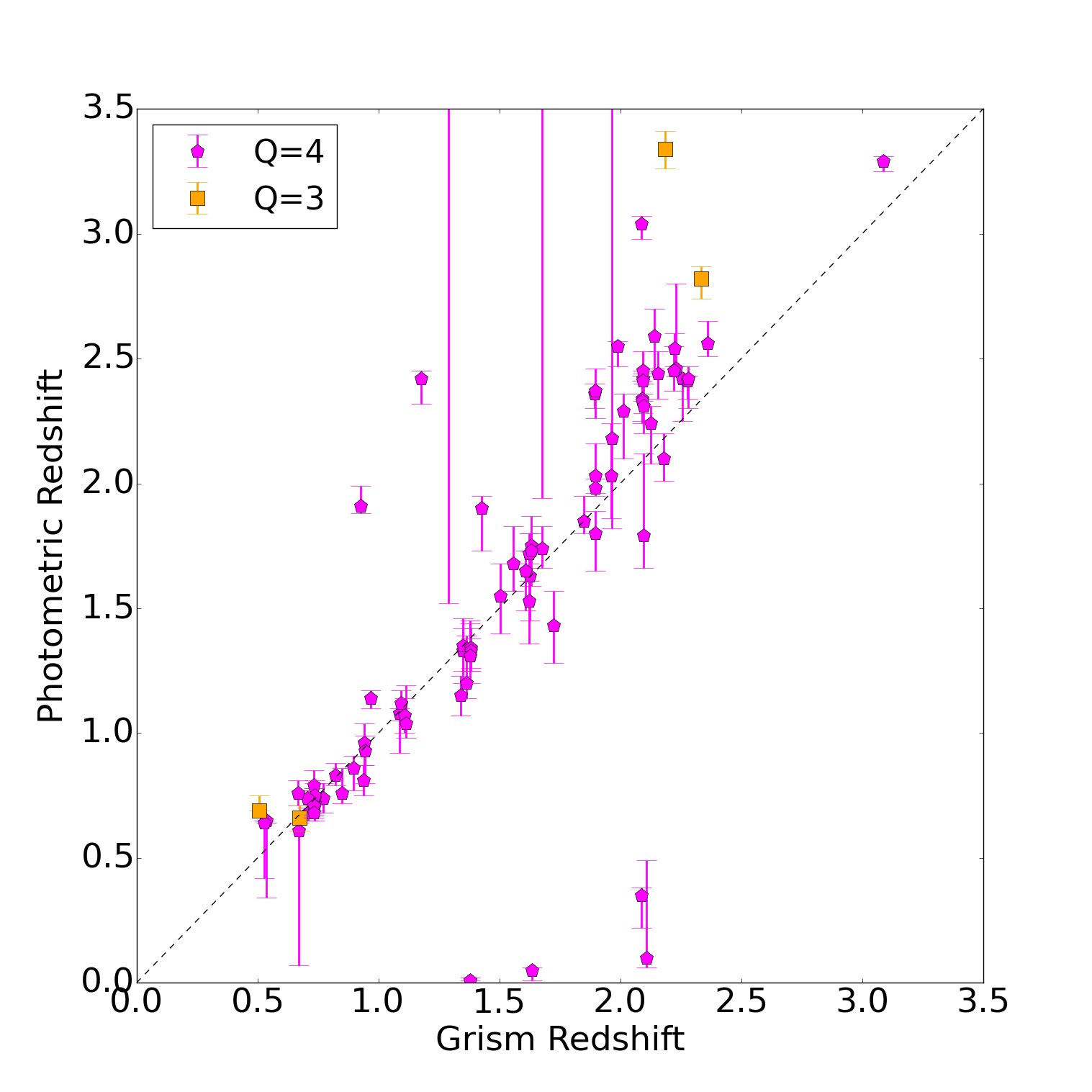}
\figcaption{
Comparison between the grism and photometric redshifts for the
86 objects with high-confidence emission lines (quality flags 4 (magenta pentagons) or 3 (orange squares)) for which photometric redshifts could be measured.
Error bars on the photometric redshifts are 1$\sigma$ (enclosing 68\% of the total
probability). There is 
good statistical agreement between photometric and grism redshifts, with
57/86 (66\%) of the grism redshifts within the photometric redshift
error bars. This suggests that the photometric redshift errors are well-estimated.
The dotted black line is shown for reference and represents
perfect agreement between photometric and grism redshifts.   
\label{fig:ELobjs_photz}}
\end{figure}

We compiled a list of singly-imaged galaxies with redshifts determined from both emission-line and absorption features, consisting of \NELQthree{}, and \NELQfour{}, quality 3 and 4 spectroscopic redshift measurements respectively,
which are color coded in Figure~\ref{fig:image} and Figure~\ref{fig:grisms} and listed in
Table~\ref{tab:ELtot}. Using the photometric catalog described above, we measure the photometric redshifts of these objects and compare them with the spectroscopic redshifts. A comparison is
displayed in Figure~\ref{fig:ELobjs_photz}. Of the objects for which photometric redshifts could be measured, we find that approximately 66\% (57/86) of the 
photometric redshifts agree within their $68\%$ confidence limits uncertainties with the
corresponding spectroscopic redshifts. This suggests that the photometric redshift errors are reliably estimated. 

\section{Gravitational Lens Model}
\label{sec:mass}

Our lens modeling method, SWUnited \citep{bradac05,bradac09}, constrains the gravitational potential of a
galaxy cluster via an iterative $\chi^2$ minimization algorithm. It takes as input a
simple initial model for the potential. After each iteration, a $\chi^2$ is calculated
from strong and weak gravitational lensing data on an adaptive,
pixelated grid. The number of grid points is increased at each iteration, and the $\chi^2$ is
recalculated. Once the minimum is found, and convergence is achieved,
derivative lensing quantity maps, such as convergence ($\kappa$),
shear ($\gamma$) and magnification ($\mu$), are produced from the
best-fit potential map.

The strong lensing constraints on the lens model are described in (\ref{sec:mul}). 
Our weak lensing catalog is based on the ACS/WFC F606W observations ($\sim30ks$)
of the cluster from the HFF program. Using the HFF image results in a factor of 
two increase in the source density of weak lensing galaxies compared to the CLASH imaging.
For the reduction and weak lensing catalog generation we make use of the pipeline
described by
\citet{schrabback+10}, which employs the KSB+ formalism \citep{kaiser1995,luppino1997,hoekstra1998} for galaxy
shape measurements as detailed by \citet{erben2001} and \citet{schrabback+07}.
A major difference compared to \citet{schrabback+10}
is the application of the pixel-based correction for charge-transfer
inefficiency (CTI) developed by \citet{massey+14}.
Our modeling of the temporally and spatially variable ACS point-spread function is based
on the principal component analysis developed by \citet{schrabback+10}, which we 
recalibrate using F606W stellar field observations taken after Servicing
Mission 4.
Further details on the recent pipeline modifications are provided by Schrabback et al. (in
prep.), including a new verification test
for the CTI correction
in the context
of cluster weak lensing studies and updates for the employed weak lensing weighting scheme. 
Weak lensing galaxies were individually assigned a redshift using the `ZBEST' keyword in the ASTRODEEP photometric catalog (see section~\ref{subsec:photometry}), or the CLASH photometric catalogs 
for galaxies outside of the HFF \HST{}+IR bands. The CLASH photometric catalogs
use the Bayesian Photometric Redshifts (BPZ) code \citep{bpz}. We use the `$z_b$,' the most likely redshift given by BPZ, to estimate the
redshifts of the weak lensing galaxies. We only use galaxies with $z_b > z_{\mathrm{cluster}} + 0.2$ to ensure that the majority of the catalog contains background galaxies. 

\begin{figure*}[ht]
\includegraphics[width=0.49\textwidth]{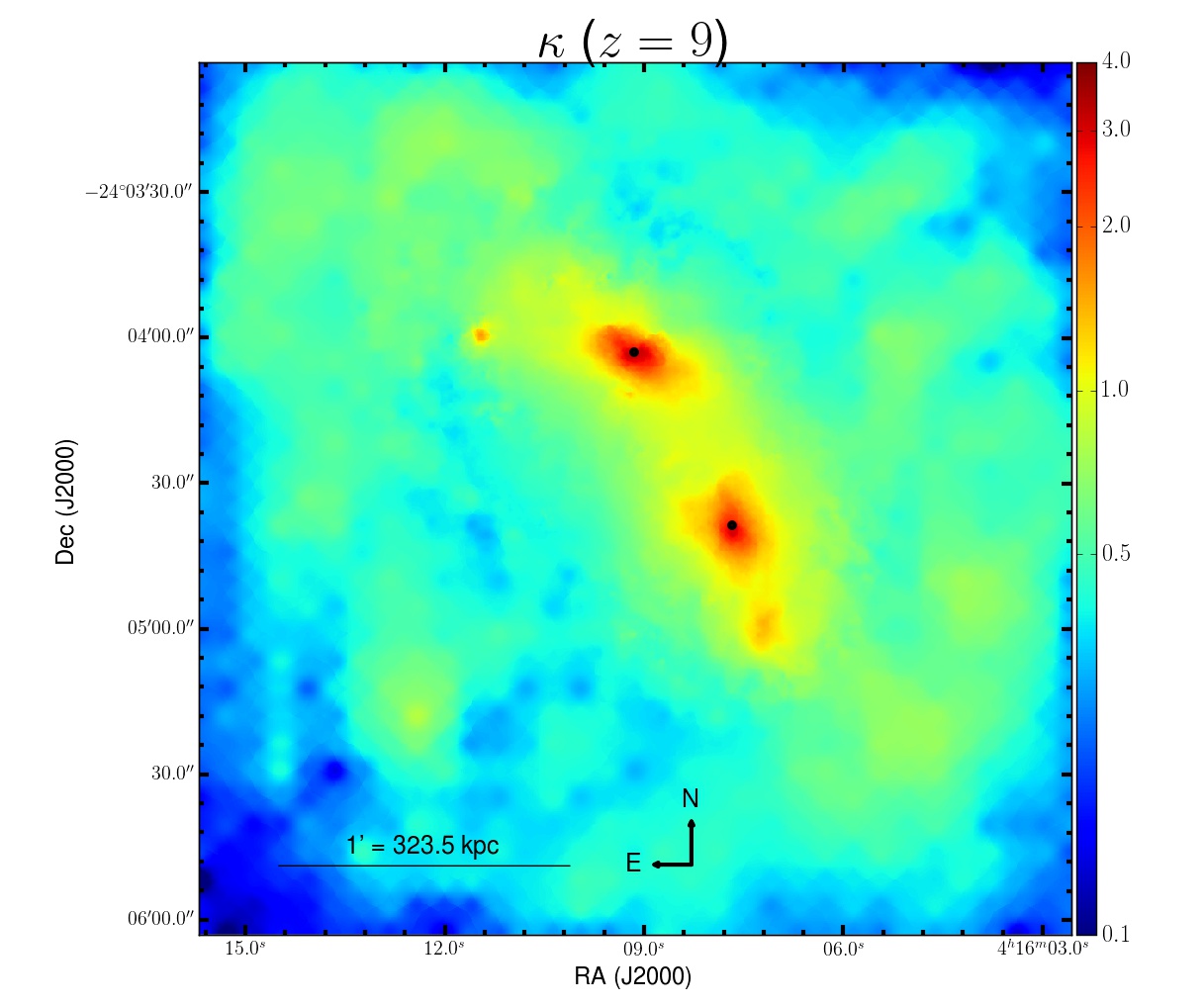}
\includegraphics[width=0.49\textwidth]{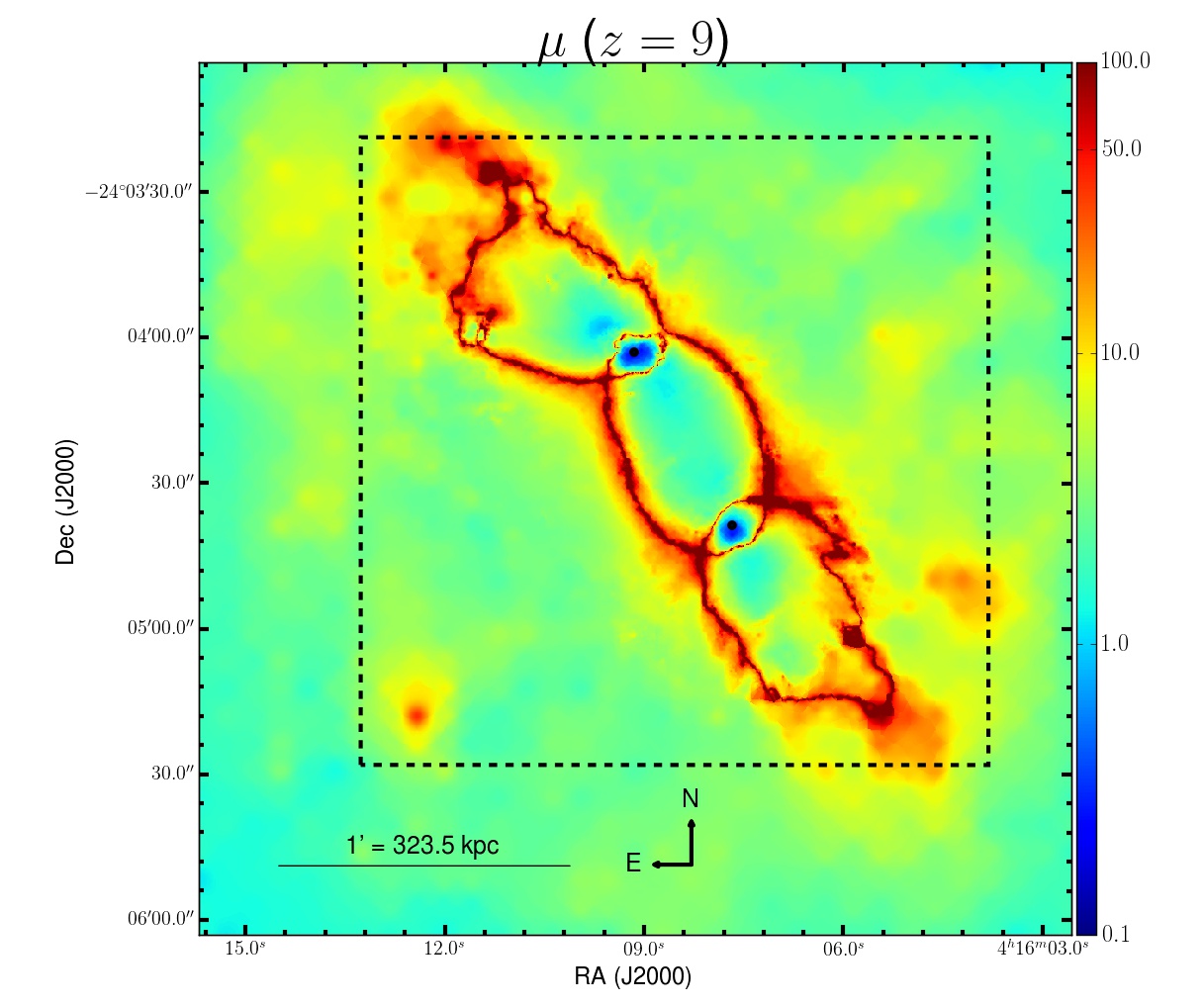} \\
\figcaption{\textbf{Left} - Convergence ($\kappa$) map of \MJ{} produced by our lens model for a source at $z_{\mathrm{s}}=9$. The convergence map reveals two primary total mass density peaks, centered approximately at the location of the two BCGs, which are marked by the two black points. \textbf{Right} - Flux magnification ($\mu$) map of \MJ{} produced by our lens model for a source at $z=9$. The approximate location of the critical curve, the curve along which magnification is maximized, can be seen. The dotted black square outlines the common area over which magnification maps were produced by all collaborative HFF modeling teams (see~\ref{subsec:compare}; Figure~\ref{fig:area}). Both the convergence and the magnification maps reveal a highly elliptical total mass distribution, as found by several other authors \cite[e.g.][]{Zitrin+13,Jauzac+14}. Both maps cover the same $3.0 \times 3.0 \, \textrm{arcmin}^2$ footprint. The two black points on each map mark the centers of the two BCGs determined from the F105W image. Note that the colorbar is log-scale. }
\label{fig:lens_maps}
\end{figure*}

Maps of the convergence and magnification for a source at $z=9$
are shown in Figure~\ref{fig:lens_maps}. The convergence map exhibits two peaks, 
roughly centered at the positions of the two brightest cluster galaxies (BCGs).  Smaller substructures 
can be seen to the northeast of the NE BCG and to the south of the SW BCG. 
The magnification map shows that the critical curve, the curve along which the magnification is maximized
is very elliptical. The magnification reaches values up to $\mu\sim10-20$ within a few arcseconds from the critical curve and values of $\mu\sim1-2$ near the edge of the HFF footprint. Typical values of the magnification 
are $\mu\sim1-5$ throughout the HFF footprint.  

\subsection{Comparison with previous work}
\label{subsec:compare}

A previous model of \MJ{} using pre-HFF data was created using the same lens modeling code used in this work. The previous model was created in response to a 
call by STScI to model the HFF clusters before the HFF images were taken. The previous model appears on the publicly accessible HFF lens modeling website as the Brada\u{c} v1 model\footnote{\url{http://www.stsci.edu/hst/campaigns/frontier-fields/Lensing-Models}}. Our
previous model was constrained using 46 total multiple images belonging to 12 distinct systems, as opposed to the \NimgUSE{} images and \NsysUSE{} systems used in the model presented here, which is also made available to the public on the HFF lens modeling website as the Brada\u{c} v3 model. Only a few modeling teams produced v2 lens models of \MJ. The v2 HFF models were submitted to STScI during the time between the two official calls for lens models. Because our team submitted lens models exclusively during the official lens modeling calls, only Brada\u{c} v1 and v3 models exist of \MJ. In the v1 model, magnification uncertainties were estimated by bootstrap-resampling the weak lensing galaxies. In this work, however, we took a different approach to estimate uncertainties, one that we expect more accurately represents the statistical uncertainties. Because the number of multiple image systems used in this model is a factor of two larger than in the v1 model, we bootstrap-resampled all of the multiple image systems used in the model that were not spectroscopically confirmed. These are the systems for which we use $z_{\textrm{Bayes}}$ in the lens model. We assessed the impact of photometric redshift uncertainty on the derived lensing quantities by resampling the redshift of each system lacking spectroscopic confirmation from their full $z_{\textrm{Bayes}}$ posteriors. We exclude values of the redshift $z<(z_{\textrm{cluster}}+0.1)$ when resampling from the $z_{\textrm{Bayes}}$ posteriors. We compare the variance in magnification due to redshift uncertainty with the variance in magnification due to bootstrap-resampling the multiple image systems, finding that the latter is dominant. We nonetheless propagate both sources of error when reporting the errors on all derived lensing quantities in this work. Systematic uncertainties are not accounted for in our error analysis.

Six other teams (CATS, Sharon, Zitrin, Williams, GLAFIC, and Diego, as they appear on the HFF lens modeling page) created new lens models of the HFF clusters, which have also been made available
to the public on the HFF lens modeling website\footnotemark[10]. The lens models were released to the public on Dec 4, 2015 while we were preparing this manuscript, but after we had completed the lens model of the cluster. All teams had access to the same multiple image constraints, including the spectroscopic constraints from GLASS described herein. The CATS (v3), Diego, Zitrin (nfw and ltm-gauss), and Williams (v3) models use the same Gold and Silver multiple images that we used to constrain our model. The CATS (v3.1), Williams (v3.1) and GLAFIC models use Gold, Silver and Bronze images, and the Sharon model uses only the Gold images. We expect that the four models that used Gold and Silver images are the most directly comparable to our lens model. 

While the determination of the Gold, Silver and Bronze samples was coordinated among the modeling teams, the determination of the redshifts of the images lacking spectroscopic confirmation, i.e., the Silver and Bronze images, was not. Each team independently determined the redshifts of the Silver images. As a result, no two models, with the exception of the Zitrin nfw and ltm-gauss models, share the exact same multiple image constraints. This should be kept in mind when comparing the models. 

\begin{figure*}
\includegraphics[width=\textwidth]{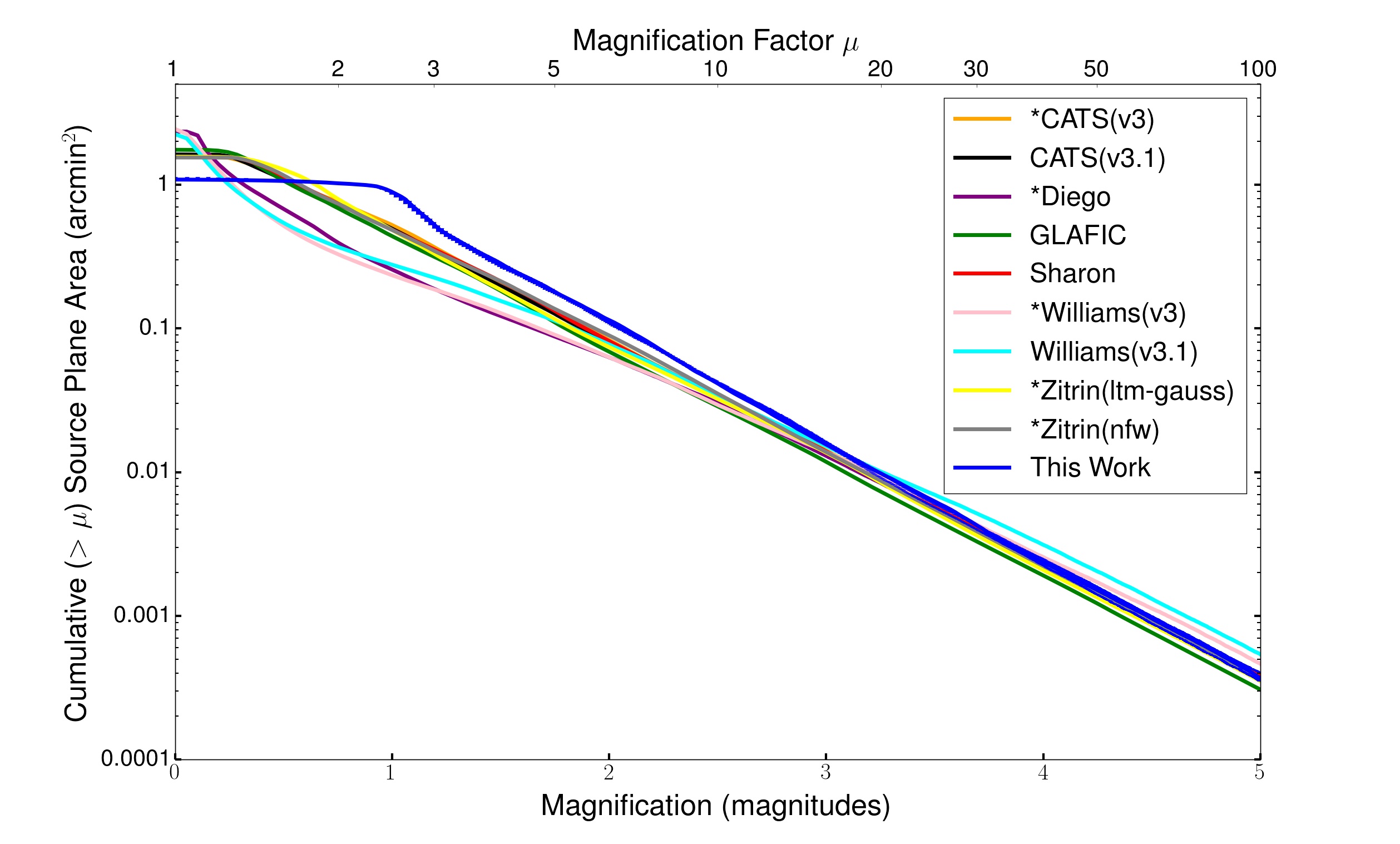}
\caption{Cumulative source plane area (``cumulative area'') versus magnification at $z=9$ for the lens model determined in this work (Bradac v3) and the other nine v3 HFF lens models. Model names preceded by ``*'' indicate models using the same sample of multiply imaged galaxies as used in this work, although the redshifts of the galaxies lacking spectroscopic confirmation have been estimated differently (see~\ref{subsec:compare} for more details). The models are in general agreement at large magnifications ($\mu\gtrsim5$). This is not surprising given that the multiple image constraints appear in the regions with large magnification near the core of the mass distribution. There is significant disagreement among the models at $1\lesssim\mu\lesssim5$. These values of magnification are indicative of the outskirts of the modeled area of the cluster, where constraints come from weak lensing alone. Only our model uses weak lensing constraints. 68\% confidence vertical error bars are shown for our model only and are approximately the thickness of the line.  }
\label{fig:area}
\end{figure*}

We first compare the cumulative magnified $z=9$ source plane area (cumulative area, hereafter) predicted by all models in Fig.~\ref{fig:area}. The cumulative area predictions among the nine models are significantly different for magnifications in the range $1\lesssim\mu\lesssim5$. As shown in Figure~\ref{fig:lens_maps}, the region with $1\lesssim\mu\lesssim5$ is primarily in the outskirts of the field, several hundred $\mathrm{kpc}$ from the critical curve. At this distance from the critical curve, weak lensing galaxies provide the only constraints to our model, as the strength of the lens is not sufficient to mutliply-image background galaxies. None of the other nine models use weak lensing constraints, but instead rely on an extrapolation of the core region to predict the magnification in the outskirts, therefore the disagreement with our model is not surprising in this regime. We assess whether the use of different sets of images could be responsible for the difference, but find that it is likely not the case. The models that similarly used the Gold and Silver samples of multiple images are still in disagreement in this regime. Further, the CATS and Williams teams constructed two models of the cluster. The v3 models of both teams use the Gold and Silver images only, whereas the v3.1 models use the Gold, Silver and Bronze samples. For each of these teams, the cumulative area predicted by the v3 and v3.1 models are very similar over a large range of magnifications ($1<\mu<100$), despite the v3.1 models using an additional \NimgBronze{} Bronze images. We also test whether the choice of our initial model could bias our model predictions. The prediction from our initial model is significantly different from the prediction from our reconstructed model. In fact, the initial model is more similar to the CATS, GLAFIC and Zitrin models for magnifications in the range $1\lesssim\mu\lesssim5$, and it is driven away from these models during the minimization.  There is general agreement among the models for cumulative areas at $\mu\gtrsim5$. This is reassuring because the region of the cluster for which $\mu\gtrsim5$ is near the critical curve, which is primarily constrained by the multiply imaged galaxies, which are numerous for this cluster and similar among the modeling teams. The image plane area used to make the cumulative area plots was the common area shared by all nine magnification maps, and is shown as the dashed box in the right panel of Figure~\ref{fig:lens_maps}. This area was set by the Williams magnification maps, which are approximately $4.65$ arcmin$^2$. 

\begin{figure}
\begin{center}
\includegraphics[width=0.5\textwidth]{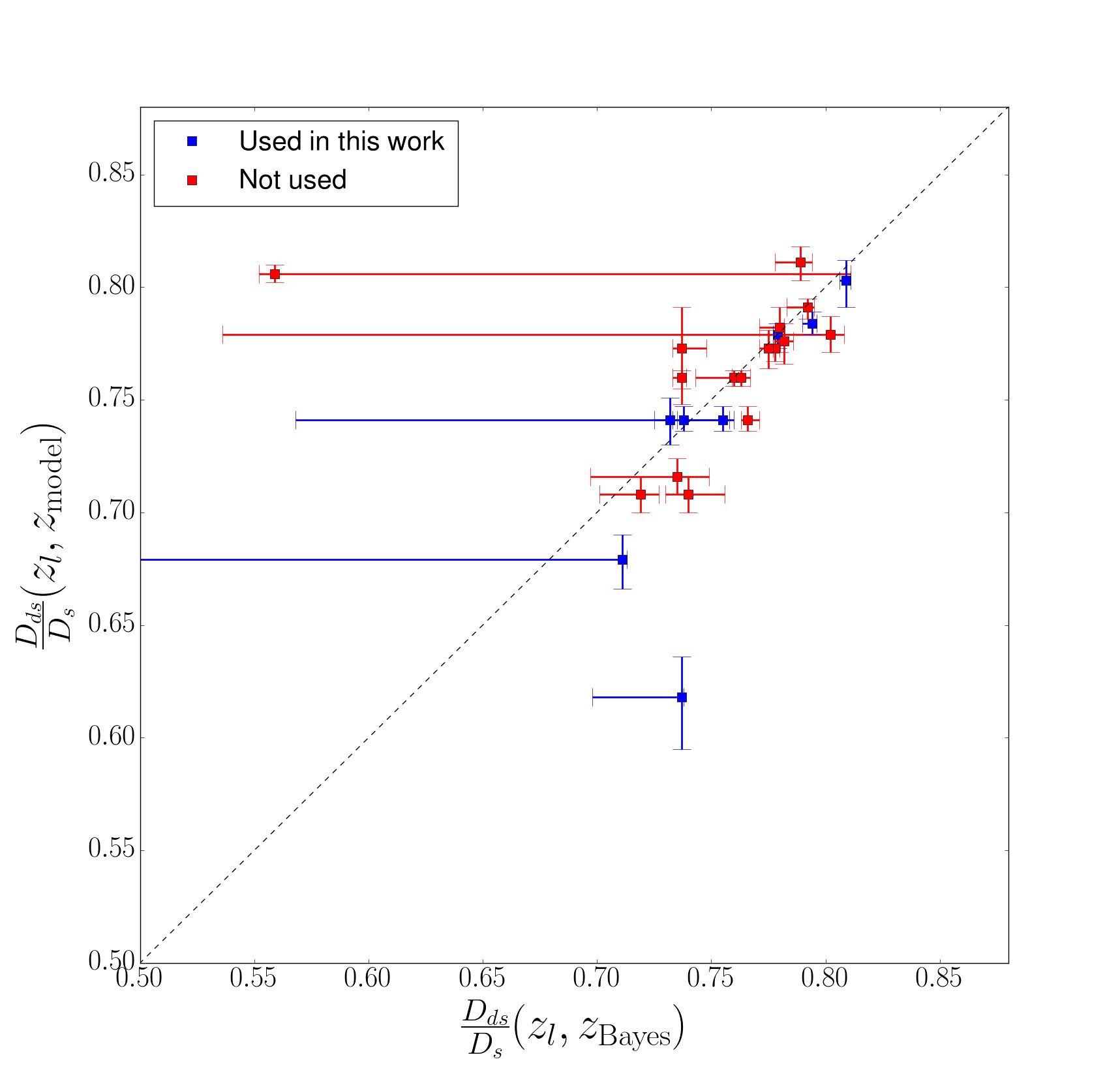}
\caption{Comparison of the ratio of angular diameter distances, the quantity with which the lensing deflection scales, for redshifts of multiply imaged galaxies determined in this work ($z_{\textrm{Bayes}}$) versus the redshifts predicted by the CATS v3.1 model ($z_{\textrm{model}}$; Jauzac et al. in prep.). The CATS v3.1 model used Gold+Silver+Bronze images, whereas we only used Gold+Silver images in our lens model, although we calculate $z_{\textrm{Bayes}}$ for all three categories to improve the statistics for this comparison. Gold+Silver images are the blue points, and the Bronze images are the red points. 
The vertical error bars were obtained by resampling from the 1$\sigma$ Gaussian errors on $z_{\textrm{model}}$. The horizontal error bars represent 68\% confidence and were obtained by resampling from $P(z_{\textrm{Bayes}})$. The asymmetric horizontal error bars arise because $P(z_{\mathrm{Bayes}})$ is multi-modal for those multiply imaged galaxies. There is overall good agreement between $z_{\textrm{Bayes}}$ and $z_{\textrm{model}}$. The dotted black line is shown for reference and represents perfect agreement. }
\label{fig:compare_z}
\end{center}
\end{figure}

Although different modeling teams used different sets of multiply imaged galaxies, it is difficult to see how this directly affects the models using the cumulative area plot. The factor with which the deflection angle scales for a source at $z=z_{\mathrm{Bayes}}$ is the ratio of the angular diameter
distances, ${\mathrm{D_{ds} \over D_s}}(z_\mathrm{l},z_{\mathrm{s}})$, where $\mathrm{D_{ds}}$ is the angular diameter distance between the lens at $z=z_{\mathrm{l}}$ and the source at $z=z_{\mathrm{s}}$, and $\mathrm{D_s}$ is the angular diameter distance between $z=0$ and $z=z_{\mathrm{s}}$. It is therefore the factor in which the source redshift
for a multiple image system directly enters the lens model. In Figure~\ref{fig:compare_z}, we compare this ratio for redshifts estimated in this work to those predicted by the CATS v3.1 lens model. We choose to compare to the CATS v3.1 model because it uses Gold, Silver and Bronze images and therefore provides the largest number of redshifts with which we can compare our photometric redshifts. The comparison is done for multiple image systems for which no spectroscopic redshift has been measured, whether in this work or previously. In this way, no spectroscopic redshift could be used as a prior for predicting the redshift. In the figure, $z_{\textrm{model}}$ is the redshift the CATS team
obtained by optimizing their lens model while leaving the redshift 
as a free parameter. Overall,
$z_{\textrm{Bayes}}$ and $z_{\textrm{model}}$ agree within the 
uncertainties. \citet{Wang+15} reached a similar conclusion 
by comparing $z_{\textrm{Bayes}}$ and $z_{\textrm{model}}$ for the 
multiple images predicted by the CATS v2 lens model of the HFF cluster Abell 2744. 
This is encouraging because similar inputs to the models
allow a more direction comparison of the results.

\section{Stellar Mass Fraction}
\label{sec:smass}
\subsection{Stellar mass density map} 

The lens model provides an estimate for the total mass density of the cluster,
composed of mostly invisible dark matter. A fractional component of the total mass density
comes from stars and can be inferred from the observed stellar light, independently from the lens model.
The \emph{Spitzer} IRAC $3.6\,\mu$m image samples close to rest-frame
$K$-band for the cluster, so we use the $3.6\,\mu$m fluxes from
cluster members to estimate the cluster stellar mass
distribution. The cluster members come from the selection by \citet{Grillo+15}, 
consisting of 109 spectroscopically confirmed 
and 66 photometrically selected cluster members. 

To create an image with $3.6\,\mu$m flux from cluster members only, we
first create a mask with value 1 for pixels that belong to cluster
members in the F160W image and 0 otherwise. We then convolve the mask
with the $3.6\,\mu$m PSF to match the IRAC angular resolution, set the
pixels below 10\% of the peak value to zero, and resample the mask
onto the IRAC pixel grid. We obtain the $3.6\,\mu$m map of cluster
members by setting all IRAC pixels not belonging to cluster members to
zero and smoothing the final surface brightness map with a two-pixel
wide Gaussian kernel.

The IRAC surface brightness map is transformed into a stellar surface mass
density map (stellar mass density map, hereafter) by first converting the 3.6$\mu$m flux into a $K$-band luminosity map.
The luminosity map is then multiplied by the stellar mass to light ratio derived by
\cite{Bell+03} using the so-called ``diet'' Salpeter stellar initial 
mass function (IMF), which has 70\% of the mass of the \cite{Sal55} IMF due to 
fewer stars at low masses. \cite{Bell+03} obtained a stellar mass to light ratio of $M_{\star}/L=0.95\pm0.26\, M_{\odot}/L_{\odot}$
in the stellar mass bin $10<\mathrm{log}(M_{\star}h^{2})<10.5$. Their
 $M_{\star}/L$ values are insensitive to the chosen stellar mass bin. The $\sim30\%$ error
 on $M_{\star}/L$ is the largest source of statistical uncertainty in the stellar mass density map.

The main source of systematic uncertainty in the stellar mass density map is the
unknown initial mass function (IMF). For example, if we adopt a
\cite{Sal55} IMF, as suggested by studies of massive early-type
galaxies, the stellar mass density increases by a factor of 1.55 everywhere.
We also assess the cluster member selection as a source of systematic uncertainty.
\citet{Grillo+15} estimate that their 
cluster member catalog is $\gtrsim95\%$ complete down to a stellar mass of $\mathrm{log}(M_{\star}/M_{\odot}) \simeq 9.8$ 
within the CLASH F160W footprint. We estimate the fraction
of stellar mass density not included in our analysis due to the incomplete 
cluster member selection at lower stellar masses. We compare the integral of the stellar mass function
obtained by \citet{Annun+14} for MACS J1206.2-0847, another CLASH cluster
at similar redshift ($z=0.44$), over the range of complete stellar masses from the \citet{Grillo+15} selection to the
integral over all stellar masses. We find that we only exclude $\sim3\%$ of the stellar mass  
within the F160W footprint from CLASH as a result of the incomplete cluster member selection.
\citet{Grillo+15} select cluster members down to 
$\mathrm{log}(M_{\star}/M_{\odot}) \simeq 8.6$, yet with $<95\%$ completeness in the stellar mass range
$8.6 \lesssim \mathrm{log}(M_{\star}/M_{\odot}) \lesssim 9.8$. Therefore, the $\sim3\%$ estimated loss in stellar mass 
is a slight overestimate of the loss due to the incomplete cluster member selection.
The presence of the
ICL, when not accounted for, can also cause the stellar mass density to be underestimated \citep[e.g.][]{Gonzalez+13}. \citet{Burke+15} recently measured the fraction of total cluster light contained in the ICL for 13/25 CLASH clusters, including \MJ{}. For \MJ{}, they found that $2.69 \pm 0.10\%$ of the total cluster light is contained in the ICL. Both the stellar mass incompleteness and ICL act to decrease
the measured stellar mass density. Both effects, however, are an order or magnitude smaller than the uncertainty due to the IMF and the stellar mass to light ratio obtained by \citet{Bell+03}.

\subsection{Stellar to total mass ratio}
\label{subsec:ratio}
We obtain the projected stellar to total mass ratio (\fstar{}, hereafter) map by dividing the stellar
mass density map obtained from photometry by the total surface
mass density map (total mass density, hereafter) obtained from our lens model. Before division, we 
match the resolutions of the two maps. The stellar mass density map resolution is controlled
by the \emph{Spitzer} IRAC 
$3.6\,\mu$m imaging resolution, which is roughly uniform across the field. 
On the other hand, the resolution of the total mass density varies considerably across the field
as a result of two processes that occur during the lens modeling procedure.  
The first effect is regularization, which globally degrades the resolution of the total mass density map.  
In order to estimate the decrease in resolution due to regularization, we used a
simulated galaxy cluster designed to match the data quality of the HFF, Hera (Meneghetti et al., in preparation).
We made our own lens model of Hera that includes the effects of regularization, 
which we compare to the correct lensing maps from the simulation (M. Meneghetti, private communication). 
The correct simulated lensing maps are of a uniform resolution that is higher than the resolution of our lensing maps. 
To determine the global resolution correction, we variably degrade the 
resolution of the simulated convergence map until we find the best match 
to our reconstructed convergence map. The second effect is non-uniformity 
in the grid introduced by the lens modeler to match the $S/N$ of the lensing measurements (\citealp{Bra++09}).
The process of increasing the resolution in this manner will be referred to as refinement hereafter. 
Each pixel in the lensing map has an associated refinement level. 
The refinement grid for \MJ{} has four levels: 1,2,4 and 8. 
Level 1 refinement represents no refinement and is reserved for the outskirts of the cluster.
Level 8 refinement is applied in a circular region centered on the multiple images used in the lens model 
with radii equal to 2.4 arcseconds. Refinement levels 2 and 4 are used to
mitigate discontinuities between level 1 and level 8 refinement. 
Refinement level 4 is used around the NE and SW BCG in circles of radii 0.6 and 0.45 arcminutes, respectively. 

\begin{figure*}[ht]
\plottwo{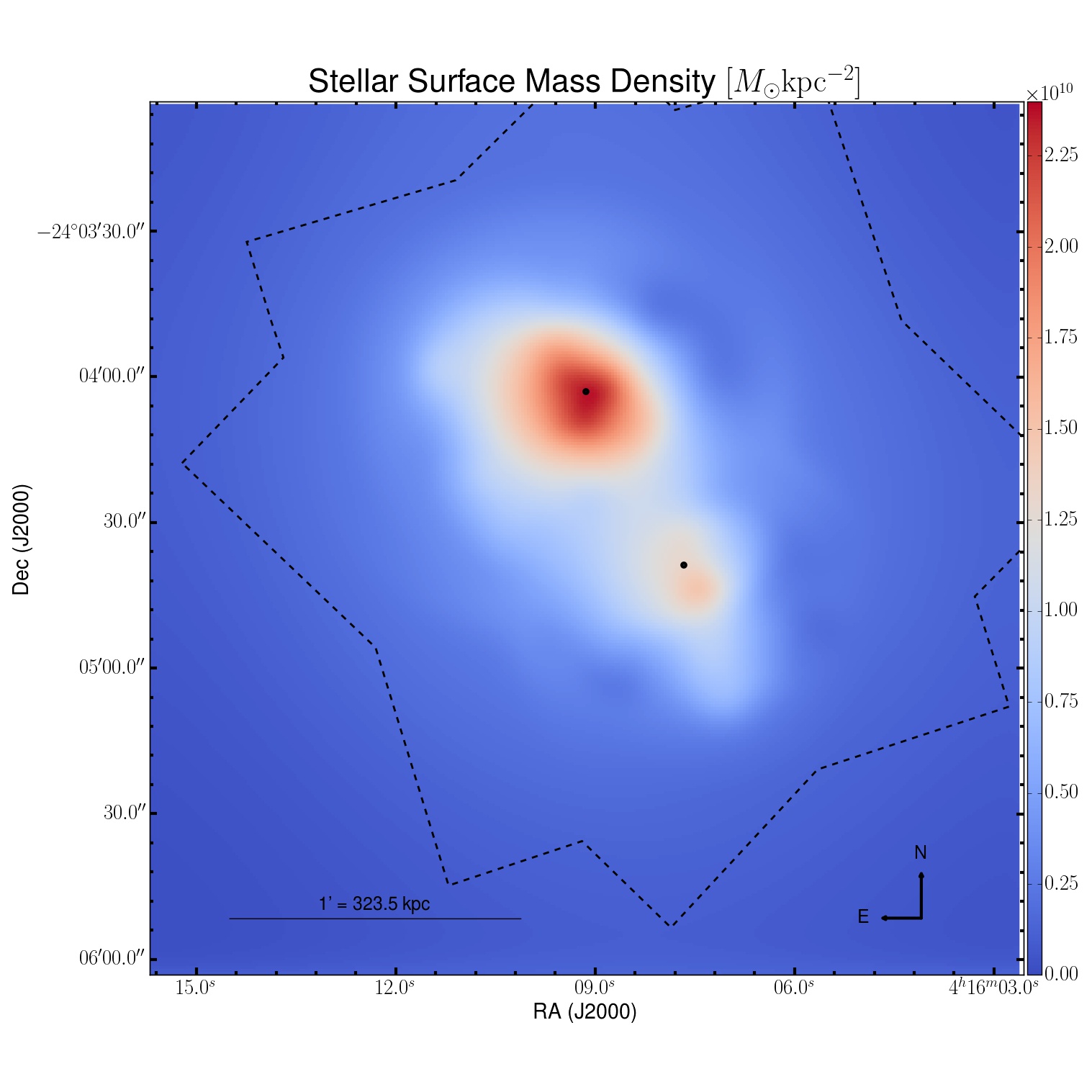}{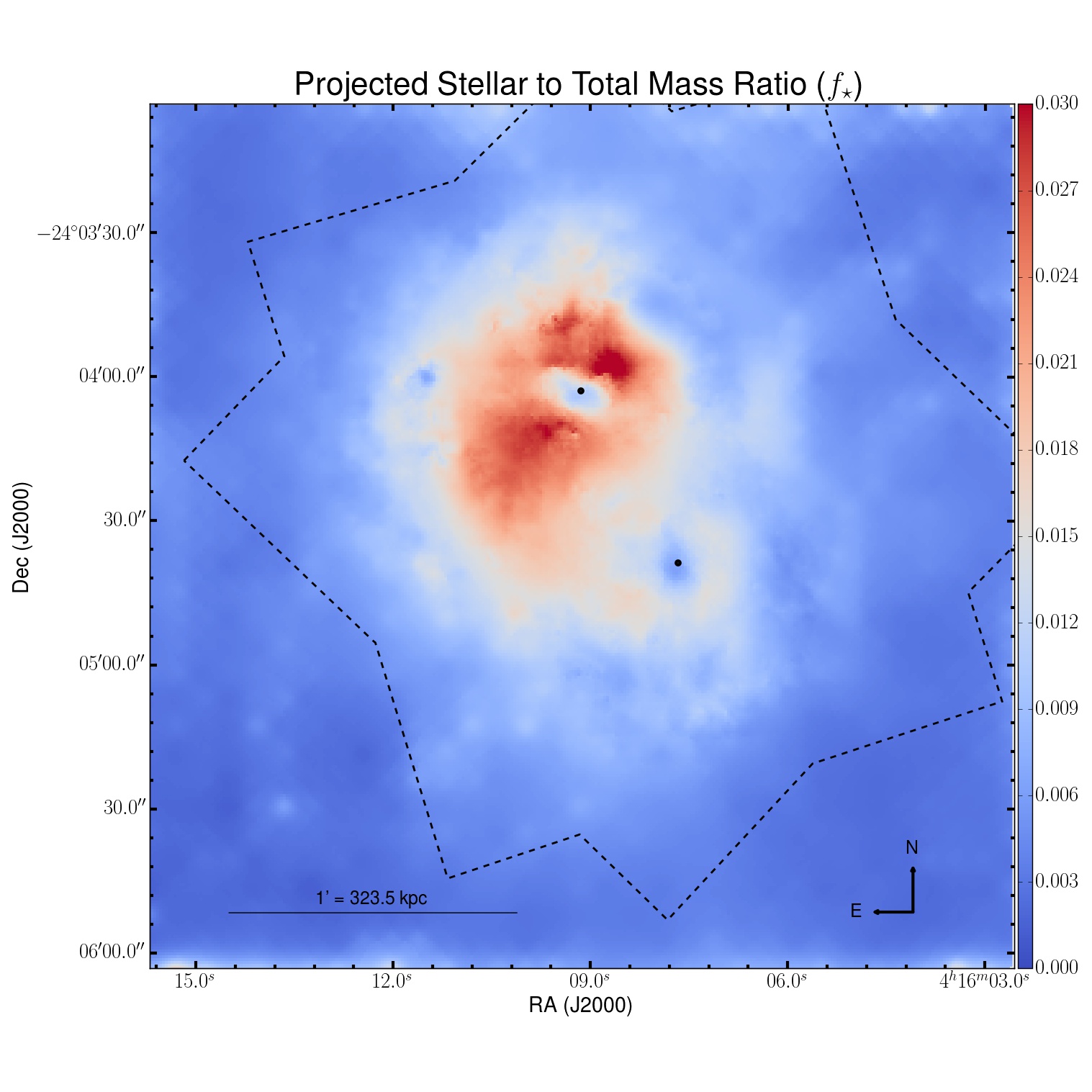}
\caption{\textbf{Left} - Stellar surface mass density (in units of $M_\odot\,\text{kpc}^{-2}$) derived from a \Spitzer{}/IRAC [3.6] image of \MJ{}. 
The resolution of this map has been matched to the resolution of the total surface mass density map. \textbf{Right} - Projected stellar to total mass ratio (\fstar{}), obtained by dividing the stellar surface mass density (left panel) by the total surface mass density obtained from our lens model. The two black points on each map mark the centers of the two BCGs determined from the F105W image (cf. Fig.~\ref{fig:lens_maps}). The dotted black line in both panels shows the dithered F160W footprint from CLASH, which comprises $\sim5.3\,\textrm{arcmin}^2$ of the entire $9\textrm{arcmin}^2$ FOV shown. The cluster member selection conducted by \citet{Grillo+15} used to make the stellar surface mass density map is complete down to $\mathrm{log}(M_{\star}/M_{\odot}) \simeq 9.8$ in this region. Outside of this region, the completeness and the uncertainty of both maps are not evaluated. 
\label{fig:stellarMassMap}}
\end{figure*}

To match the resolution of the stellar mass density map to that of the total mass density map from our lens model, 
we convolve the stellar mass density map with a Gaussian kernel of spatially varying width. 
We vary the kernel width according to the level of refinement at each pixel. 
The kernel width in refinement level 2 regions is always
half the width of the kernel used in refinement level 1 regions. Likewise, the
kernel width in refinement levels 4 and 8 are always ${1 \over 4}$ and ${1 \over 8}$ 
of the width of the kernel used in refinement level 1 regions, respectively. 
We vary the refinement level 1 kernel width from 0 to 1 arcminutes and assess the squared difference 
in the convergence in our reconstruction and the simulated map, finding a best fit kernel width of $0.75$ arcminutes in the refinement level 1 region.
The stellar mass density map is convolved with the four different kernels in the four refinement regions determined by this value alone.
The resulting resolution-corrected stellar mass density map is shown in the left panel of Figure~\ref{fig:stellarMassMap}.

The \fstar{} map is shown in the right panel of
Figure~\ref{fig:stellarMassMap}. There is significant variation in \fstar{}
throughout the cluster. While \fstar{} reaches as high as $\sim0.03$ in some places, 
the global mean within the stellar mass-complete region of the map is $0.009\pm 0.003$ (stat.; 68\% confidence), 
after adding in a $2.69 \pm 0.10\%$ ICL contribution to the stellar mass density determined by \citet{Burke+15}. 

The IMF is largest source of systematic uncertainty in \fstar{}. The choice of a \cite{Sal55} IMF over the diet-Salpter IMF assumed in this work would lead to an increase in the stellar mass density map, and therefore in \fstar{}, of $\sim50\%$. Another source of systematic error in \fstar{} is the choice of cosmology, which affects both the stellar mass density and the total mass density. Cosmology impacts the stellar mass density through the distance modulus, which is used to convert the observed surface brightness of cluster members to physical surface brightness. Comparing the distance modulus calculated using our fiducial one-significant-figure concordance cosmology and a two-significant-figure cosmology from \cite[e.g.]{Planck2015XIII}, we see a difference of 
$\sim30\%$. The choice of cosmology has a much smaller effect on the total mass density. Cosmology impacts the total mass density through the critical surface mass density, $\Sigma_c$. The difference in $\Sigma_c$ is $\lesssim3\%$. The effect of cosmology on $\Sigma_c$ is not as significant as the effect on the distance modulus because it enters $\Sigma_c$ in a ratio of angular diameter distances.    

\begin{figure*}[ht]
\includegraphics[width=0.5\textwidth]{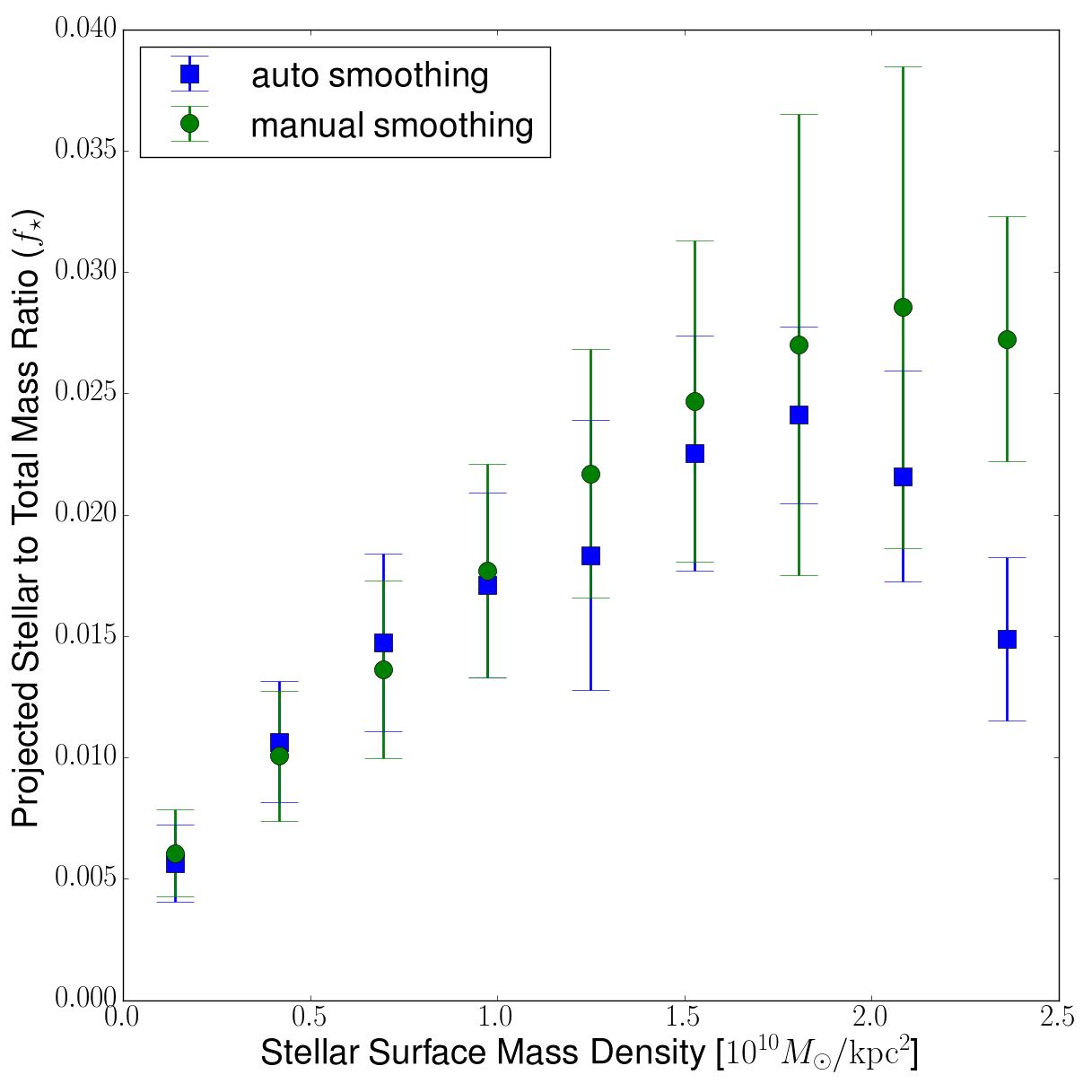}
\caption{Projected stellar mass to total mass ratio (\fstar{}) versus stellar surface mass density (in units of $10^{10} M_\odot\,\text{kpc}^{-2}$) for the ``auto'' and ``manual'' smoothing approaches (described in ~\ref{subsec:ratio}) used to match the resolution of the stellar mass and total mass density maps. The data points and error bars represent the mean and standard deviation of all points falling in each of the equally spaced 9 bins in stellar surface mass density. Only data from within the stellar mass-complete region shown in Figure~\ref{fig:stellarMassMap} are displayed in this figure. The two different approaches provide an estimate of the systematic error associated with the resolution matching procedure. The trend in \fstar{} for stellar surface mass densities for $<2\times10^{10} M_\odot\,\text{kpc}^{-2}$ is bolstered by the agreement between the approaches in this regime. However, at the highest stellar surface mass densities in the cluster, which are found only near the cores of the two BCGs, the two approaches disagree. Thus \fstar{} is too uncertain at stellar surface mass densities $\gtrsim2\times10^{10} M_\odot\,\text{kpc}^{-2}$ to conclude a downturn, as observed for the auto approach. 
\label{fig:ratio_vs_sm}}
\end{figure*}

We also assess the potential systematic error resulting from smoothing the stellar mass density map to match the resolution of the total mass density map as described above. We show the dependence of \fstar{} on the stellar surface mass density in Figure~\ref{fig:ratio_vs_sm} for two different smoothing approaches. The ``auto smoothing'' method is the one described above. The ``manual smoothing'' approach differs from the previous approach in how we estimate the optimal kernel width at each refinement level. In the auto smoothing method, the kernel width in each refinement region is simply $1/\mathrm{l}$ times the kernel width in refinement region 1, where $\mathrm{l}$ is the refinement level. In the manual smoothing method, however, we determine the optimal kernel width in each refinement region separately. For each refinement level, we mask out the part of the convergence maps not refined at that level before comparing the squared difference in the simulated and reconstructed convergence maps for a range of kernel widths between 0 and 1 arcminutes. In the regions corresponding to refinement levels 1 and 2, we find similar kernel widths using both smoothing methods. However, we find much smaller kernel widths in regions corresponding to refinement levels 4 and 8 when using the manual smoothing method. The stellar mass density map is smoothed significantly less near the BCGs in the manual smoothing approach. This effect is illustrated in Figure~\ref{fig:ratio_vs_sm}. In the auto smoothing approach, there is a downturn in \fstar{} for stellar surface mass densities $\gtrsim2\times10^{10} M_\odot\,\text{kpc}^{-2}$. The downturn is significantly less pronounced in the manual smoothing approach, where the peaks of the stellar mass density are more preserved due to less smoothing. Though a downturn in \fstar{} for stellar surface mass densities associated with the two BCGs could indicate interesting astrophysics, such as decreased star formation efficiency in the BCGs or a varying IMF, it is not robust against the choice of smoothing. The disparity in \fstar{} between the auto and manual smoothing approaches at stellar surface mass densities $\gtrsim2\times10^{10} M_\odot\,\text{kpc}^{-2}$ observed in Figure~\ref{fig:ratio_vs_sm} provides an estimate of the systematic error in \fstar{} as a result of smoothing the stellar mass density map. Overall, however, the choice of smoothing approach only affects $\langle f_{\star} \rangle$ by $0.001$, sub-dominant to the statistical error on $\langle f_{\star} \rangle$. The trend in \fstar{} with stellar surface mass density for stellar surface mass densities $<2\times10^{10} M_\odot\,\text{kpc}^{-2}$ is insensitive to the smoothing approach. This trend holds over $\sim98\%$ of the area of the stellar mass complete region of the \fstar{} map because values of the stellar surface mass densities exceeding $2\times10^{10} M_\odot\,\text{kpc}^{-2}$ are rare, only being observed near the peaks of the two BCGs, as can be seen in the left panel of Figure~\ref{fig:stellarMassMap}. Thus our conclusion that there is considerable variation in \fstar{} throughout the majority of the \HST\ WFC3/IR FOV is also insensitive to the systematics associated with smoothing. 

We compare our value of $\langle f_{\star} \rangle$ to values obtained in the recent literature. \citet{Bahcall+14} measured \fstar{} for $>10^{5}$ groups and clusters in the MaxBCG cluster catalog (\citealp{Koester+07}). Their cluster sample is taken from a photometric redshift range of $0.1 < z < 0.3$. Their \fstar{} measurements cover a large range of scales ($25 \mathrm{kpc} - 30 h^{-1} \mathrm{Mpc}$) and total masses ($M_{200} \sim 10^{13} - 10^{15} M_{\odot}$), where $M_{200}$ is the total mass within $r_{200}$, the radius within which the mean density is 200 times the matter density of the universe. On all scales larger than a few hundred $\mathrm{kpc}$, \citet{Bahcall+14} measure a constant value of $\langle f_{\star} \rangle  =0.010 \pm 0.004 $  (68\% confidence). For the most massive clusters in their sample, which are most analogous to \MJ{}, they report a similar value of $\sim1\%$. \citet{Bahcall+14} used stellar mass to light ratios calculated with i-band magnitudes from the Sloan Digital Sky Survey (SDSS; \citealp{York+00}) while employing a Chabrier IMF. In order to directly compare our results to \citet{Bahcall+14}, we recalculated $\langle f_{\star} \rangle$ using F105W, the band with the smallest K-correction to SDSS i-band at the redshift of \MJ{}, and then scaled the resulting light map using the same stellar mass to light ratio that they used, $M_{\star}/L=2.5$. After recalculating, we obtain a value of  $\langle f_{\star} \rangle = 0.012^{+0.005}_{-0.003}$ (stat.; 68\% confidence), in agreement with the large scale value obtained by \citet{Bahcall+14}. \citet{Gonzalez+13} measured \fstar{} for 12 clusters at $z\sim0.1$ over the mass range $M_{500} = 1 - 5 \times 10^{14} M_{\odot}$. They measure values of \fstar{} ranging from $5\%$ at the lower mass end of their sample to $1.5\%$ at the upper mass end. \citet{Umetsu+14} recently measured $M_{500}=7.0\pm1.3\times10^{14} M_{\odot}$ for \MJ{}. We recalculated \fstar{} in the same band (WFPC2 F814W) and using the same stellar mass to light ratio as \citet{Gonzalez+13} ($M_{\star}/L=2.65$), finding a value of $ \langle f_\star{} \rangle=0.014^{+0.005}_{-0.004}$ (stat.; 68\% confidence). While \MJ{} is at higher redshift ($z_{\mathrm{cluster}}=0.396$) and has higher mass than the clusters studied by \citet{Gonzalez+13}, our measured value of $ \langle f_\star{} \rangle$ for \MJ{} is comparable to the values measured by \citet{Gonzalez+13}  at the highest masses. We also note that the area in which we calculate \fstar{} is smaller than $r_{200}$, the radius in which \citet{Gonzalez+13} measured $ \langle f_\star{} \rangle$ for their cluster sample. \citet{Balestra+15} measure $r_{200}$=1.82$ \,\mathrm{Mpc}$ for \MJ{}. Adopting this value of $r_{200}$, the region in which we measure \fstar{} is $\sim0.4r_{200}$. We assumed a 30\% error on the values of $M_{\star}/L$ (cf. \citealp{Bell+03}) when computing $\langle f_{\star} \rangle$ to compare to the values of \fstar{} reported by \citet{Bahcall+14} and \citet{Gonzalez+13}. 

\vspace{1cm}

\vspace{1cm}

\clearpage

\section{Conclusions}
\label{sec:conc}

The massive galaxy cluster \MJ{} is a powerful gravitational lens with excellent constraints 
for lens modeling. A coordinated search for multiple images of strongly-lensed galaxies performed by
several lens modeling teams found
$\sim200$ candidate multiple images consisting of $\sim100$ source galaxies. In order to provide the best constraints 
to the many lens modeling teams, including our own, we inspected
each of these candidate multiple images in the GLASS spectroscopy. 
Using GLASS spectroscopic measurements together with constraints obtained through the collaborative HFF modeling effort,
we produced a gravitational lens model of \MJ{}. 
We then compared the projected stellar mass density map derived from IRAC photometry to the 
total mass density map obtained from our lens model to study the
projected stellar to total mass fraction throughout the cluster field. Our main results are 
summarized here:

\begin{enumerate}

\item We have measured spectroscopic redshifts for \Nimgconfirm{} multiple images (quality flag 3 (probable) and 4 (secure)), confirming \Nsysnew{} multiple image systems for the first time. The spectroscopically confirmed images were used to constrain our gravitational lens models and the nine other lens models discussed in this work. These lens models, including our own, are publicly available\footnotemark[10]. 

\item We performed a visual search for 
faint emission and absorption lines, establishing a spectroscopic redshift catalog of weakly-lensed galaxies 
throughout the primary cluster field. We compared our photometric redshifts with 
grism spectroscopic redshifts and found good agreement, giving us more confidence
in the photometric redshifts (and their errors) of the multiple images. We compared our photometric redshifts with redshifts determined
from the v3.1 CATS lens model of \MJ{} for the multiple images used in their lens model. We find general agreement with their redshifts.

\item The cumulative magnified source plane area (cumulative area) predicted by our lens model was compared with the nine other lens models of \MJ{} constrained using products from the same imaging and spectroscopy data. The cumulative area predictions agree among the models for $\mu\gtrsim5$ (mostly near the cluster core), but the results diverge among the models for the outlying regions with $\mu\lesssim5$. We attribute the model differences to the lack of constrains in this region, with the exception of our model which uses weak lensing constraints derived from the HFF imaging data. Despite the conservative approach of including only the Gold (spec-$z$) and Silver (high confidence phot-$z$) multiple image systems identified by the HFF modeling collaboration, systematics from misidentification and redshift estimation of the Silver systems may influence the lens model. However, the inclusion of grism spectroscopic redshifts helps the lens modeling by providing stronger constraints and revising the incorrect redshifts used in the previous models.

\item We obtained a stellar surface mass density map from deep \Spitzer{}/IRAC imaging data, using cluster members selected by \citet{Grillo+15}. We compare the stellar surface mass density map to the total surface mass density produced from our lens model, producing a map of the projected stellar mass ratio, \fstar{}. There is significant variation in \fstar{} throughout the cluster. \fstar{} increases with stellar surface mass density up to stellar surface mass density of $\sim2\times10^{10} M_\odot\,\text{kpc}^{-2}$, above which our results are inconclusive. The global mean projected stellar mass fraction is $\langle f_{\star} \rangle = 0.009 \pm 0.003$ (stat.; 68\% confidence) using a diet-Salpeter IMF. We compare our results with recent measurements of $\langle f_{\star} \rangle$ in the literature taken over a wide range of total cluster masses and redshifts. After correcting for different IMFs and filters used to convert stellar light to mass, we find that our measured value of $\langle f_{\star} \rangle$ is broadly consistent with the literature values.  

\end{enumerate}

\medskip

\acknowledgments

AH acknowledges support by NASA Headquarters under the
NASA Earth and Space Science Fellowship Program - Grant ASTRO14F-
0007. This work utilizes gravitational lensing models produced by PIs Brada\u{c}, 
Natarajan \& Kneib (CATS), Merten \& Zitrin, Sharon, and Williams, and the GLAFIC and Diego groups. 
The lens models were obtained from the Mikulski Archive for Space Telescopes (MAST).
This work is based in part on observations made with the NASA/ESA Hubble
Space Telescope, obtained at STScI. The data
were obtained from MAST. 
We acknowledge support through
grants HST-13459, HST-GO13177, HST-AR13235. MB, KH, and AH acknowledge
support for this work through a \Spitzer{} award issued by
JPL/Caltech. MB and AH also acknowledge support from the special funding as
part of the \HST{} Frontier Fields program conducted by STScI. STScI
is operated by AURA, Inc. under NASA contract NAS 5-26555. 
TT acknowledges support by the Packard Foundation through a Packard
Research Fellowship, and thanks the Osservatorio Astronomico di
Monteporzio Catone and the American Academy in Rome for their kind
hospitality during the writing of this manuscript. BV acknowledges the
support from the World Premier International Research Center
Initiative (WPI), MEXT, Japan and the Kakenhi Grant-in-Aid for Young
Scientists (B)(26870140) from the Japan Society for the Promotion of
Science (JSPS).

\bibliographystyle{apj}
\bibliography{bibtexlibrary}

\begin{thebibliography}{}
\expandafter\ifx\csname natexlab\endcsname\relax\def\natexlab#1{#1}\fi

\bibitem[{{Annunziatella} {et~al.}(2014){Annunziatella}, {Biviano}, {Mercurio},
  {Nonino}, {Rosati}, {Balestra}, {Presotto}, {Girardi}, {Gobat}, {Grillo},
  {Kelson}, {Medezinski}, {Postman}, {Scodeggio}, {Brescia}, {Demarco},
  {Fritz}, {Koekemoer}, {Lemze}, {Lombardi}, {Sartoris}, {Umetsu}, {Vanzella},
  {Bradley}, {Coe}, {Donahue}, {Infante}, {Kuchner}, {Maier}, {Reg{\H o}s},
  {Verdugo}, \& {Ziegler}}]{Annun+14}
{Annunziatella}, M., {Biviano}, A., {Mercurio}, A., {et~al.} 2014, \aap, 571,
  A80

\bibitem[{{Bahcall} \& {Kulier}(2014)}]{Bahcall+14}
{Bahcall}, N.~A., \& {Kulier}, A. 2014, \mnras, 439, 2505

\bibitem[{{Balestra} {et~al.}(2015){Balestra}, {Mercurio}, {Sartoris},
  {Girardi}, {Grillo}, {Nonino}, {Rosati}, {Biviano}, {Ettori}, {Forman},
  {Jones}, {Koekemoer}, {Medezinski}, {Ogrean}, {Tozzi}, {Umetsu}, {Vanzella},
  {van Weeren}, {Zitrin}, {Annunziatella}, {Caminha}, {Broadhurst}, {Coe},
  {Donahue}, {Fritz}, {Frye}, {Kelson}, {Lombardi}, {Maier}, {Meneghetti},
  {Monna}, {Postman}, {Scodeggio}, {Seitz}, \& {Ziegler}}]{Balestra+15}
{Balestra}, I., {Mercurio}, A., {Sartoris}, B., {et~al.} 2015, ArXiv e-prints,
  arXiv:1511.02522

\bibitem[{{Bell} {et~al.}(2003){Bell}, {McIntosh}, {Katz}, \&
  {Weinberg}}]{Bell+03}
{Bell}, E.~F., {McIntosh}, D.~H., {Katz}, N., \& {Weinberg}, M.~D. 2003, \apjs,
  149, 289

\bibitem[{{Ben{\'{\i}}tez}(2000)}]{bpz}
{Ben{\'{\i}}tez}, N. 2000, \apj, 536, 571

\bibitem[{Bertin \& Arnouts(1996)}]{Bertin:1996p12964}
Bertin, E., \& Arnouts, S. 1996, Astronomy and Astrophysics Supplement, 117,
  393

\bibitem[{{Brada{\v c}} {et~al.}(2005){Brada{\v c}}, {Erben}, {Schneider},
  {Hildebrandt}, {Lombardi}, {Schirmer}, {Miralles}, {Clowe}, \&
  {Schindler}}]{bradac05}
{Brada{\v c}}, M., {Erben}, T., {Schneider}, P., {et~al.} 2005, \aap, 437, 49

\bibitem[{{Brada{\v c}} {et~al.}(2006){Brada{\v c}}, {Clowe}, {Gonzalez},
  {Marshall}, {Forman}, {Jones}, {Markevitch}, {Randall}, {Schrabback}, \&
  {Zaritsky}}]{Bra++06}
{Brada{\v c}}, M., {Clowe}, D., {Gonzalez}, A.~H., {et~al.} 2006, \apj, 652,
  937

\bibitem[{{Brada{\v c}} {et~al.}(2009{\natexlab{a}}){Brada{\v c}}, {Treu},
  {Applegate}, {Gonzalez}, {Clowe}, {Forman}, {Jones}, {Marshall}, {Schneider},
  \& {Zaritsky}}]{bradac09}
{Brada{\v c}}, M., {Treu}, T., {Applegate}, D., {et~al.} 2009{\natexlab{a}},
  \apj, 706, 1201

\bibitem[{{Brada{\v c}} {et~al.}(2009{\natexlab{b}}){Brada{\v c}}, {Treu},
  {Applegate}, {Gonzalez}, {Clowe}, {Forman}, {Jones}, {Marshall}, {Schneider},
  \& {Zaritsky}}]{Bra++09}
---. 2009{\natexlab{b}}, \apj, 706, 1201

\bibitem[{Brammer {et~al.}(2014)Brammer, Pirzkal, McCullough, \&
  MacKenty}]{Brammer:2014p34990}
Brammer, G.~B., Pirzkal, N., McCullough, P.~R., \& MacKenty, J.~W. 2014, STScI
  ISR

\bibitem[{Brammer {et~al.}(2015)Brammer, Ryan, \& Pirzkal}]{Brammer:2015}
Brammer, G.~B., Ryan, R., \& Pirzkal, N. 2015, STScI ISR 2015-17

\bibitem[{Brammer {et~al.}(2012)Brammer, van Dokkum, Franx, Fumagalli, Patel,
  Rix, Skelton, Kriek, Nelson, Schmidt, Bezanson, Cunha, Erb, Fan, Schreiber,
  Illingworth, Labb{\'e}, Leja, Lundgren, Magee, Marchesini, McCarthy,
  Momcheva, Muzzin, Quadri, Steidel, Tal, Wake, Whitaker, \&
  Williams}]{Brammer:2012p12977}
Brammer, G.~B., van Dokkum, P.~G., Franx, M., {et~al.} 2012, The Astrophysical
  Journal Supplement, 200, 13

\bibitem[{{Burke} {et~al.}(2015){Burke}, {Hilton}, \& {Collins}}]{Burke+15}
{Burke}, C., {Hilton}, M., \& {Collins}, C. 2015, \mnras, 449, 2353

\bibitem[{{Clowe} {et~al.}(2006){Clowe}, {Brada{\v{c}}}, {Gonzalez},
  {Markevitch}, {Randall}, {Jones}, \& {Zaritsky}}]{clowe06}
{Clowe}, D., {Brada{\v{c}}}, M., {Gonzalez}, A.~H., {et~al.} 2006, \apjl, 648,
  L109

\bibitem[{Coe {et~al.}(2013)Coe, Zitrin, Carrasco, Shu, Zheng, Postman,
  Bradley, Koekemoer, Bouwens, Broadhurst, Monna, Host, Moustakas, Ford,
  Moustakas, Wel, Donahue, Rodney, Ben{\'\i}tez, Jouvel, Seitz, Kelson, \&
  Rosati}]{Coe:2013p26313}
Coe, D., Zitrin, A., Carrasco, M., {et~al.} 2013, The Astrophysical Journal,
  762, 32

\bibitem[{Dahlen {et~al.}(2013)Dahlen, Mobasher, Faber, Ferguson, Barro,
  Finkelstein, Finlator, Fontana, Gruetzbauch, Johnson, Pforr, Salvato,
  Wiklind, Wuyts, Acquaviva, Dickinson, Guo, Huang, Huang, Newman, Bell,
  Conselice, Galametz, Gawiser, Giavalisco, Grogin, Hathi, Kocevski, Koekemoer,
  Koo, Lee, McGrath, Papovich, Peth, Ryan, Somerville, Weiner, \&
  Wilson}]{Dahlen:2013p33380}
Dahlen, T., Mobasher, B., Faber, S.~M., {et~al.} 2013, The Astrophysical
  Journal, 775, 93

\bibitem[{{Diego} {et~al.}(2015{\natexlab{a}}){Diego}, {Broadhurst}, {Molnar},
  {Lam}, \& {Lim}}]{Diego+15}
{Diego}, J.~M., {Broadhurst}, T., {Molnar}, S.~M., {Lam}, D., \& {Lim}, J.
  2015{\natexlab{a}}, \mnras, 447, 3130

\bibitem[{{Diego} {et~al.}(2015{\natexlab{b}}){Diego}, {Broadhurst}, {Zitrin},
  {Lam}, {Lim}, {Ford}, \& {Zheng}}]{Diego+15b}
{Diego}, J.~M., {Broadhurst}, T., {Zitrin}, A., {et~al.} 2015{\natexlab{b}},
  \mnras, 451, 3920

\bibitem[{{Ebeling} {et~al.}(2001){Ebeling}, {Edge}, \& {Henry}}]{Ebeling+01}
{Ebeling}, H., {Edge}, A.~C., \& {Henry}, J.~P. 2001, \apj, 553, 668

\bibitem[{{Erben} {et~al.}(2001){Erben}, {Van Waerbeke}, {Bertin}, {Mellier},
  \& {Schneider}}]{erben2001}
{Erben}, T., {Van Waerbeke}, L., {Bertin}, E., {Mellier}, Y., \& {Schneider},
  P. 2001, \aap, 366, 717

\bibitem[{{Fioc} \& {Rocca-Volmerange}(1997)}]{Fioc+97}
{Fioc}, M., \& {Rocca-Volmerange}, B. 1997, \aap, 326, 950

\bibitem[{{Gonzaga}(2012)}]{Gonzaga:2014p26307}
{Gonzaga}, S. 2012, {The DrizzlePac Handbook} (STScI)

\bibitem[{{Gonzalez} {et~al.}(2013){Gonzalez}, {Sivanandam}, {Zabludoff}, \&
  {Zaritsky}}]{Gonzalez+13}
{Gonzalez}, A.~H., {Sivanandam}, S., {Zabludoff}, A.~I., \& {Zaritsky}, D.
  2013, \apj, 778, 14

\bibitem[{{Grillo} {et~al.}(2015){Grillo}, {Suyu}, {Rosati}, {Mercurio},
  {Balestra}, {Munari}, {Nonino}, {Caminha}, {Lombardi}, {De Lucia}, {Borgani},
  {Gobat}, {Biviano}, {Girardi}, {Umetsu}, {Coe}, {Koekemoer}, {Postman},
  {Zitrin}, {Halkola}, {Broadhurst}, {Sartoris}, {Presotto}, {Annunziatella},
  {Maier}, {Fritz}, {Vanzella}, \& {Frye}}]{Grillo+15}
{Grillo}, C., {Suyu}, S.~H., {Rosati}, P., {et~al.} 2015, \apj, 800, 38

\bibitem[{{Hoekstra} {et~al.}(1998){Hoekstra}, {Franx}, {Kuijken}, \&
  {Squires}}]{hoekstra1998}
{Hoekstra}, H., {Franx}, M., {Kuijken}, K., \& {Squires}, G. 1998, \apj, 504,
  636

\bibitem[{{Jauzac} {et~al.}(2014){Jauzac}, {Cl{\'e}ment}, {Limousin},
  {Richard}, {Jullo}, {Ebeling}, {Atek}, {Kneib}, {Knowles}, {Natarajan},
  {Eckert}, {Egami}, {Massey}, \& {Rexroth}}]{Jauzac+14}
{Jauzac}, M., {Cl{\'e}ment}, B., {Limousin}, M., {et~al.} 2014, \mnras, 443,
  1549

\bibitem[{{Jauzac} {et~al.}(2015{\natexlab{a}}){Jauzac}, {Richard}, {Jullo},
  {Cl{\'e}ment}, {Limousin}, {Kneib}, {Ebeling}, {Natarajan}, {Rodney}, {Atek},
  {Massey}, {Eckert}, {Egami}, \& {Rexroth}}]{Jauzac+15b}
{Jauzac}, M., {Richard}, J., {Jullo}, E., {et~al.} 2015{\natexlab{a}}, \mnras,
  452, 1437

\bibitem[{{Jauzac} {et~al.}(2015{\natexlab{b}}){Jauzac}, {Jullo}, {Eckert},
  {Ebeling}, {Richard}, {Limousin}, {Atek}, {Kneib}, {Cl{\'e}ment}, {Egami},
  {Harvey}, {Knowles}, {Massey}, {Natarajan}, {Neichel}, \&
  {Rexroth}}]{Jauzac+15}
{Jauzac}, M., {Jullo}, E., {Eckert}, D., {et~al.} 2015{\natexlab{b}}, \mnras,
  446, 4132

\bibitem[{{Johnson} {et~al.}(2014){Johnson}, {Sharon}, {Bayliss}, {Gladders},
  {Coe}, \& {Ebeling}}]{Johnson+14}
{Johnson}, T.~L., {Sharon}, K., {Bayliss}, M.~B., {et~al.} 2014, \apj, 797, 48

\bibitem[{Kaiser {et~al.}(1995)Kaiser, Squires, \& Broadhurst}]{kaiser1995}
Kaiser, N., Squires, G., \& Broadhurst, T. 1995, ApJ, 449, 460

\bibitem[{{Kawamata} {et~al.}(2015){Kawamata}, {Oguri}, {Ishigaki},
  {Shimasaku}, \& {Ouchi}}]{Kawamata+15}
{Kawamata}, R., {Oguri}, M., {Ishigaki}, M., {Shimasaku}, K., \& {Ouchi}, M.
  2015, ArXiv e-prints, arXiv:1510.06400

\bibitem[{{Koester} {et~al.}(2007){Koester}, {McKay}, {Annis}, {Wechsler},
  {Evrard}, {Bleem}, {Becker}, {Johnston}, {Sheldon}, {Nichol}, {Miller},
  {Scranton}, {Bahcall}, {Barentine}, {Brewington}, {Brinkmann}, {Harvanek},
  {Kleinman}, {Krzesinski}, {Long}, {Nitta}, {Schneider}, {Sneddin}, {Voges},
  \& {York}}]{Koester+07}
{Koester}, B.~P., {McKay}, T.~A., {Annis}, J., {et~al.} 2007, \apj, 660, 239

\bibitem[{{Luppino} \& {Kaiser}(1997)}]{luppino1997}
{Luppino}, G.~A., \& {Kaiser}, N. 1997, \apj, 475, 20

\bibitem[{{Mann} \& {Ebeling}(2012)}]{Mann+2012}
{Mann}, A.~W., \& {Ebeling}, H. 2012, \mnras, 420, 2120

\bibitem[{{Massey} {et~al.}(2014){Massey}, {Schrabback}, {Cordes}, {Marggraf},
  {Israel}, {Miller}, {Hall}, {Cropper}, {Prod'homme}, \& {Matias
  Niemi}}]{massey+14}
{Massey}, R., {Schrabback}, T., {Cordes}, O., {et~al.} 2014, \mnras, 439, 887

\bibitem[{{Merten} {et~al.}(2015){Merten}, {Meneghetti}, {Postman}, {Umetsu},
  {Zitrin}, {Medezinski}, {Nonino}, {Koekemoer}, {Melchior}, {Gruen},
  {Moustakas}, {Bartelmann}, {Host}, {Donahue}, {Coe}, {Molino}, {Jouvel},
  {Monna}, {Seitz}, {Czakon}, {Lemze}, {Sayers}, {Balestra}, {Rosati},
  {Ben{\'{\i}}tez}, {Biviano}, {Bouwens}, {Bradley}, {Broadhurst}, {Carrasco},
  {Ford}, {Grillo}, {Infante}, {Kelson}, {Lahav}, {Massey}, {Moustakas},
  {Rasia}, {Rhodes}, {Vega}, \& {Zheng}}]{2015ApJ...806....4M}
{Merten}, J., {Meneghetti}, M., {Postman}, M., {et~al.} 2015, \apj, 806, 4

\bibitem[{{Momcheva} {et~al.}(2015){Momcheva}, {Brammer}, {van Dokkum},
  {Skelton}, {Whitaker}, {Nelson}, {Fumagalli}, {Maseda}, {Leja}, {Franx},
  {Rix}, {Bezanson}, {Da Cunha}, {Dickey}, {F{\"o}rster Schreiber},
  {Illingworth}, {Kriek}, {Labb{\'e}}, {Ulf Lange}, {Lundgren}, {Magee},
  {Marchesini}, {Oesch}, {Pacifici}, {Patel}, {Price}, {Tal}, {Wake}, {van der
  Wel}, \& {Wuyts}}]{Momcheva+15}
{Momcheva}, I.~G., {Brammer}, G.~B., {van Dokkum}, P.~G., {et~al.} 2015, ArXiv
  e-prints, arXiv:1510.02106

\bibitem[{{Newman} {et~al.}(2013){Newman}, {Treu}, {Ellis}, \&
  {Sand}}]{New++13}
{Newman}, A.~B., {Treu}, T., {Ellis}, R.~S., \& {Sand}, D.~J. 2013, \apj, 765,
  25

\bibitem[{{Ogrean} {et~al.}(2015){Ogrean}, {van Weeren}, {Jones}, {Clarke},
  {Sayers}, {Mroczkowski}, {Nulsen}, {Forman}, {Murray}, {Pandey-Pommier},
  {Randall}, {Churazov}, {Bonafede}, {Kraft}, {David}, {Andrade-Santos},
  {Merten}, {Zitrin}, {Umetsu}, {Goulding}, {Roediger}, {Bagchi}, {Bulbul},
  {Donahue}, {Ebeling}, {Johnston-Hollitt}, {Mason}, {Rosati}, \&
  {Vikhlinin}}]{Ogrean+15}
{Ogrean}, G.~A., {van Weeren}, R.~J., {Jones}, C., {et~al.} 2015, \apj, 812,
  153

\bibitem[{{Oke}(1974)}]{Oke74}
{Oke}, J.~B. 1974, \apjs, 27, 21

\bibitem[{{Planck Collaboration}(2015)}]{Planck2015XIII}
{Planck Collaboration}. 2015, ArXiv e-prints, arXiv:1502.01589

\bibitem[{Postman {et~al.}(2012)Postman, Coe, Ben{\'\i}tez, Bradley,
  Broadhurst, Donahue, Ford, Graur, Graves, Jouvel, Koekemoer, Lemze,
  Medezinski, Molino, Moustakas, Ogaz, Riess, Rodney, Rosati, Umetsu, Zheng,
  Zitrin, Bartelmann, Bouwens, Czakon, Golwala, Host, Infante, Jha,
  Jimenez-Teja, Kelson, Lahav, Lazkoz, Maoz, McCully, Melchior, Meneghetti,
  Merten, Moustakas, Nonino, Patel, Reg{\"o}s, Sayers, Seitz, \&
  Wel}]{Postman:2012p27556}
Postman, M., Coe, D., Ben{\'\i}tez, N., {et~al.} 2012, The Astrophysical
  Journal Supplement, 199, 25

\bibitem[{{Press}(1997)}]{1997upa..conf...49P}
{Press}, W.~H. 1997, in Unsolved Problems in Astrophysics, ed. J.~N. {Bahcall}
  \& J.~P. {Ostriker}, 49--60

\bibitem[{{Richard} {et~al.}(2014){Richard}, {Jauzac}, {Limousin}, {Jullo},
  {Cl{\'e}ment}, {Ebeling}, {Kneib}, {Atek}, {Natarajan}, {Egami}, {Livermore},
  \& {Bower}}]{Richard+14}
{Richard}, J., {Jauzac}, M., {Limousin}, M., {et~al.} 2014, \mnras, 444, 268

\bibitem[{{Rodney} {et~al.}(2015{\natexlab{a}}){Rodney}, {Patel}, {Scolnic},
  {Foley}, {Molino}, {Brammer}, {Jauzac}, {Bradac}, {Coe}, {Broadhurst},
  {Diego}, {Graur}, {Hjorth}, {Hoag}, {Jha}, {Johnson}, {Kelly}, {Lam},
  {McCully}, {Medezinski}, {Meneghetti}, {Merten}, {Richard}, {Riess},
  {Sharon}, {Strolger}, {Treu}, {Wang}, {Williams}, \& {Zitrin}}]{Rod++15}
{Rodney}, S.~A., {Patel}, B., {Scolnic}, D., {et~al.} 2015{\natexlab{a}}, ArXiv
  e-prints, arXiv:1505.06211

\bibitem[{{Rodney} {et~al.}(2015{\natexlab{b}}){Rodney}, {Strolger}, {Kelly},
  {Bradac}, {Brammer}, {Filippenko}, {Foley}, {Graur}, {Hjorth}, {Jha},
  {McCully}, {Molino}, {Riess}, {Schmidt}, {Selsing}, {Sharon}, {Treu},
  {Weiner}, \& {Zitrin}}]{Rodney+15b}
{Rodney}, S.~A., {Strolger}, L.-G., {Kelly}, P.~L., {et~al.}
  2015{\natexlab{b}}, ArXiv e-prints, arXiv:1512.05734

\bibitem[{{Salpeter}(1955)}]{Sal55}
{Salpeter}, E.~E. 1955, \apj, 121, 161

\bibitem[{{Sand} {et~al.}(2008){Sand}, {Treu}, {Ellis}, {Smith}, \&
  {Kneib}}]{San++08}
{Sand}, D.~J., {Treu}, T., {Ellis}, R.~S., {Smith}, G.~P., \& {Kneib}, J. 2008,
  \apj, 674, 711

\bibitem[{Schmidt {et~al.}(2014)Schmidt, Treu, Brammer, Brada{\v c}, Wang,
  Dijkstra, Dressler, Fontana, Gavazzi, Henry, Hoag, Jones, Kelly, Malkan,
  Mason, Pentericci, Poggianti, Stiavelli, Trenti, von~der Linden, \&
  Vulcani}]{Sch++14}
Schmidt, K.~B., Treu, T., Brammer, G.~B., {et~al.} 2014, The Astrophysical
  Journal Letters, 782, L36

\bibitem[{{Schmidt} {et~al.}(2015){Schmidt}, {Treu}, {Brada{\v c}}, {Vulcani},
  {Huang}, {Hoag}, {Maseda}, {Guaita}, {Pentericci}, {Brammer}, {Dijkstra},
  {Dressler}, {Fontana}, {Henry}, {Jones}, {Mason}, {Trenti}, \&
  {Wang}}]{Sch++15}
{Schmidt}, K.~B., {Treu}, T., {Brada{\v c}}, M., {et~al.} 2015, ArXiv e-prints,
  arXiv:1511.04205

\bibitem[{{Schrabback} {et~al.}(2007){Schrabback}, {Erben}, {Simon},
  {Miralles}, {Schneider}, {Heymans}, {Eifler}, {Fosbury}, {Freudling},
  {Hetterscheidt}, {Hildebrandt}, \& {Pirzkal}}]{schrabback+07}
{Schrabback}, T., {Erben}, T., {Simon}, P., {et~al.} 2007, \aap, 468, 823

\bibitem[{{Schrabback} {et~al.}(2010){Schrabback}, {Hartlap}, {Joachimi},
  {Kilbinger}, {Simon}, {Benabed}, {Brada{\v c}}, {Eifler}, {Erben},
  {Fassnacht}, {High}, {Hilbert}, {Hildebrandt}, {Hoekstra}, {Kuijken},
  {Marshall}, {Mellier}, {Morganson}, {Schneider}, {Semboloni}, {van Waerbeke},
  \& {Velander}}]{schrabback+10}
{Schrabback}, T., {Hartlap}, J., {Joachimi}, B., {et~al.} 2010, \aap, 516, A63

\bibitem[{{Sharon} {et~al.}(2014){Sharon}, {Gladders}, {Rigby}, {Wuyts},
  {Bayliss}, {Johnson}, {Florian}, \& {Dahle}}]{Sha++14}
{Sharon}, K., {Gladders}, M.~D., {Rigby}, J.~R., {et~al.} 2014, \apj, 795, 50

\bibitem[{{Treu} \& {Ellis}(2014)}]{Treu+Ellis14}
{Treu}, T., \& {Ellis}, R.~S. 2014, ArXiv e-prints, arXiv:1412.6916

\bibitem[{{Treu} {et~al.}(2015){Treu}, {Schmidt}, {Brammer}, {Vulcani}, {Wang},
  {Brada{\v c}}, {Dijkstra}, {Dressler}, {Fontana}, {Gavazzi}, {Henry}, {Hoag},
  {Huang}, {Jones}, {Kelly}, {Malkan}, {Mason}, {Pentericci}, {Poggianti},
  {Stiavelli}, {Trenti}, \& {von der Linden}}]{Treu+15a}
{Treu}, T., {Schmidt}, K.~B., {Brammer}, G.~B., {et~al.} 2015, \apj, 812, 114

\bibitem[{{Treu} {et~al.}(2016){Treu}, {Brammer}, {Diego}, {Grillo}, {Kelly},
  {Oguri}, {Rodney}, {Rosati}, {Sharon}, {Zitrin}, {Balestra}, {Brada{\v c}},
  {Broadhurst}, {Caminha}, {Halkola}, {Hoag}, {Ishigaki}, {Johnson}, {Karman},
  {Kawamata}, {Mercurio}, {Schmidt}, {Strolger}, {Suyu}, {Filippenko}, {Foley},
  {Jha}, \& {Patel}}]{Treu+16}
{Treu}, T., {Brammer}, G., {Diego}, J.~M., {et~al.} 2016, \apj, 817, 60

\bibitem[{{Umetsu} {et~al.}(2014){Umetsu}, {Medezinski}, {Nonino}, {Merten},
  {Postman}, {Meneghetti}, {Donahue}, {Czakon}, {Molino}, {Seitz}, {Gruen},
  {Lemze}, {Balestra}, {Ben{\'{\i}}tez}, {Biviano}, {Broadhurst}, {Ford},
  {Grillo}, {Koekemoer}, {Melchior}, {Mercurio}, {Moustakas}, {Rosati}, \&
  {Zitrin}}]{Umetsu+14}
{Umetsu}, K., {Medezinski}, E., {Nonino}, M., {et~al.} 2014, \apj, 795, 163

\bibitem[{{Wang} {et~al.}(2015){Wang}, {Hoag}, {Huang}, {Treu}, {Brada{\v c}},
  {Schmidt}, {Brammer}, {Vulcani}, {Jones}, {Ryan}, {Amor{\'{\i}}n},
  {Castellano}, {Fontana}, {Merlin}, \& {Trenti}}]{Wang+15}
{Wang}, X., {Hoag}, A., {Huang}, K.-H., {et~al.} 2015, \apj, 811, 29

\bibitem[{{Werner} {et~al.}(2004){Werner}, {Roellig}, {Low}, {Rieke}, {Rieke},
  {Hoffmann}, {Young}, {Houck}, {Brandl}, {Fazio}, {Hora}, {Gehrz}, {Helou},
  {Soifer}, {Stauffer}, {Keene}, {Eisenhardt}, {Gallagher}, {Gautier}, {Irace},
  {Lawrence}, {Simmons}, {Van Cleve}, {Jura}, {Wright}, \&
  {Cruikshank}}]{Werner+04}
{Werner}, M.~W., {Roellig}, T.~L., {Low}, F.~J., {et~al.} 2004, \apjs, 154, 1

\bibitem[{{York} {et~al.}(2000){York}, {Adelman}, {Anderson}, {Anderson},
  {Annis}, {Bahcall}, {Bakken}, {Barkhouser}, {Bastian}, {Berman}, {Boroski},
  {Bracker}, {Briegel}, {Briggs}, {Brinkmann}, {Brunner}, {Burles}, {Carey},
  {Carr}, {Castander}, {Chen}, {Colestock}, {Connolly}, {Crocker}, {Csabai},
  {Czarapata}, {Davis}, {Doi}, {Dombeck}, {Eisenstein}, {Ellman}, {Elms},
  {Evans}, {Fan}, {Federwitz}, {Fiscelli}, {Friedman}, {Frieman}, {Fukugita},
  {Gillespie}, {Gunn}, {Gurbani}, {de Haas}, {Haldeman}, {Harris}, {Hayes},
  {Heckman}, {Hennessy}, {Hindsley}, {Holm}, {Holmgren}, {Huang}, {Hull},
  {Husby}, {Ichikawa}, {Ichikawa}, {Ivezi{\'c}}, {Kent}, {Kim}, {Kinney},
  {Klaene}, {Kleinman}, {Kleinman}, {Knapp}, {Korienek}, {Kron}, {Kunszt},
  {Lamb}, {Lee}, {Leger}, {Limmongkol}, {Lindenmeyer}, {Long}, {Loomis},
  {Loveday}, {Lucinio}, {Lupton}, {MacKinnon}, {Mannery}, {Mantsch}, {Margon},
  {McGehee}, {McKay}, {Meiksin}, {Merelli}, {Monet}, {Munn}, {Narayanan},
  {Nash}, {Neilsen}, {Neswold}, {Newberg}, {Nichol}, {Nicinski}, {Nonino},
  {Okada}, {Okamura}, {Ostriker}, {Owen}, {Pauls}, {Peoples}, {Peterson},
  {Petravick}, {Pier}, {Pope}, {Pordes}, {Prosapio}, {Rechenmacher}, {Quinn},
  {Richards}, {Richmond}, {Rivetta}, {Rockosi}, {Ruthmansdorfer}, {Sandford},
  {Schlegel}, {Schneider}, {Sekiguchi}, {Sergey}, {Shimasaku}, {Siegmund},
  {Smee}, {Smith}, {Snedden}, {Stone}, {Stoughton}, {Strauss}, {Stubbs},
  {SubbaRao}, {Szalay}, {Szapudi}, {Szokoly}, {Thakar}, {Tremonti}, {Tucker},
  {Uomoto}, {Vanden Berk}, {Vogeley}, {Waddell}, {Wang}, {Watanabe},
  {Weinberg}, {Yanny}, {Yasuda}, \& {SDSS Collaboration}}]{York+00}
{York}, D.~G., {Adelman}, J., {Anderson}, Jr., J.~E., {et~al.} 2000, \aj, 120,
  1579

\bibitem[{Zheng {et~al.}(2012)Zheng, Postman, Zitrin, Moustakas, Shu, Jouvel,
  H{\o}st, Molino, Bradley, Coe, Moustakas, Carrasco, Ford, Ben{\'\i}tez,
  Lauer, Seitz, Bouwens, Koekemoer, Medezinski, Bartelmann, Broadhurst,
  Donahue, Grillo, Infante, Jha, Kelson, Lahav, Lemze, Melchior, Meneghetti,
  Merten, Nonino, Ogaz, Rosati, Umetsu, \& van~der Wel}]{Zheng:2012p33879}
Zheng, W., Postman, M., Zitrin, A., {et~al.} 2012, Nature, 489, 406

\bibitem[{{Zitrin} {et~al.}(2013){Zitrin}, {Meneghetti}, {Umetsu},
  {Broadhurst}, {Bartelmann}, {Bouwens}, {Bradley}, {Carrasco}, {Coe}, {Ford},
  {Kelson}, {Koekemoer}, {Medezinski}, {Moustakas}, {Moustakas}, {Nonino},
  {Postman}, {Rosati}, {Seidel}, {Seitz}, {Sendra}, {Shu}, {Vega}, \&
  {Zheng}}]{Zitrin+13}
{Zitrin}, A., {Meneghetti}, M., {Umetsu}, K., {et~al.} 2013, \apjl, 762, L30

\bibitem[{{Zitrin} {et~al.}(2014){Zitrin}, {Zheng}, {Broadhurst}, {Moustakas},
  {Lam}, {Shu}, {Huang}, {Diego}, {Ford}, {Lim}, {Bauer}, {Infante}, {Kelson},
  \& {Molino}}]{Zit++14}
{Zitrin}, A., {Zheng}, W., {Broadhurst}, T., {et~al.} 2014, \apjl, 793, L12

\bibitem[{{Zitrin} {et~al.}(2015){Zitrin}, {Fabris}, {Merten}, {Melchior},
  {Meneghetti}, {Koekemoer}, {Coe}, {Maturi}, {Bartelmann}, {Postman},
  {Umetsu}, {Seidel}, {Sendra}, {Broadhurst}, {Balestra}, {Biviano}, {Grillo},
  {Mercurio}, {Nonino}, {Rosati}, {Bradley}, {Carrasco}, {Donahue}, {Ford},
  {Frye}, \& {Moustakas}}]{Zitrin+15}
{Zitrin}, A., {Fabris}, A., {Merten}, J., {et~al.} 2015, \apj, 801, 44

\end{thebibliography}
\begin{appendix}
\section{Redshift PDFs for the Silver sample multiply imaged systems not used in the lens model }
\label{sec:apA}
\begin{figure*}
  \centering
  \includegraphics[width=0.49\textwidth]{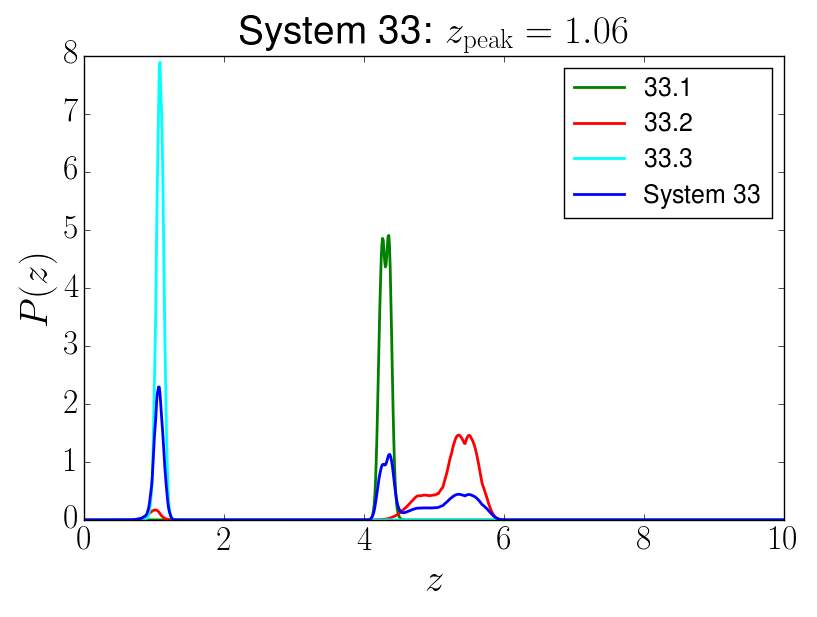} 
  \includegraphics[width=0.49\textwidth]{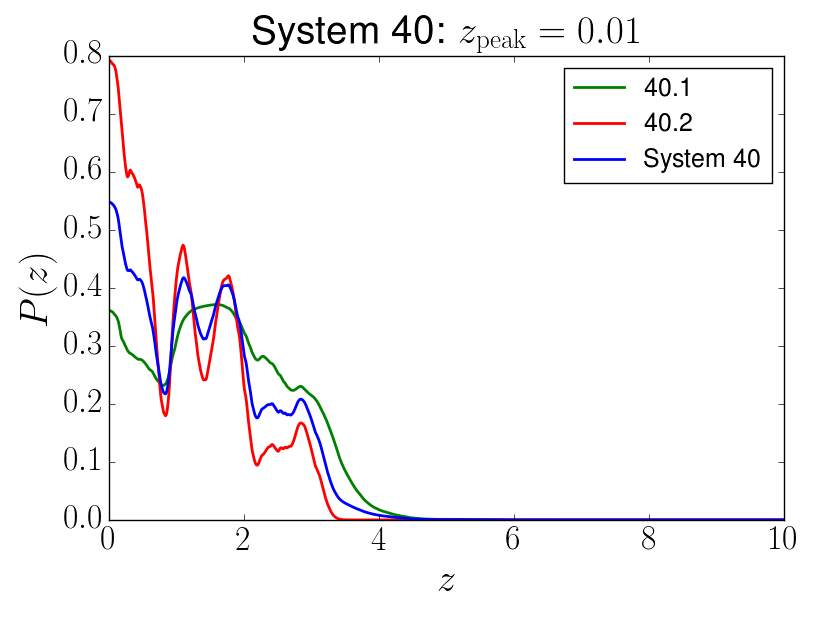} 
    \includegraphics[width=0.49\textwidth]{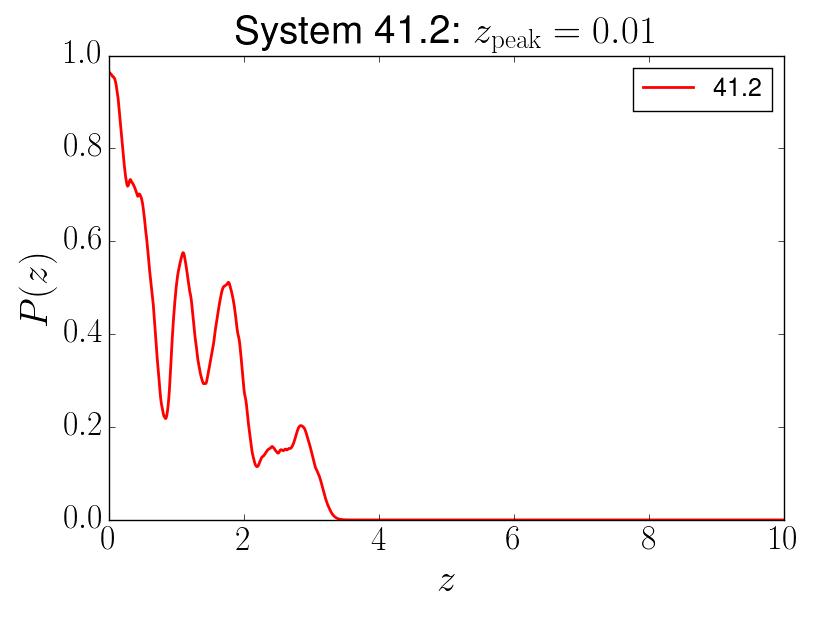} 
      \includegraphics[width=0.49\textwidth]{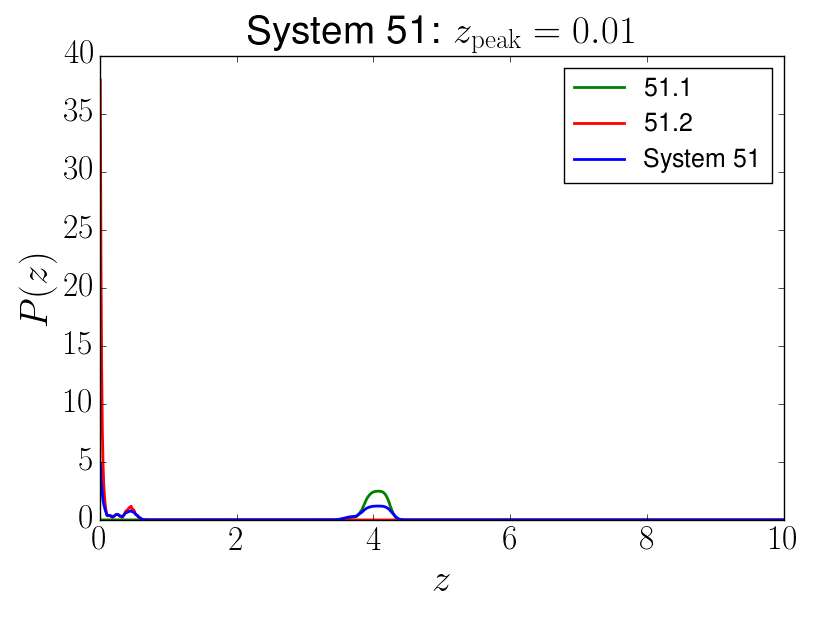} 
  \caption{Probability density functions (PDFs) for the redshifts of multiple image systems in the Silver sample, which are too poorly constrained to use in the lens model. In all but the bottom left panel, the blue line represents the combined redshift ($z_{\mathrm{Bayes}}$) PDF, derived using the hierarchical Bayesian method developed by \citet{Dahlen:2013p33380}. In system 41 (bottom left panel), only image 41.2 was detected in the photometric catalog, so multiple redshifts could not be combined. What is shown is the single PDF for the photometric redshift of 41.2, which is very poorly constrained. }
  \label{fig:zbayes33} 
\end{figure*}

In Figure~\ref{fig:zbayes33} we show the PDFs of the individual and, where available, combined redshifts of the multiply imaged systems in the Silver sample that we were unable to use in our lens model. We believe these systems are real, but we were unable to sufficiently constrain their redshifts to include them in the lens model. 

\clearpage
\section{GLASS spectra of multiply imaged galaxies}
\begin{figure*}[b]
  \centering
  \includegraphics[width=0.95\textwidth]{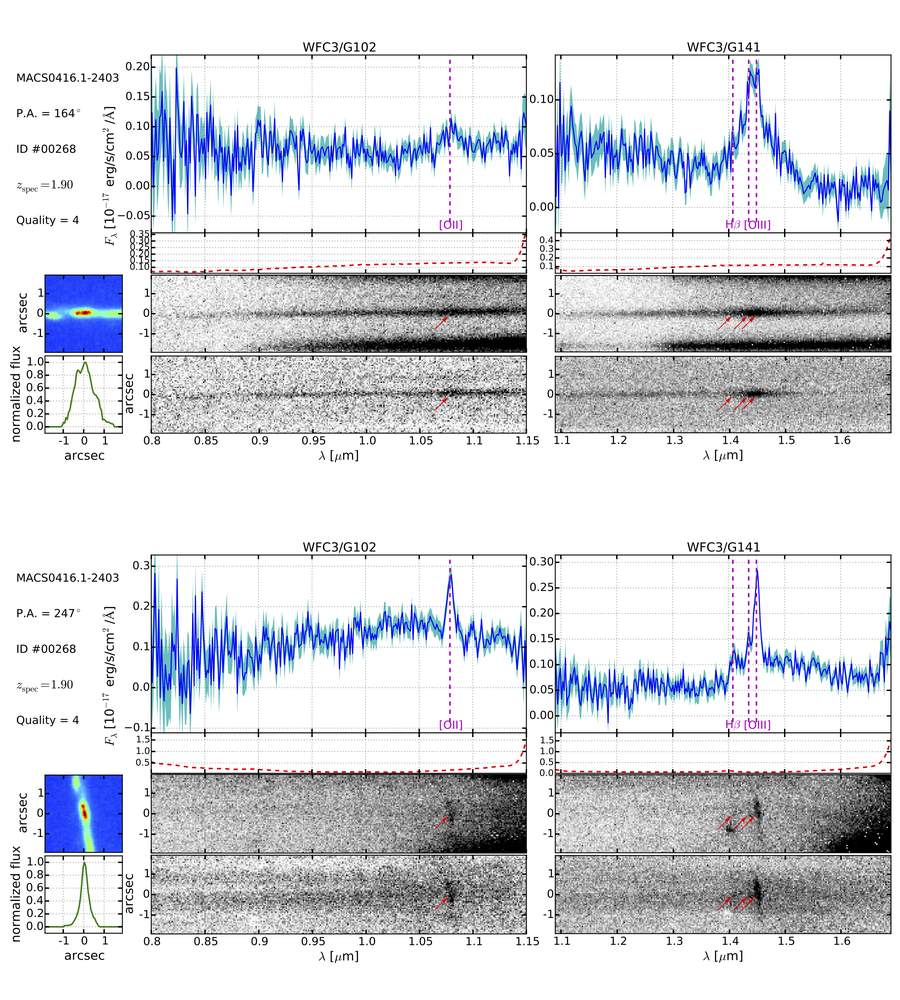}
  \caption{Grism spectroscopic confirmation for ID \#268 (arc 2.1) observed at the two P.A.s shown in the two sub-figures.
  In each sub-figure, the two panels on top show the 1-dimensional spectra, where the observed flux and contamination model are
  denoted by blue solid and red dashed lines respectively. The cyan shaded region represents the noise level. The six panels
  at the bottom show the 2-dimensional postage stamp created from the coadded HFF+CLASH+GLASS image, the 1-dimensional collapsed image, 
  and the interlaced 2-dimensional spectra without (top) and with (bottom) the contamination 
  subtracted. In the 1- and 2-dimensional spectra, the identified emission lines are denoted by vertical dashed lines in magenta 
  and arrows in red respectively.}
  \label{fig:ELarc2.1} 
\end{figure*}

\begin{figure*}
  \centering
  \includegraphics[width=\textwidth]{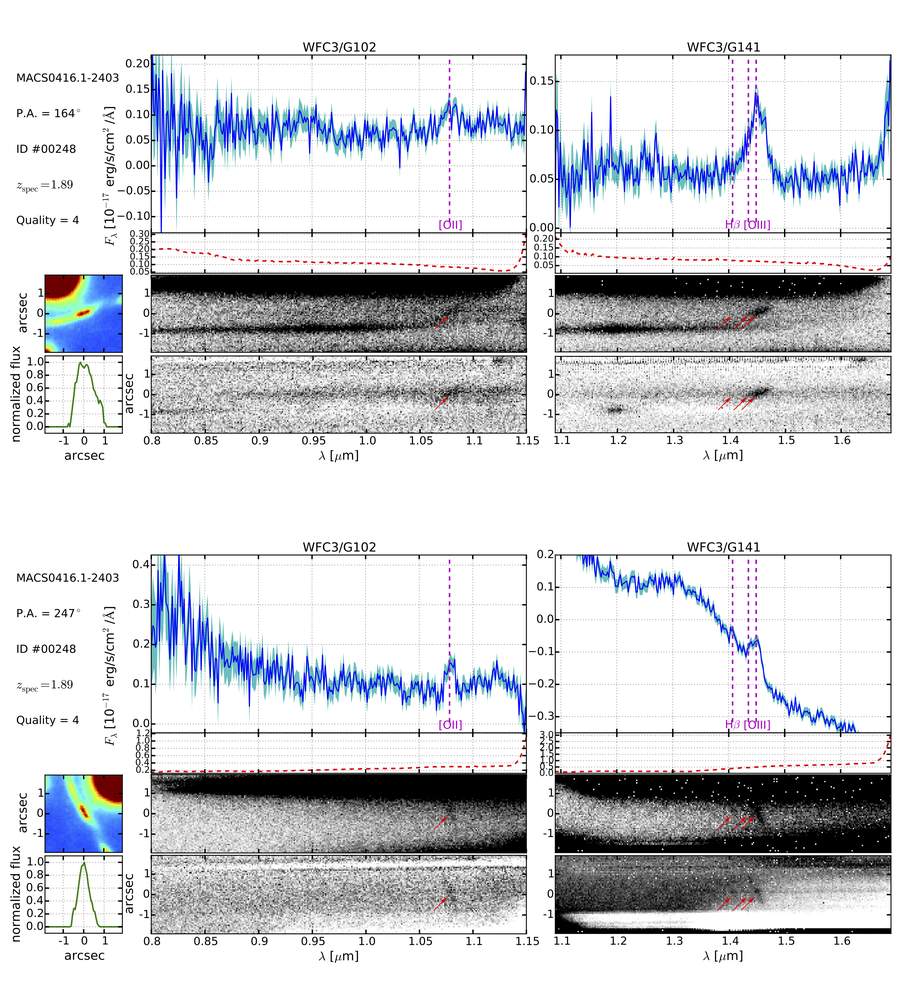}
  \caption{Same as Figure~\ref{fig:ELarc2.1}, except that object ID \#248 (arc 2.2) is shown.}
  \label{fig:ELarc2.2}
\end{figure*}

\begin{figure*}
  \centering
  \includegraphics[width=\textwidth]{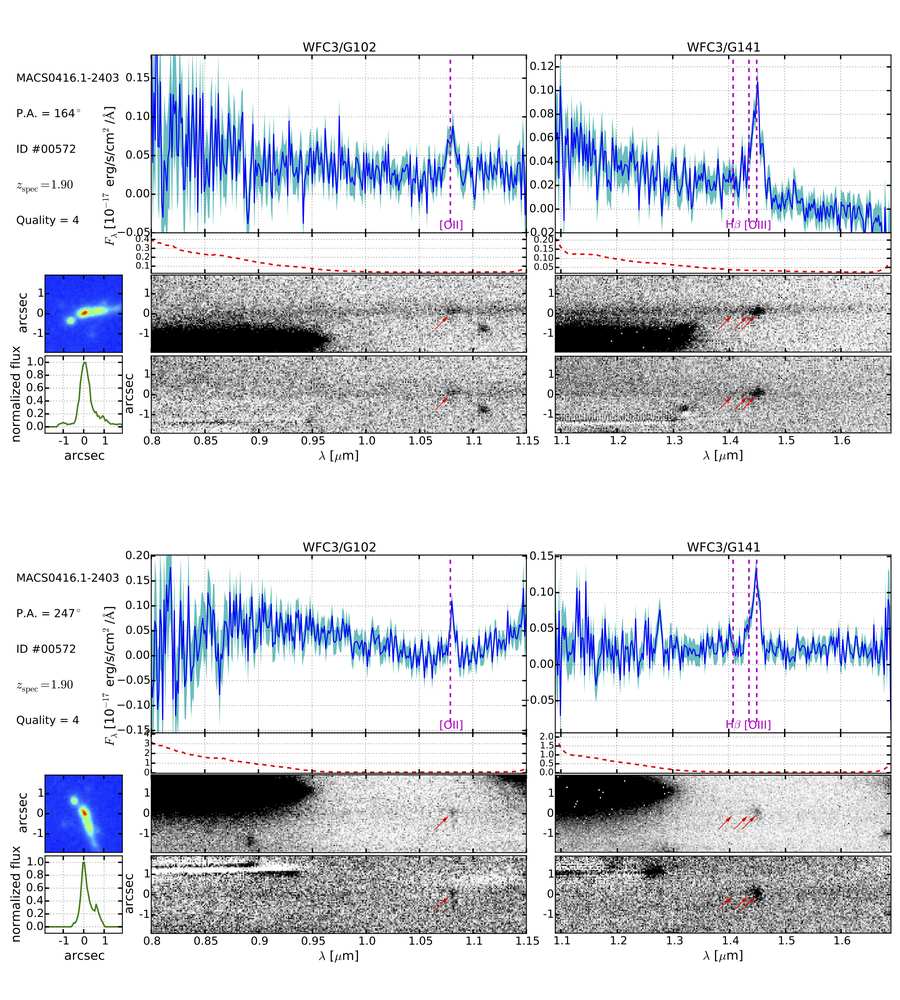}
  \caption{Same as Figure~\ref{fig:ELarc2.1}, except that object ID \#572 (arc 2.3) is shown. }
  \label{fig:ELarc2.3}
\end{figure*}

\begin{figure*}
  \centering
  \includegraphics[width=\textwidth]{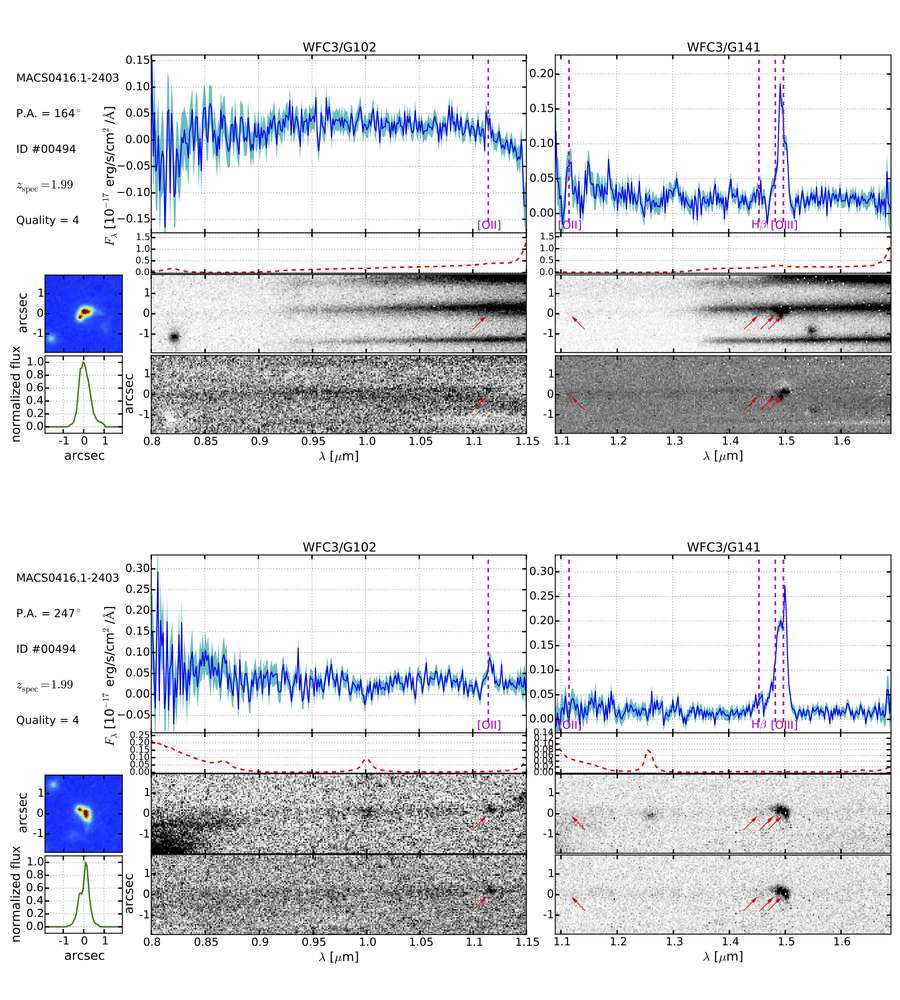}\\
  \caption{Same as Figure~\ref{fig:ELarc2.1}, except that object ID \#494 (arc 3.1/4.1) is shown.}
  \label{fig:ELarc3.1}
\end{figure*}

\begin{figure*}
  \centering
  \includegraphics[width=\textwidth]{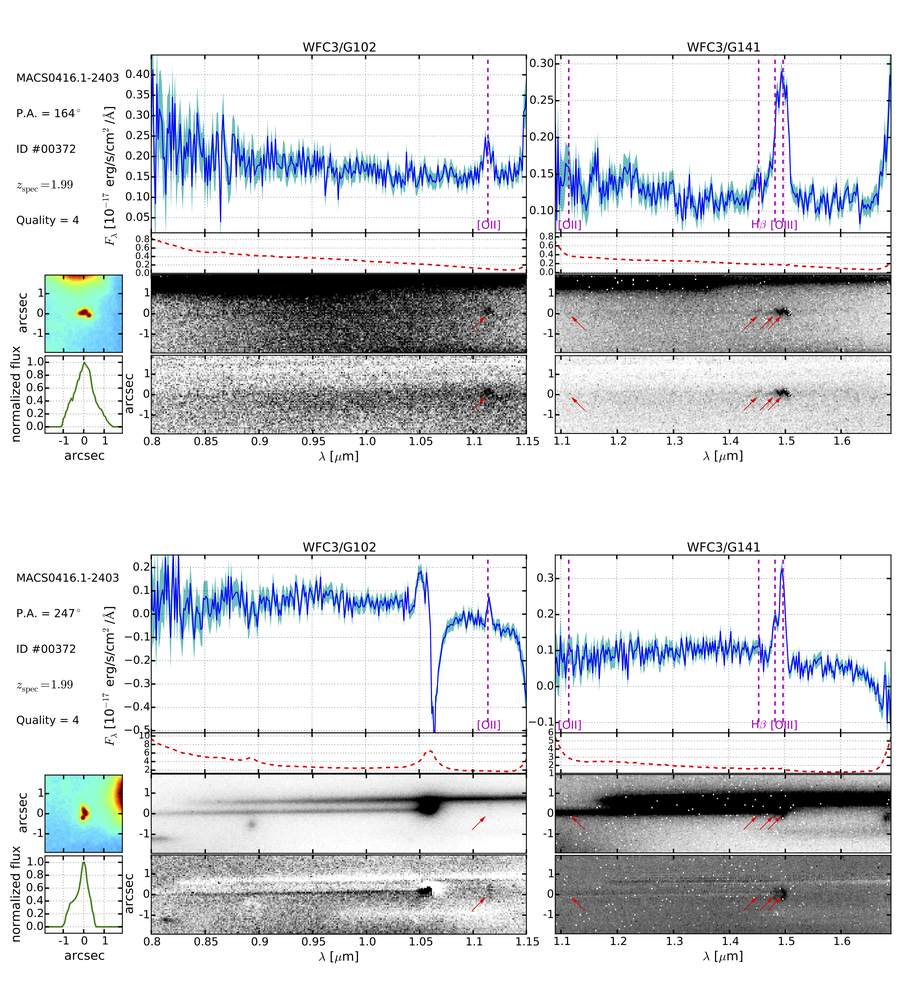}
  \caption{Same as Figure~\ref{fig:ELarc2.1}, except that object ID \#372 (arc 3.2/4.2) is shown.}
  \label{fig:ELarc3.2}
\end{figure*}

\begin{figure*}
  \centering
  \includegraphics[width=\textwidth]{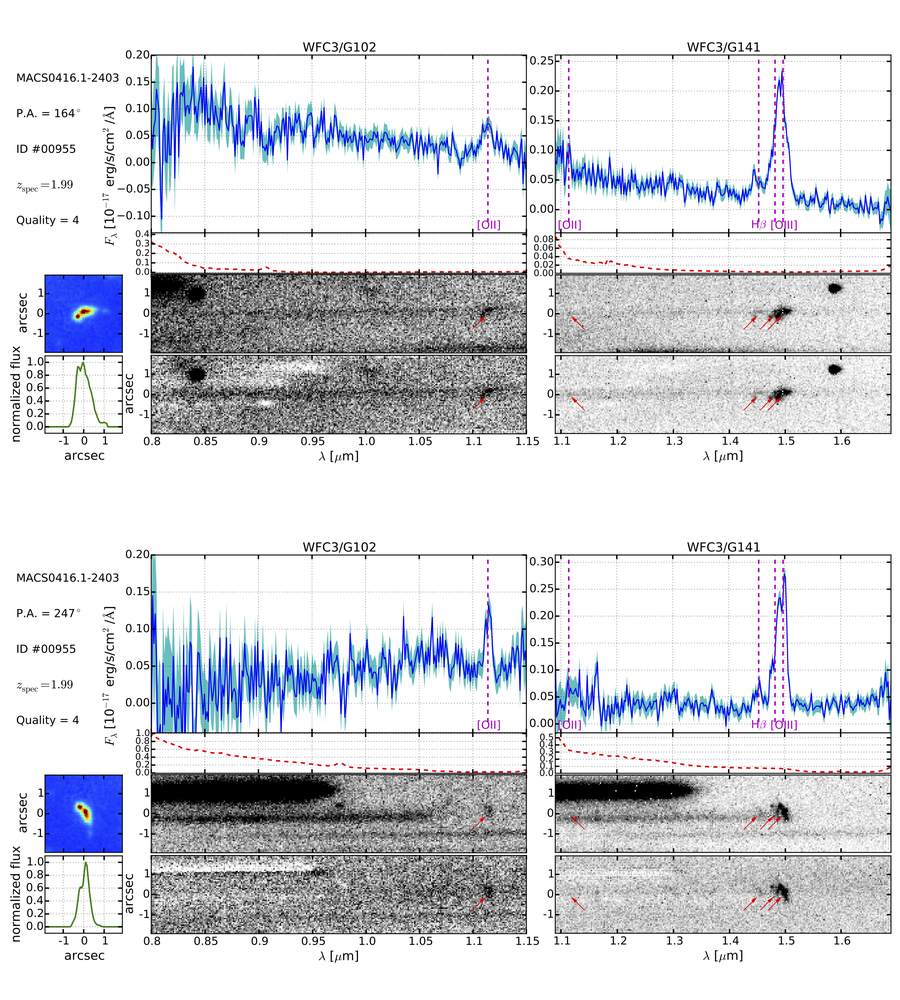}\\
  \caption{Same as Figure~\ref{fig:ELarc2.1}, except that object ID \#955 (arc 3.3/4.3) is shown.}
  \label{fig:ELarc3.3}
\end{figure*}

\begin{figure*}
  \centering
  \includegraphics[width=\textwidth]{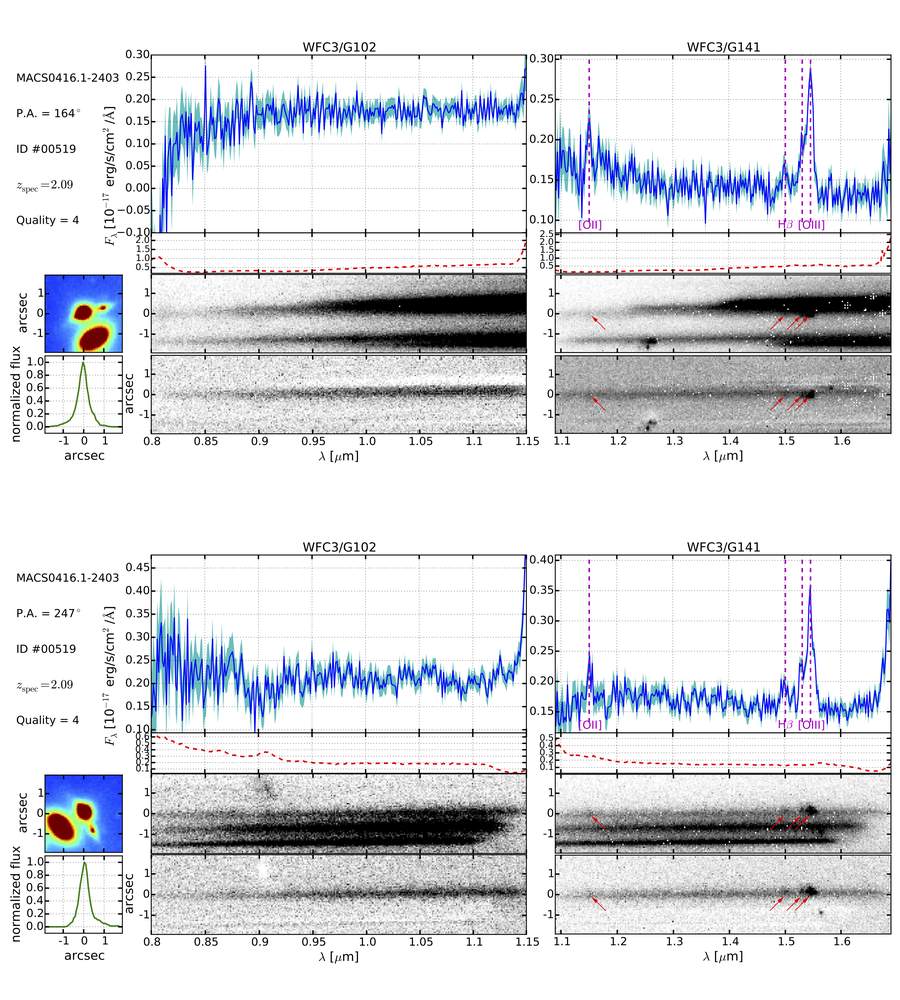}\\
  \caption{Same as Figure~\ref{fig:ELarc2.1}, except that object ID \#519 (arc 5.1) is shown. The multiple image is heavily contaminated by a foreground cluster member, as can be seen in the two-dimensional HFF postage stamp. However, after subtracting the contamination, the emission lines are clearly detected.  }
  \label{fig:ELarc5.1}
\end{figure*}

\begin{figure*}
  \centering
 \includegraphics[width=\textwidth]{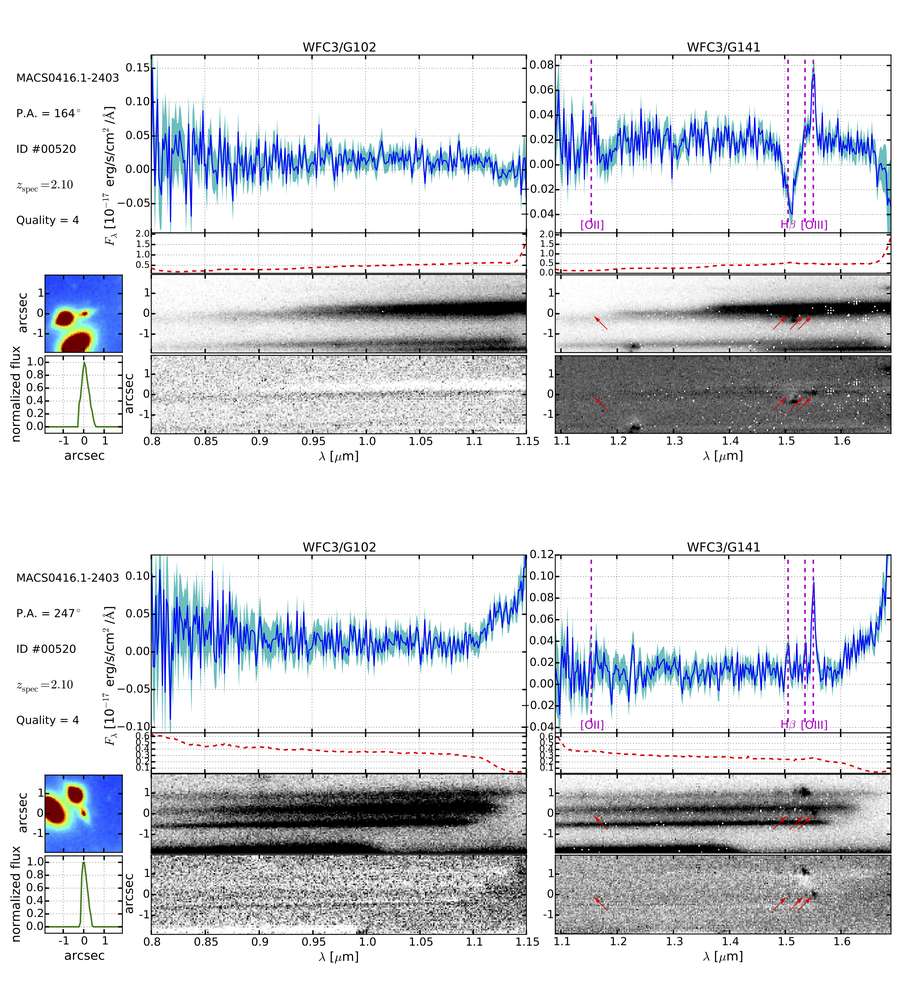}
  \caption{Same as Figure~\ref{fig:ELarc2.1}, except that object ID \#520 (arc 5.2) is shown. }
  \label{fig:ELarc5.2}
\end{figure*}

\begin{figure*}
  \centering
  \includegraphics[width=\textwidth]{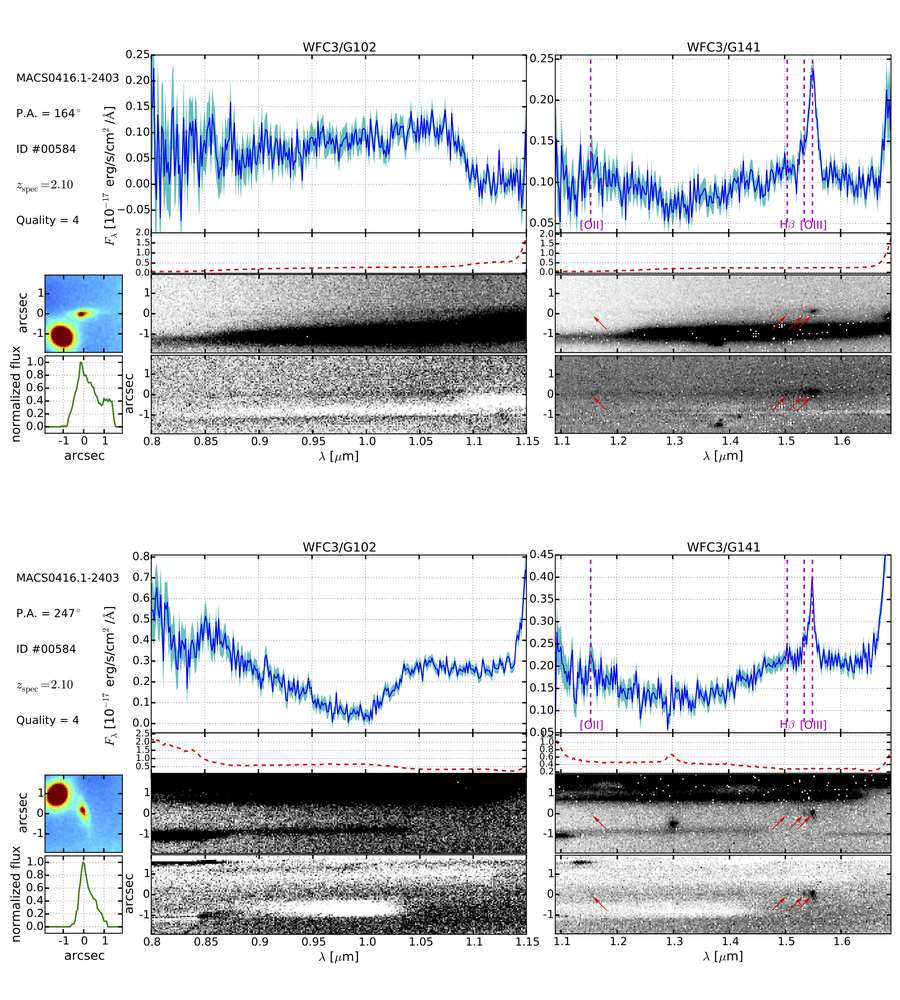}
  \caption{Same as Figure~\ref{fig:ELarc2.1}, except that object ID \#584 (arc 5.3) is shown.}
  \label{fig:ELarc5.3}
\end{figure*}

\begin{figure*}
  \centering
  \includegraphics[width=\textwidth]{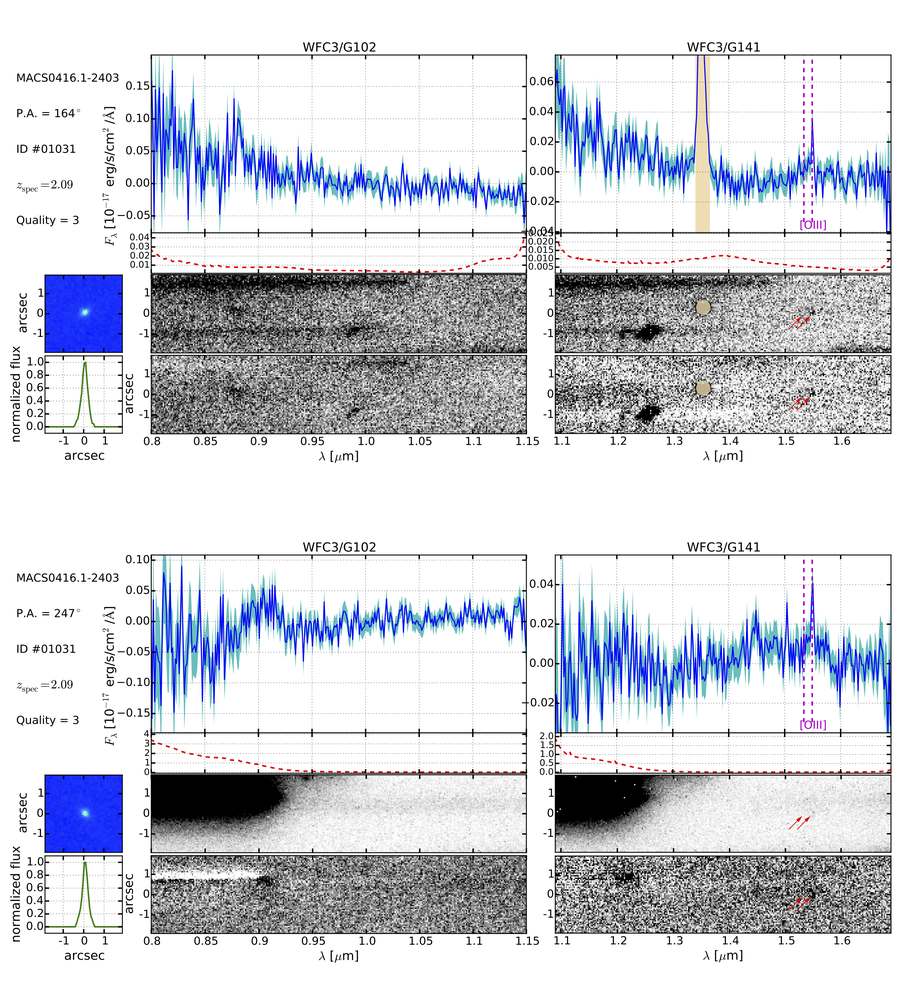}
  \caption{Same as Figure~\ref{fig:ELarc2.1}, except that object ID \#1031 (arc 5.4) is shown. The tan colored region in the one- and two-dimensional P.A.=164$^\circ$ G141 spectra covers a contaminant that was not identified by the contamination model.  }
  \label{fig:ELarc5.4}
\end{figure*}

\begin{figure*}
  \centering
  \includegraphics[width=\textwidth]{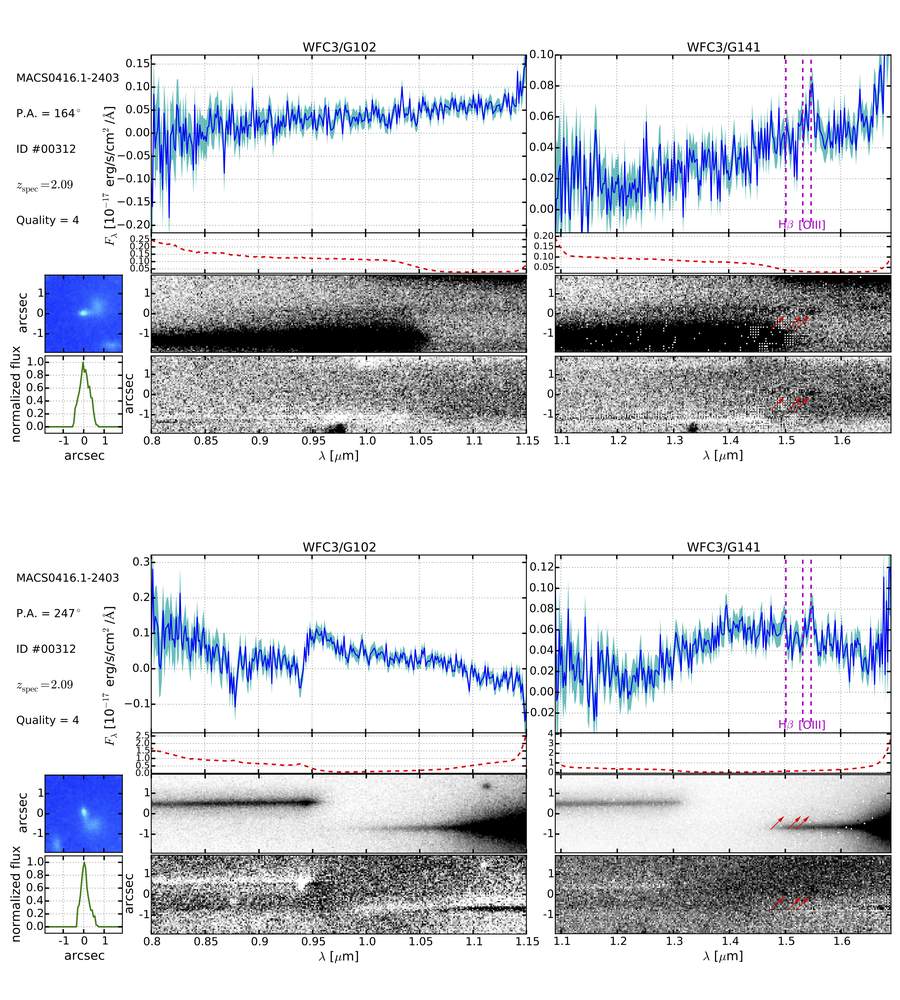}
  \caption{Same as Figure~\ref{fig:ELarc2.1}, except that object ID \#312 (arc 7.1) is shown.}
  \label{fig:ELarc7.1}
\end{figure*}

\begin{figure*}
  \centering
  \includegraphics[width=\textwidth]{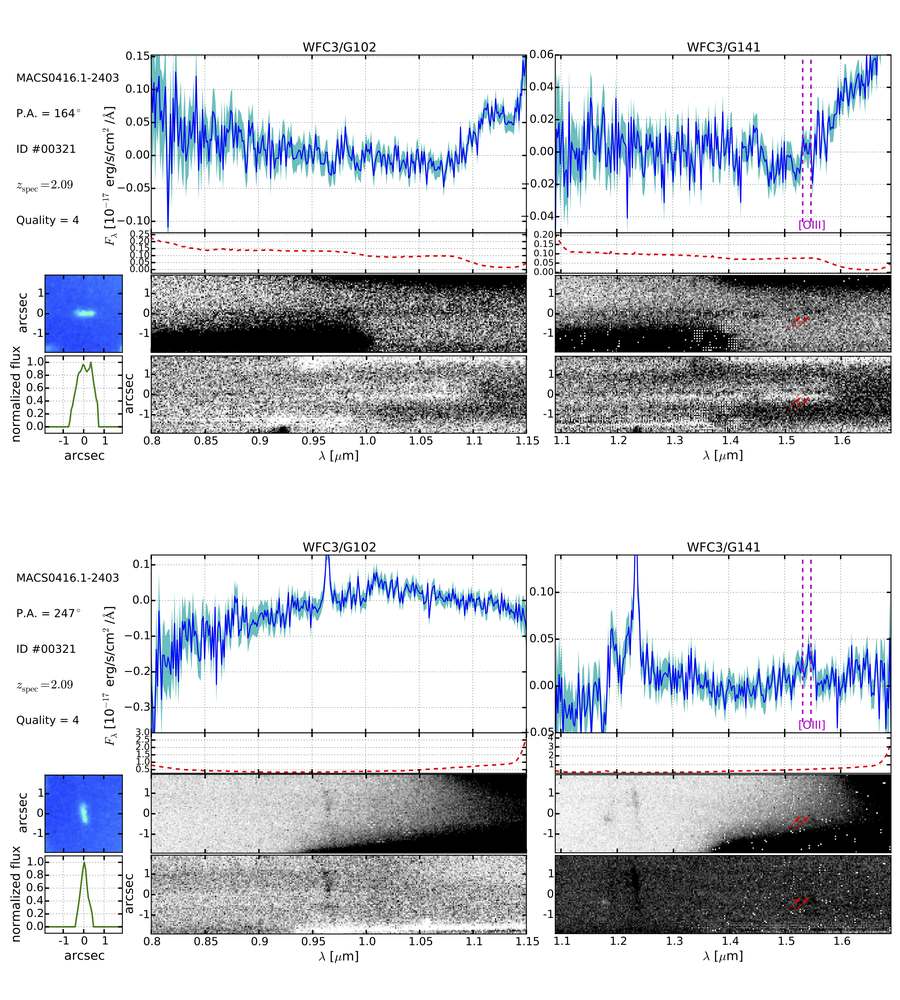}
  \caption{Same as Figure~\ref{fig:ELarc2.1}, except that object ID \#321 (arc 7.2) is shown.}
  \label{fig:ELarc7.2}
\end{figure*}

\begin{figure*}
  \centering
  \includegraphics[width=\textwidth]{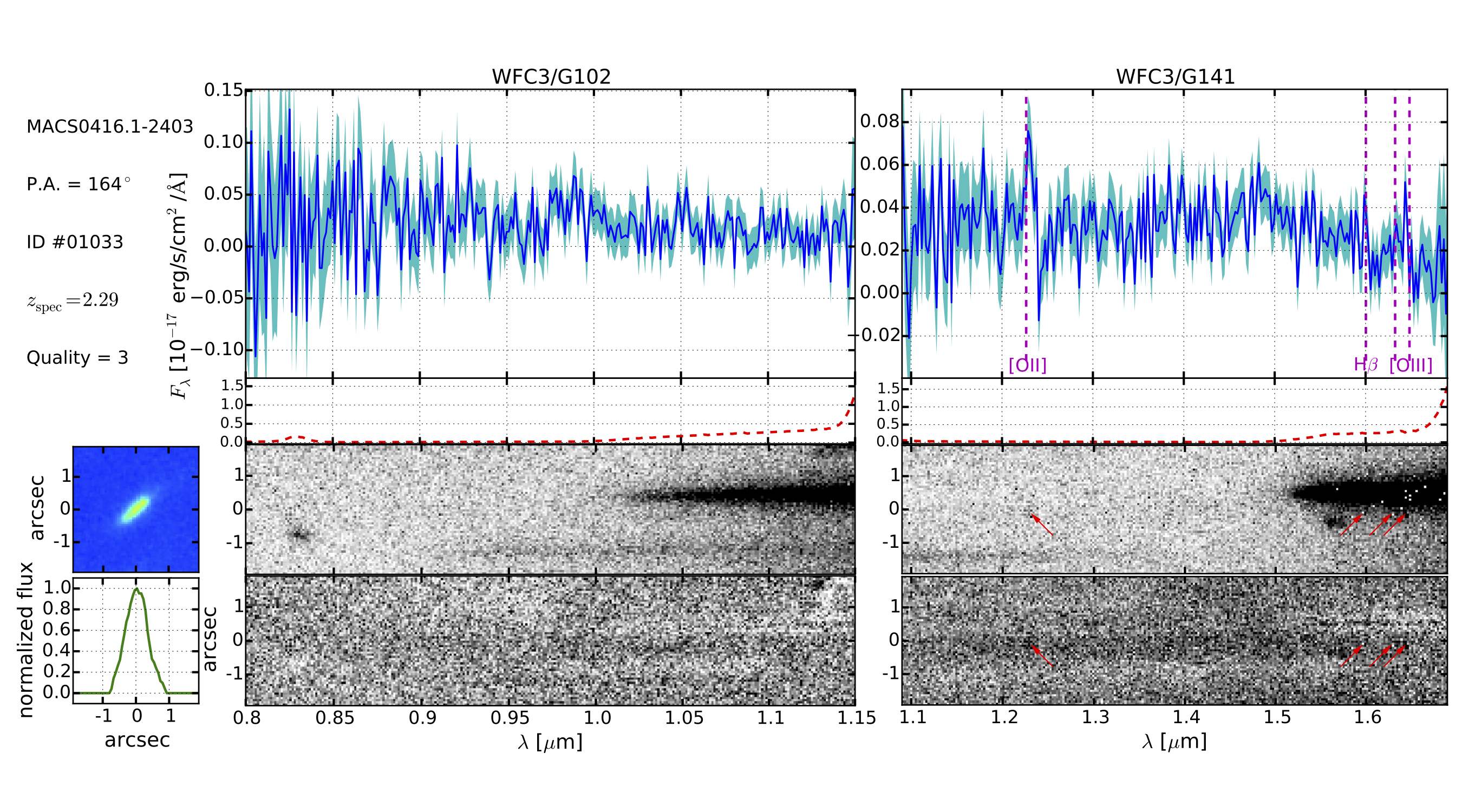}
  \caption{Same as Figure~\ref{fig:ELarc2.1}, except that object ID \#1033 (arc 10.1) is shown for a single PA. PA=247 is heavily contaminated, making the identification of emission lines extremely difficult. }
  \label{fig:ELarc10.1}
\end{figure*}

\begin{figure*}
  \centering
  \includegraphics[width=\textwidth]{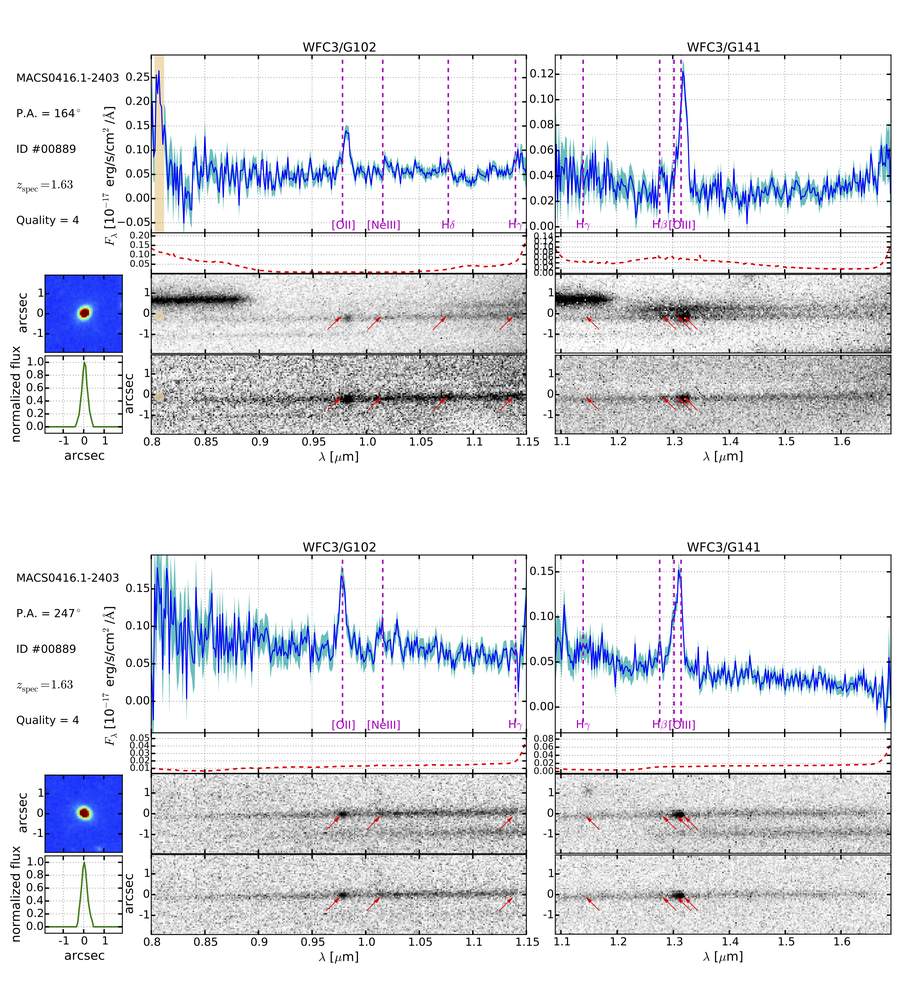}
  \caption{Same as Figure~\ref{fig:ELarc2.1}, except that object ID \#889 (arc 14.1) is shown. The tan colored region in the one- and two-dimensional PA=164$^\circ$ G102 spectra covers a contaminant that was not identified by the contamination model.}
  \label{fig:ELarc14.1}
\end{figure*}

\begin{figure*}
  \centering
  \includegraphics[width=\textwidth]{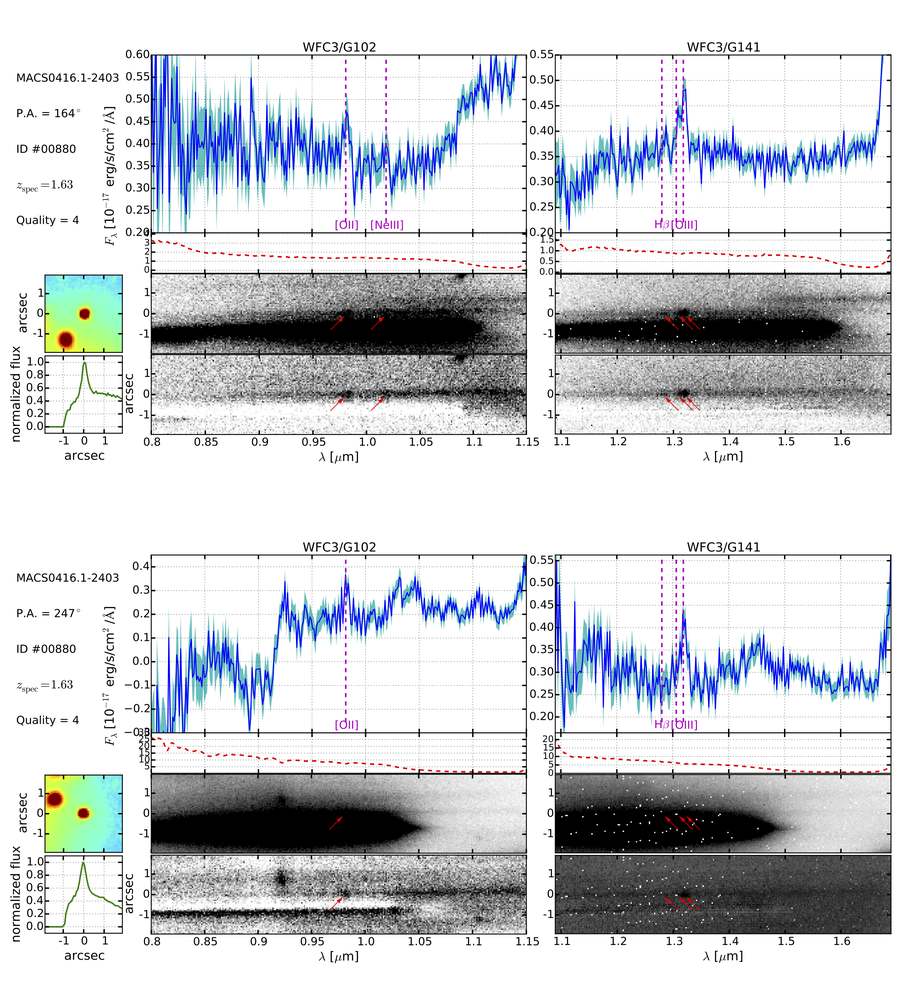}
  \caption{Same as Figure~\ref{fig:ELarc2.1}, except that object ID \#880 (arc 14.2) is shown.}
  \label{fig:ELarc14.2}
\end{figure*}

\begin{figure*}
  \centering
  \includegraphics[width=\textwidth]{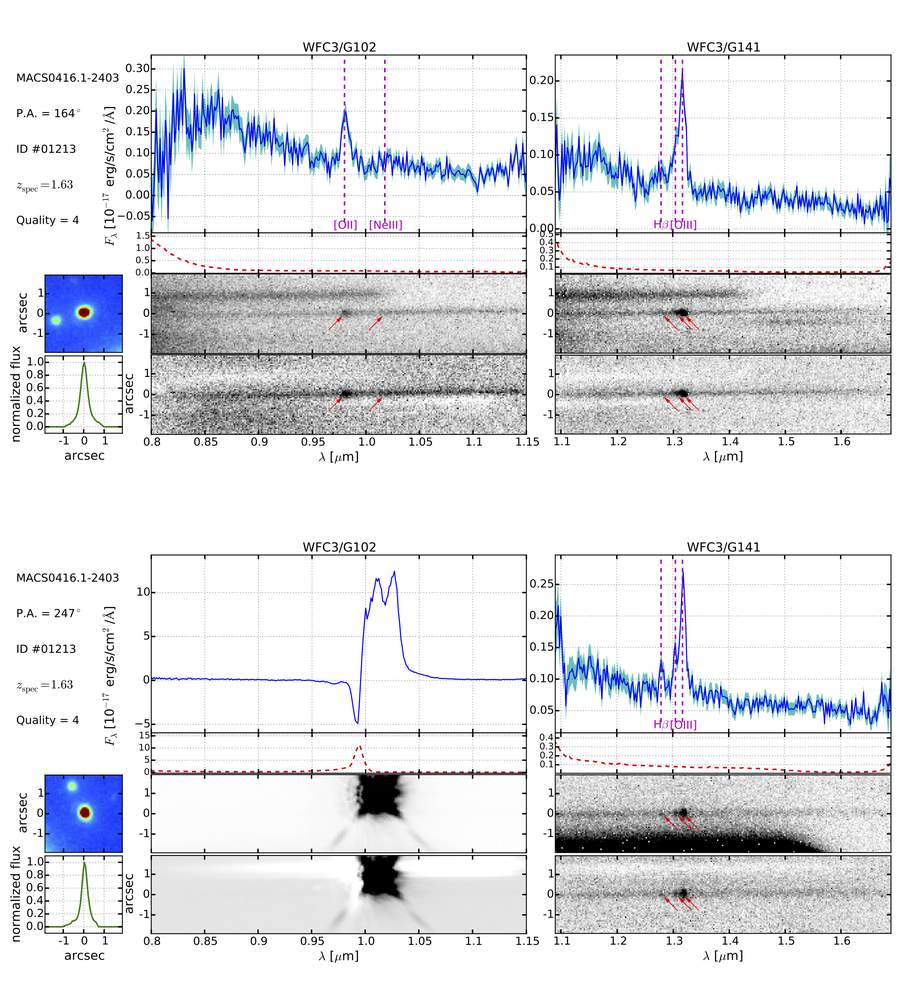}
  \caption{Same as Figure~\ref{fig:ELarc2.1}, except that object ID \#1213 (arc 14.3) is shown. The  P.A.=$247^{\circ}$  G102 spectra is entirely contaminated by a bright star.}
  \label{fig:ELarc14.3}
\end{figure*}

\begin{figure*}
  \centering
  \includegraphics[width=\textwidth]{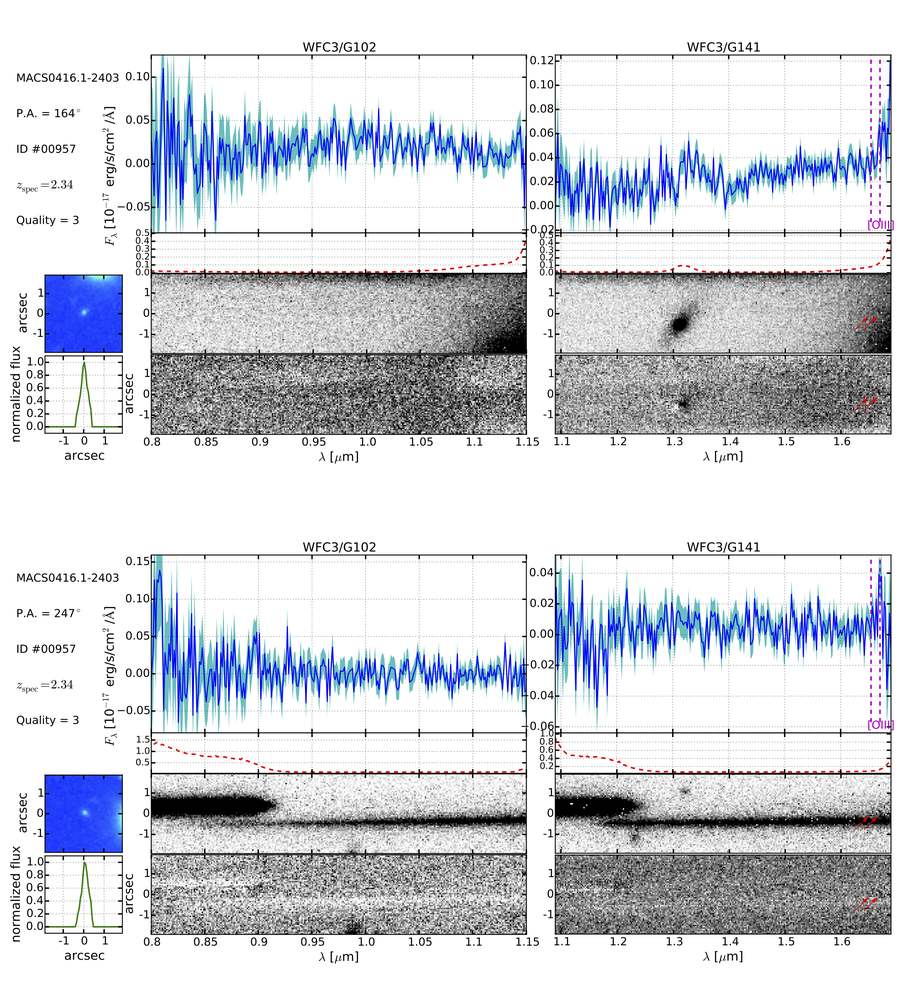}
  \caption{Same as Figure~\ref{fig:ELarc2.1}, except that object ID \#957 (arc 15.1) is shown.}
  \label{fig:ELarc15.1}
\end{figure*}

\begin{figure*}
  \centering
  \includegraphics[width=\textwidth]{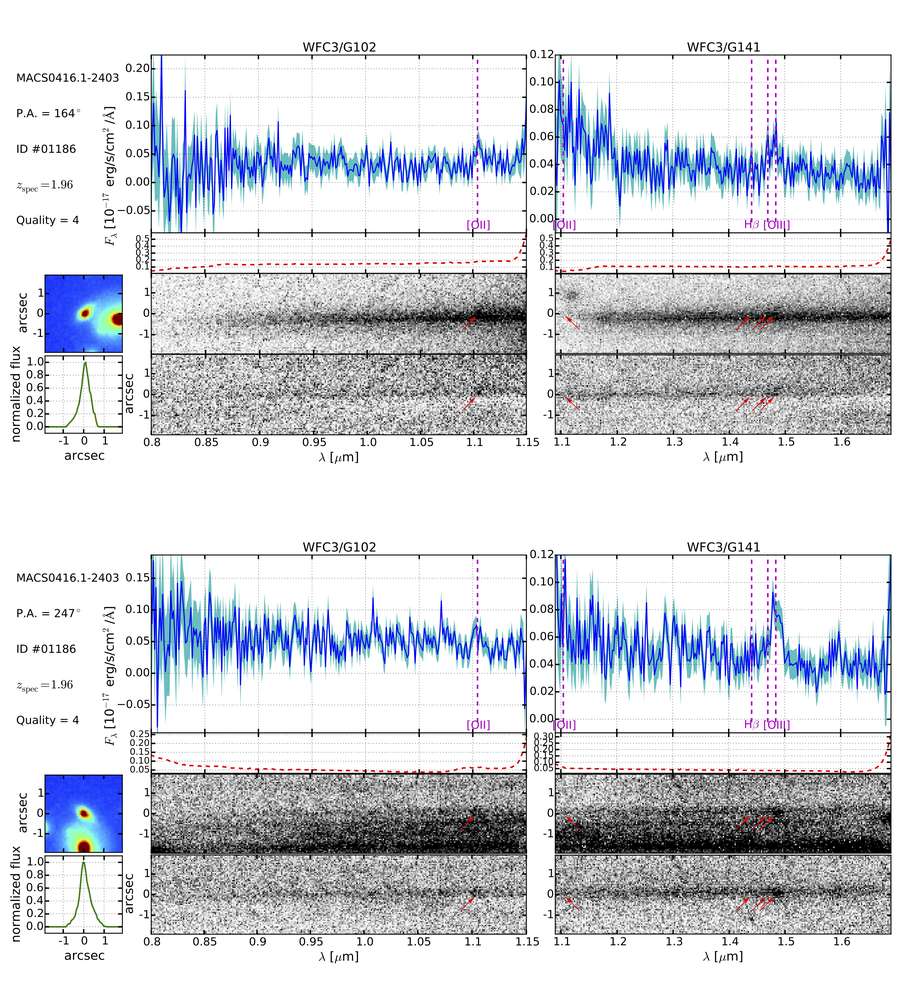}
  \caption{Same as Figure~\ref{fig:ELarc2.1}, except that object ID \#1186 (arc 16.1) is shown.}
  \label{fig:ELarc16.1}
\end{figure*}

\begin{figure*}
  \centering
  \includegraphics[width=\textwidth]{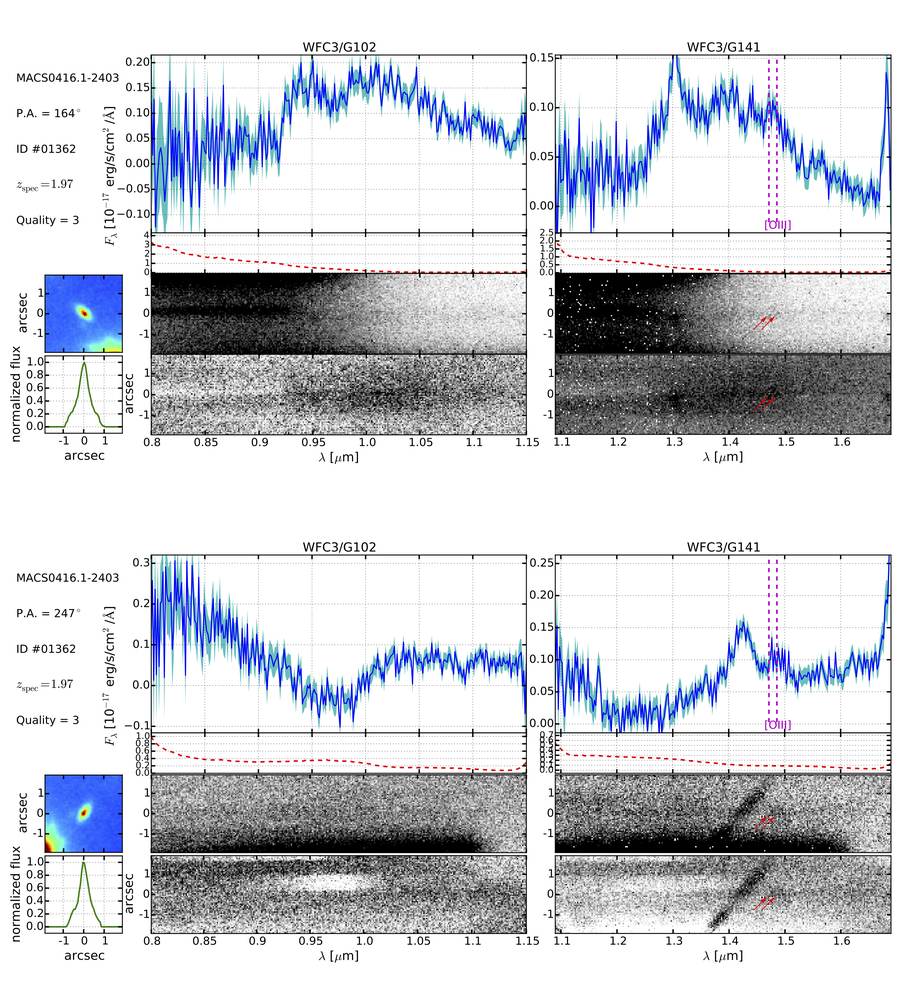}
  \caption{Same as Figure~\ref{fig:ELarc2.1}, except that object ID \#1362 (arc 16.3) is shown.}
  \label{fig:ELarc16.3}
\end{figure*}

\begin{figure*}
  \centering
  \includegraphics[width=\textwidth]{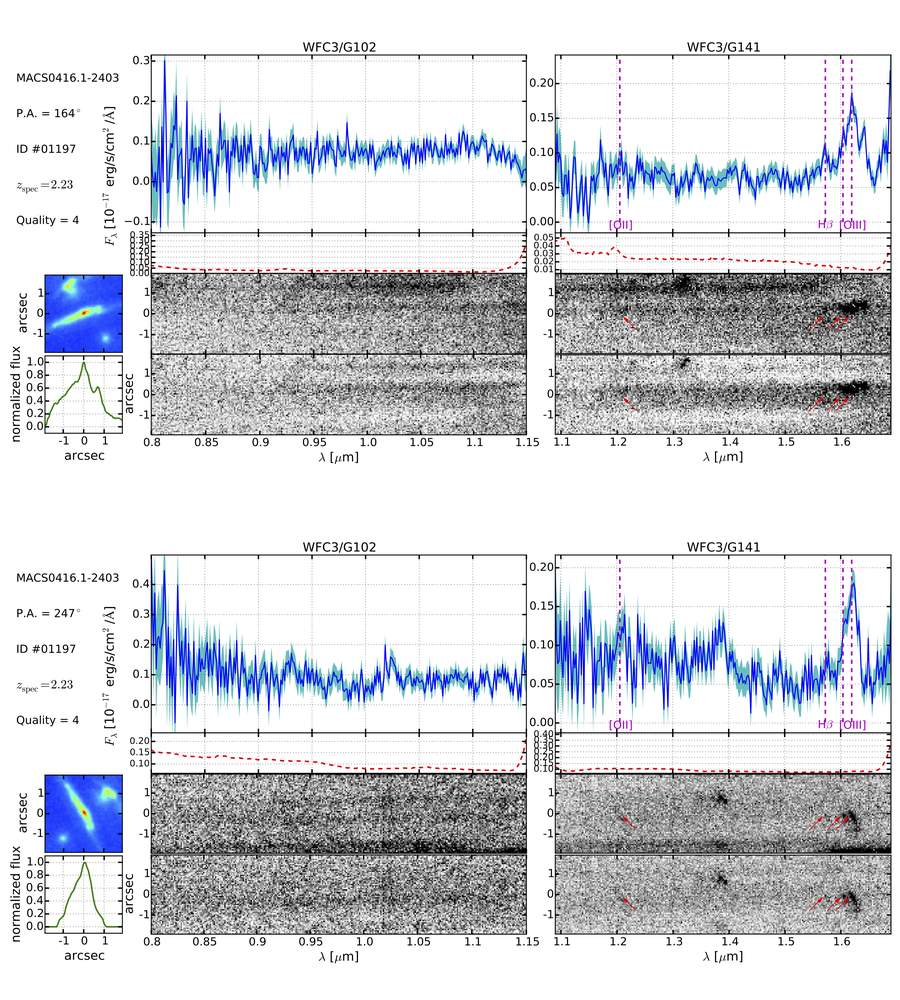}
  \caption{Same as Figure~\ref{fig:ELarc2.1}, except that object ID \#1197 (arc 17.3) is shown.}
  \label{fig:ELarc17.3}
\end{figure*}

\begin{figure*}
  \centering
  \includegraphics[width=\textwidth]{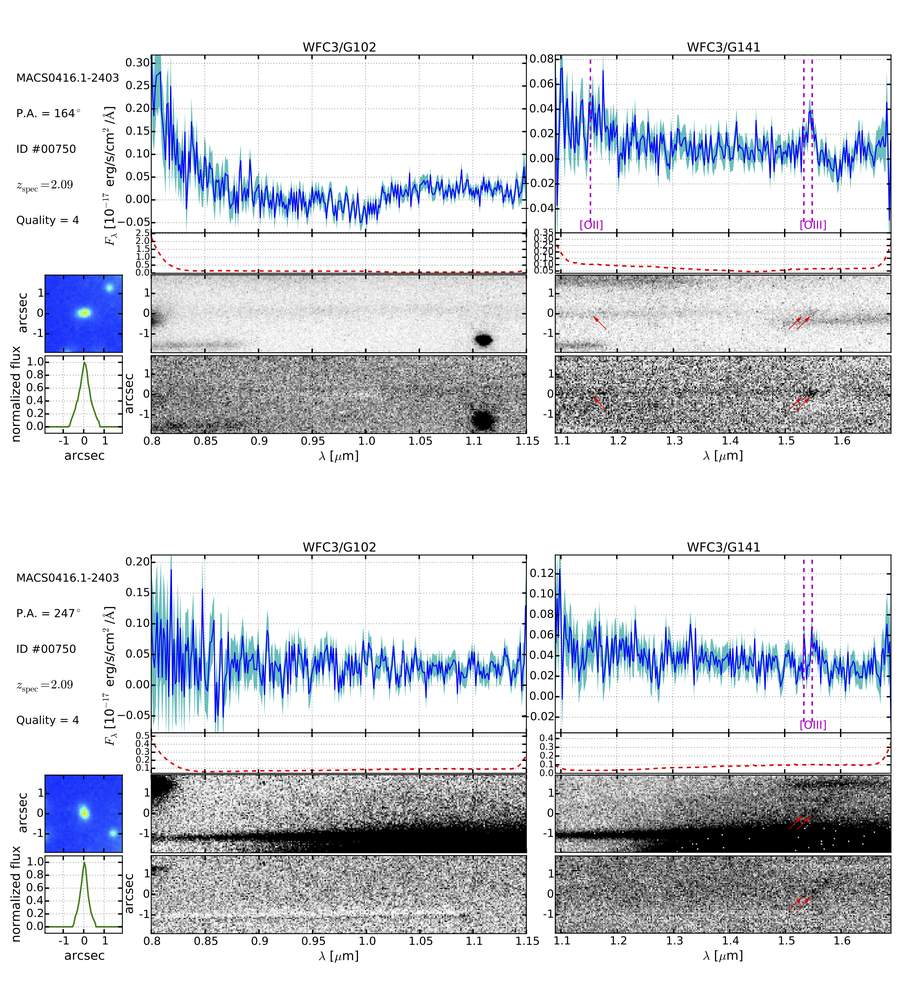}
  \caption{Same as Figure~\ref{fig:ELarc2.1}, except that object ID \#750 (arc 23.1) is shown.}
  \label{fig:ELarc23.1}
\end{figure*}

\begin{figure*}
  \centering
  \includegraphics[width=\textwidth]{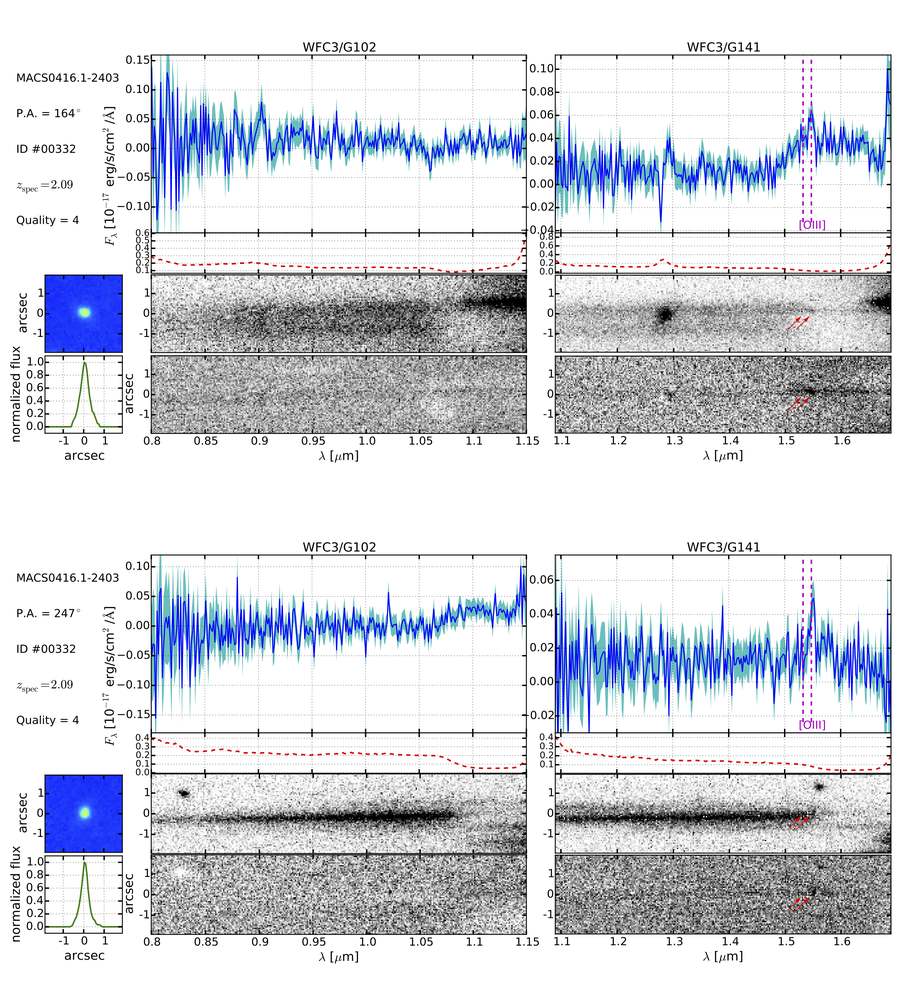}
  \caption{Same as Figure~\ref{fig:ELarc2.1}, except that object ID \#332 (arc 23.3) is shown.}
  \label{fig:ELarc23.3}
\end{figure*}

\begin{figure*}
  \centering
  \includegraphics[width=\textwidth]{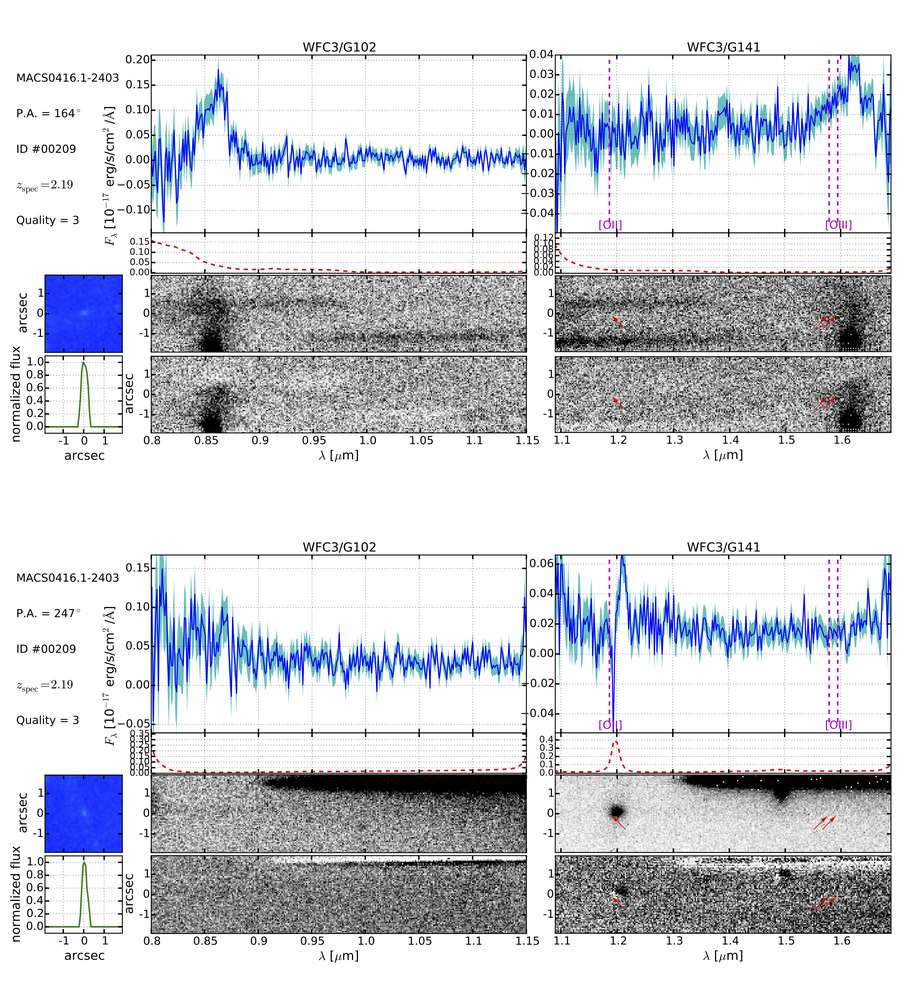}
  \caption{Same as Figure~\ref{fig:ELarc2.1}, except that object ID \#209 (arc 26.1) is shown.}
  \label{fig:ELarc26.1}
\end{figure*}

\begin{figure*}
  \centering
  \includegraphics[width=\textwidth]{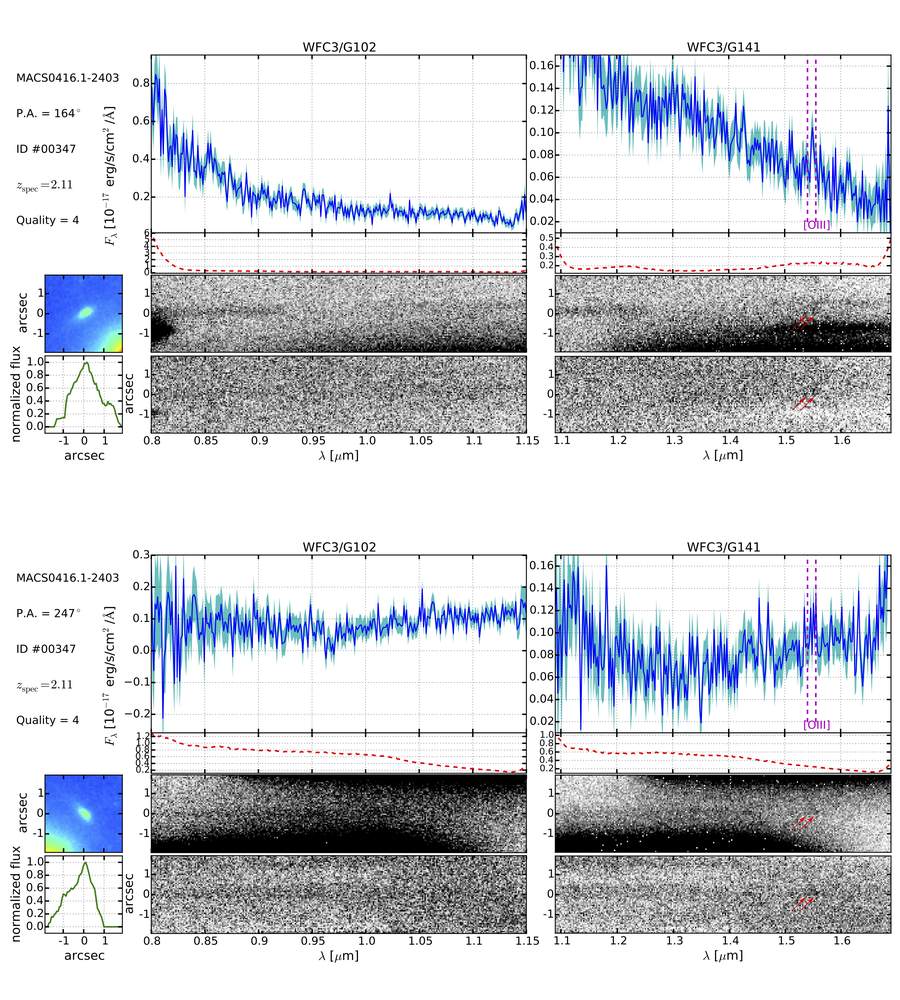}
  \caption{Same as Figure~\ref{fig:ELarc2.1}, except that object ID \#347 (arc 27.2) is shown.}
  \label{fig:ELarc27.2}
\end{figure*}

\begin{figure*}
  \centering
  \includegraphics[width=\textwidth]{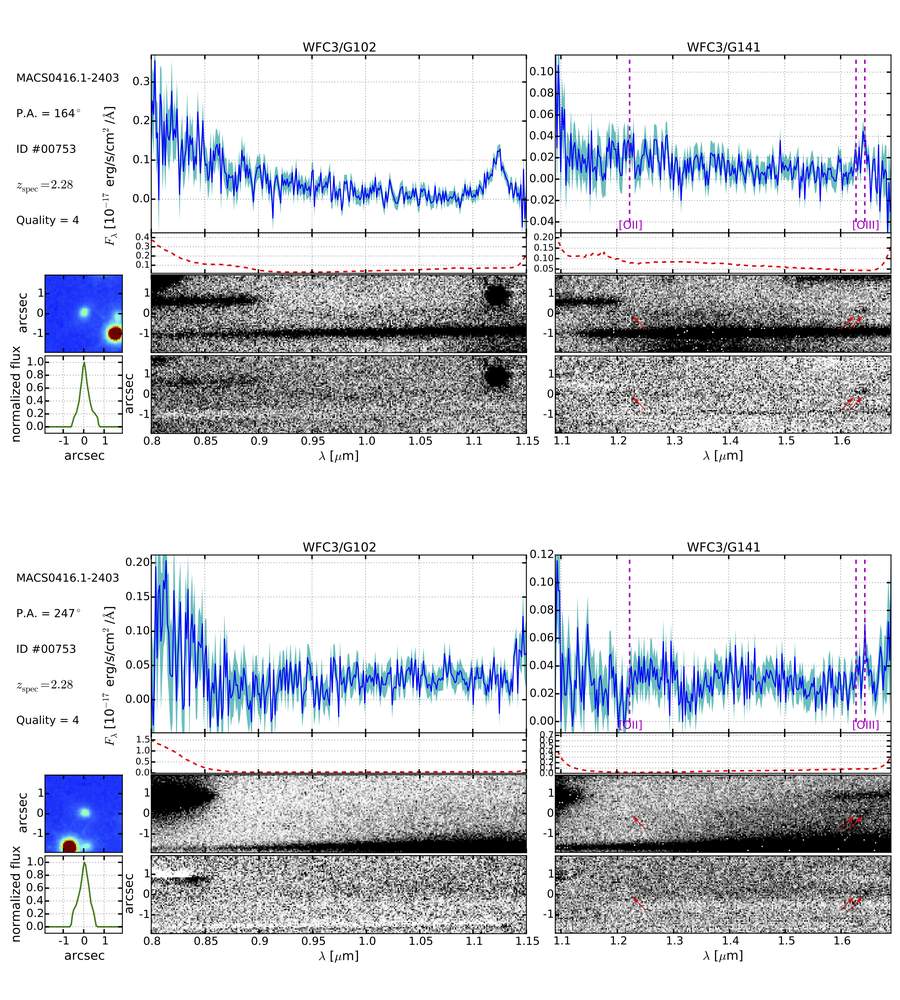}
  \caption{Same as Figure~\ref{fig:ELarc2.1}, except that object ID \#753 (arc 29.3) is shown.}
  \label{fig:ELarc29.3}
\end{figure*}

In Figures~\ref{fig:ELarc2.1}-\ref{fig:ELarc29.3} we show the GLASS spectra confirming the redshifts of known multiply imaged galaxies in \MJ{}. 
\label{sec:apB}

\clearpage

\clearpage
\clearpage
\clearpage
\clearpage
\clearpage
\clearpage
\clearpage
\clearpage
\clearpage
\clearpage
\clearpage
\clearpage
\clearpage
\clearpage
\clearpage
%
%
\clearpage
\clearpage

\clearpage
\clearpage
\clearpage
\clearpage
\clearpage

\end{appendix}

\setcounter{table}{1}
\makeatletter 
\renewcommand{\thetable}{\arabic{table}}
\clearpage
\LongTables
\newcolumntype{L}[1]{>{\raggedright\let\newline\\\arraybackslash\hspace{5pt}}m{#1}}
\tabletypesize{\scriptsize} \tabcolsep=0.08cm
\begin{deluxetable*}{L{4em}cccccccccccccc} \tablecolumns{14}
\tablewidth{0pt}
\tablecaption{Multiply lensed arc systems identified in the \MJ{} field\label{tab:mult_images}}
\tablehead{
  \colhead{ID$_\textrm{arc}$} &
  \colhead{ID$_\textrm{GLASS}$} &
  \colhead{R.A.} &
  \colhead{Dec.} &
  \colhead{Cref.} &
  \colhead{$z_{\textrm{spec}}(1)$} &
  \colhead{$z$ref(1)} &
  \colhead{$z_{\textrm{spec}}(2)$} &
  \colhead{$z$ref(2)} &
  \colhead{$z_{\textrm{grism}}$} &
  \colhead{$z_{\textrm{Bayes}}$} &
  \colhead{F140W} &
  \colhead{Magnification}\tablenotemark{1} &
  \colhead{Sample} \\
  & & (deg.) & (deg.) & & & & & & & & (mag.) & & 
}
\startdata
1.1 & 245 & 64.040750 & -24.061592 & MJ14 & 1.896 & MJ14 & 1.89 & AZ13 & \nodata & $1.9^{+0.1}_{-0.1}$ & $23.92 \pm 0.02$ & $5.40^{+0.48}_{-0.10}$ & Gold \\ 
1.2 & 244 & 64.043479 & -24.063542 & MJ14 & 1.896 & MJ14 & 1.896 & JR14 & \nodata & \nodata & $23.43 \pm 0.02$ & $51.02^{+236.36}_{+28.13}$ & Gold \\ 
1.3 & 571 & 64.047354 & -24.068669 & MJ14 & 1.896 & MJ14 & 1.896 & JR14 & \nodata & \nodata & $25.09 \pm 0.03$ & $4.03^{+0.12}_{-0.04}$ & Gold \\ 
\hline\noalign{\smallskip}
2.1 & 268 & 64.041183 & -24.061881 & MJ14 & 1.8925 & MJ14 & 1.8925 & JR14 & 1.90 & $2.4^{+0.1}_{-0.2}$ & $23.01 \pm 0.01$ & $8.26^{+0.97}_{-0.61}$ & Gold \\ 
2.2 & 248 & 64.043004 & -24.063036 & MJ14 & 1.8925 & MJ14 & 1.8925 & JR14 & 1.89 & \nodata & $23.08 \pm 0.01$ & $8.49^{+0.34}_{-0.70}$ & Gold \\ 
2.3 & 572 & 64.047475 & -24.068850 & MJ14 & 1.8925 & MJ14 & 1.8925 & JR14 & 1.90 & \nodata & $22.97 \pm 0.01$ & $4.13^{+0.29}_{-0.37}$ & Gold \\ 
\hline\noalign{\smallskip}
3.1 & 494 & 64.030783 & -24.067117 & MJ14 & 1.9885 & MJ14 & 1.9885 & JR14 & 1.99 & $2.5^{+0.1}_{-0.3}$ & $23.32 \pm 0.01$ & $4.13^{+0.03}_{-0.09}$ & Gold \\ 
3.2 & 372 & 64.035254 & -24.070981 & MJ14 & 1.9885 & MJ14 & 1.9885 & JR14 & 1.99 & \nodata & $23.13 \pm 0.01$ & $2.27^{-0.02}_{-0.06}$ & Gold \\
\enddata
\end{deluxetable*}
\clearpage
\LongTables
\tabletypesize{\scriptsize} \tabcolsep=0.11cm
\begin{deluxetable*}{ccccccccccc} \tablecolumns{11}
\tablewidth{0pt}
\tablecaption{GLASS spectroscopic results for multiply and singly lensed sources \label{tab:ELtot}}
\tablehead{
 \colhead{ID$_\textrm{GLASS}$\tablenotemark{1}} &
 \colhead{ID$_\textrm{arc}$} &
 \colhead{R.A.} &
 \colhead{Dec.} &
 \colhead{F140W} &
 \colhead{$z_\textrm{phot}$\tablenotemark{2}} &
 \colhead{$z_\textrm{grism}$} &
 \colhead{Quality} &
 \colhead{N$_\mathrm{lines}$\tablenotemark{3}} &
 \colhead{Line(s)} &
 \colhead{Magnification} \\
 \colhead{} &
 \colhead{} &
 \colhead{(deg.)} &
 \colhead{(deg.)} &
 \colhead{(mag.)} &
 \colhead{} &
 \colhead{} &
 \colhead{} &
 \colhead{} &
 \colhead{} &
 \colhead{}
}
268 & 2.1 & 64.04116 & -24.06185 & $23.01\pm0.01$ & \nodata & 1.90 & 4 & 2 &  [O\,{\sc iii}] [O\,{\sc ii}] & $8.26^{+0.97}_{-0.61}$ \\ 
248 & 2.2 & 64.04304 & -24.06305 & $23.08\pm0.01$ & $2.4^{+0.1}_{-0.1}$ & 1.89 & 4 & 2 &  [O\,{\sc ii}] [O\,{\sc iii}] & $8.49^{+0.34}_{-0.70}$ \\ 
572 & 2.3 & 64.04748 & -24.06885 & $22.97\pm0.01$ & $2.4^{+0.2}_{-0.3}$ & 1.90 & 4 & 2 &  [O\,{\sc iii}] [O\,{\sc ii}] & $4.13^{+0.29}_{-0.37}$ \\ 
494 & 3.1;4.1\tablenotemark{3} & 64.03080 & -24.06720 & $23.32\pm0.01$ & \nodata & 1.99 & 4 & 3 &  [O\,{\sc iii}] H$\beta$ [O\,{\sc ii}] & $4.13^{+0.03}_{-0.09}$ \\ 
372 & 3.2;4.2 & 64.03518 & -24.07098 & $23.13\pm0.01$ & $2.5^{+0.1}_{-0.1}$ & 1.99 & 4 & 2 &  [O\,{\sc iii}] H$\beta$ & $2.27^{-0.02}_{-0.06}$ \\ 
955 & 3.3;4.3 & 64.04185 & -24.07581 & $23.21\pm0.01$ & \nodata & 1.99 & 4 & 3 &  [O\,{\sc iii}] H$\beta$ [O\,{\sc ii}] & $3.20^{+0.12}_{-0.01}$ \\ 
519 & 5.1 & 64.03245 & -24.06848 & $21.58\pm0.00$ & \nodata & 2.09 & 4 & 2 &  [O\,{\sc iii}] H$\beta$ & $14.37^{+0.24}_{-1.09}$ \\
\enddata
\tablecomments{GLASS spectroscopic results for all mulitply-lensed and the first 10 singly-lensed galaxies. The full catalog is given in the electronic edition. $z_\textrm{phot}$ lists the best-fit and 68\% confidence limits on the photometric redshift from the ASTRODEEP photometric catalog. $z_\textrm{grism}$ is the grism spectroscopic redshift we measure in this work. The nominal uncertainty on $z_\textrm{grism}$ is $\sim0.01$ for objects with N$_\mathrm{lines}\ge1$. However, the redshift uncertainty is more variable for objects with N$_\mathrm{lines}=0$, which were confirmed by fitting the continuum grism spectra to SED templates (G. Brammer, in prep.). The full GLASS redshift PDFs are available online$^8$.}
\tablenotetext{1}{Objects with ID$_\textrm{GLASS}$ in the format 9XXXX come from an extraction based on a different alignment than the one used to extract the other GLASS IDs. }
\tablenotetext{2}{The images for which $z_\textrm{phot}$ is not shown are either contaminated or not detected in the photometric catalog.}
\tablenotetext{3}{Systems 3 and 4 are believed to be different subtructures of the same source galaxy due to the similar redshift and spatial position of the multiple images in each system.}
\tablenotetext{4}{The uncertainty in $z_\textrm{grism}$ for ID$_\textrm{arc}=12.3$ is $\sim0.02$, rather than the nominal $\sim0.01$ because the redshift was obtained by fitting its continuum emission in the grism to SED templates rather than through the identification of emission lines. }
\tablenotetext{5}{28.1 and 28.2 were detected as the same object in the segmentation map used to extract the grism data. Like 12.3, the redshifts of 28.1 and 28.2 were obtained by fitting the continuum emission in the grism to SED templates. }\end{deluxetable*}
\clearpage


\end{document}